%% file: main.tex
\documentclass[onecolumn]{article}
\usepackage{amsmath}
\usepackage{amssymb}
\usepackage[backend=biber, style=phys, eprint=true, maxnames=4]{biblatex}
\DeclareFieldFormat{titlecase}{#1}
\usepackage{braket}
\usepackage{graphicx}
\usepackage{quantikz}
\usepackage{adjustbox}
\usepackage{multirow}
\usepackage{authblk}
\usepackage{bm}
\usepackage{hyperref}

\usepackage{booktabs}
\usepackage{multirow}
\usepackage{tabularx}

\newcommand{\id}{\mathbb{I}}
\newcommand{\btheta}{\boldsymbol{\theta}}
\newcommand{\zeroket}{\ket{\mathbf{0}}}

\title{Noisy Quantum Simulation: Performance and Resource Considerations for the Tavis-Cummings and Heisenberg Models}
\author[1]{Alisa Haukisalmi}
\author[1]{Daniel Paz Ramos}
\author[1]{Matti Raasakka\footnote{corresponding author: matti.raasakka@aalto.fi}}
\author[1]{Andrea Marchesin}
\author[2]{Lauri Ylinen}
\author[1]{Ilkka Tittonen}

\affil[1]{Micro and Quantum Systems research group, Dept. of Electronics and Nanoengineering, Aalto University}
\affil[2]{Dept. of Mathematics and Statistics, University of Jyväskylä}
\date{}

\begin{document}
\sloppy
\maketitle

% Abstract
\input{abstract}

% Main text
\input{body}

% References
\clearpage
\printbibliography

% Appendix
\appendix
\include{appendix}

\end{document}

%% file: abstract.tex
\begin{abstract}

Fault-tolerant quantum computers promise the simulation of complex quantum systems beyond the reach of classical computation. In contrast, current noisy intermediate-scale quantum (NISQ) devices are constrained by hardware noise. Consequently, quantum simulation methods remain limited in their near-term applicability. Two prominent techniques addressing these challenges are zero-noise extrapolation (ZNE) and incremental structural learning (ISL). In this work, ZNE and ISL are benchmarked for simulating the Trotterized time evolution of two models: the Tavis-Cummings model (TCM) and the Heisenberg spin chain (HSC), using a classically simulated noisy hardware backend. The methods are evaluated on the basis of the accuracy of expectation values relative to noiseless simulations and their resource demands such as circuit depths and shot counts. The impact of noise on optimization routines in ISL, previously underexplored, is also investigated.

Results indicate that ISL performs more favorably in HSC systems, consistently surpassing ZNE in expectation value accuracy. Conversely, for the TCM, ISL generally yields lower accuracies despite reduced Trotter circuit depths, with weak interactions often leading to pronounced phase lags or flat expectation curves. Notably, when performing ISL optimization under noiseless conditions, the protocol is generally able to reduce dephasing errors, but average accuracies still vary on the simulated Hamiltonian. Our findings highlight the sensitivity of quantum simulation protocols to the structure of the Hamiltonian encoding system dynamics. Trends across systems suggest that ISL optimization benefits from Trotter circuits with stronger interactions, and that ansatz construction favors isotropic couplings. Moreover, although ISL introduces approximation errors, it demonstrates greater robustness than ZNE in systems with deeper Trotter circuits.

\end{abstract}

%% file: body.tex
\section{Introduction}

Quantum computing has emerged as a powerful paradigm with potential applications across many scientific and engineering domains, where it may outperform classical approaches for specific tasks~\cite{Shafique_2024,NAP_2019}. Areas of potential quantum advantage include quantum-secure cryptography~\cite{Montanaro_2016}, accelerated training of certain machine learning models~\cite{Ramezani_2020}, efficient solvers for combinatorial optimization problems~\cite{abbas_2024}, and the simulation of quantum systems~\cite{bassman_simulating_2021}. Among these, quantum simulation holds particular importance due to its direct relevance to quantum mechanics, enabling the study of many-body dynamics, quantum chemistry, and condensed matter phenomena that are classically intractable~\cite{byrnes_simulating_2006,bauer_quantum_2020,Smith_2019}.

Classical simulation of quantum dynamics becomes exponentially demanding as system size increases, due to the rapid growth of the Hilbert space. To address this limitation, Richard Feynman proposed in 1982 the engineering of quantum systems to simulate other quantum systems~\cite{feynman_simulating_1982}. He conjectured that such quantum devices could naturally and compactly represent quantum states of interest, leveraging quantum mechanical properties such as superposition and entanglement. Quantum simulation methods have since been broadly classified into analog quantum simulation, where engineered hardware directly mimics a target physical system, and digital quantum simulation (DQS), where time evolution is implemented using programmable quantum circuits composed of logic gates acting on qubits~\cite{georgescu_quantum_2014}. Devices that operate by manipulating qubits with a universal set of quantum logic gates, are correspondingly known as digital quantum computers (DQCs), and enable DQS.

In 1996, Seth Lloyd established foundational methods for DQS, demonstrating that the time evolution of local Hamiltonians can be efficiently simulated using Trotterization~\cite{lloyd_universal_1996}. This method employs the Trotter product formula~\cite{trotter_product_1959}, approximating the exponential of a complex Hamiltonian as a product of exponentials of simpler local terms. As a result, continuous time evolution is discretized into a sequence of time steps, with approximation accuracy dependent on step size. Trotterization has since become a standard approach for simulating Hamiltonian dynamics on DQCs~\cite{nielsen_quantum_2010, georgescu_quantum_2014}.

Despite the algorithmic flexibility and hardware independence offered by DQS, its practical realization on current noisy intermediate-scale quantum (NISQ) hardware is severely constrained by decoherence, limited qubit counts, and shallow achievable circuit depths~\cite{Alexeev_2021, Corcoles_2020, preskill_quantum_2018}. These limitations restrict both the system sizes and timescales that can be meaningfully explored, since increased noise directly deteriorates the fidelity of simulation results. This problem is particularly severe for Trotterized simulations, as circuit depth scales linearly with the number of time steps, making fine-grained, long-time evolution especially vulnerable to noise accumulation~\cite{cerezo_variational_2021, Burdine_2024}. Thus, techniques specifically designed to enhance simulation accuracy under realistic hardware constraints have become essential.

Quantum error mitigation (QEM) constitutes a class of such techniques, aiming to reduce noise-induced biases in estimated expectation values using post-processing methods, without requiring hardware-level error correction~\cite{Cai_2023}. Among these methods, zero-noise extrapolation (ZNE) has gained considerable attention due to its ease of implementation~\cite{Cai_2023}. In particular, ZNE estimates noise-free expectation values by extrapolating measurement statistics collected from equivalent circuits under systematically varied noise levels~\cite{temme_error_2017, endo_practical_2018, endo_hybrid_2021}. Given its simplicity, ZNE serves as a suitable baseline for evaluating more complex algorithmic approaches.

Complementarily, variational quantum simulation (VQS) methods directly adapt quantum simulation methods to the limitations of NISQ devices~\cite{endo_hybrid_2021}. VQS encompasses a subset of hybrid quantum–classical algorithms known as variational quantum algorithms (VQAs), which encode computational tasks into parameterized quantum circuits optimized via classical algorithms~\cite{cerezo_variational_2021}. The adaptive circuit structure and parameter flexibility inherent in VQAs have demonstrated resilience against noise, allowing the implementation of shallow-depth circuits tailored specifically to hardware limitations~\cite{Fontana_2021, cerezo_variational_2021, Khanal_2023, Kolotouros_2024}. In VQS, trial quantum states are optimized iteratively to approximate a discrete-step time evolution of quantum systems. Prominent approaches include variational approximations to Trotterized dynamics~\cite{cirstoiu_variational_2020, jaderberg_minimum_2020, bharti_noisy_2022}, differential equation-based methods~\cite{li_efficient_2017, bharti_noisy_2022}, and extensions suitable for mixed states~\cite{yuan_theory_2019} or open-system dynamics~\cite{endo_hybrid_2021}.

This work specifically investigates incremental structural learning (ISL)~\cite{jaderberg_minimum_2020}, a VQS approach that iteratively transforms Trotterized quantum circuits into shallower, optimized approximations. The method relies on approximate recompilation, wherein parameterized circuits are constructed to maximize fidelity with respect to an input circuit. While ISL has been studied individually~\cite{jaderberg_minimum_2020, fitzpatrick_evaluating_2021}, and variational methods have been broadly compared to standard Trotterization~\cite{miessen_quantum_2021}, a direct comparison between ZNE and ISL under realistic hardware noise remains absent. Furthermore, the robustness of approximate recompilation routines in noisy environments is not yet well understood.

To fill these gaps, this study presents a comparative analysis of ISL and ZNE applied to the Trotterized time evolution of Hamiltonians with different interaction structures. The protocols are evaluated on emulated NISQ hardware and compared against both ideal Trotterized evolution and ISL optimization performed on noiseless hardware. Two representative quantum simulation models are considered: the Tavis–Cummings model (TCM)~\cite{tavis_exact_1968} and the Heisenberg spin chain (HSC)~\cite{heisenberg_model_1928}. These models differ significantly in their coupling pattern and time-evolution dynamics, enabling us to evaluate how algorithmic performance depends on Hamiltonian characteristics.

Our analysis focuses on two main metrics: accuracy, defined as deviations of expectation values from noiseless benchmarks, and resource requirements, measured by circuit depth and the number of circuit executions. By evaluating both protocols across representative systems of varying sizes and associated noise levels, this study seeks to clarify their respective strengths, limitations, and overall practicality for quantum simulation within the NISQ regime.

The remainder of this paper is organized as follows. Section~\ref{sec:methods} outlines the methods employed for DQS, details the implementation of ZNE and ISL, and describes the quantum simulation models alongside their mapping to NISQ hardware. Section~\ref{sec:results} presents numerical comparisons of ISL and ZNE in TCM and HSC systems, highlighting their relative performance under different simulation scenarios. Finally, Section~\ref{sec:conclusions} summarizes the main findings and offers concluding insights along with suggestions for future research directions.

\section{Methods} \label{sec:methods}
This section describes the computational and theoretical methods employed in this study. First, the methods used for DQS are introduced, including the standard Trotterization procedure, with the application of error mitigation through ZNE, and through a variational approach following the ISL protocol. Next the two considered physical models, the TCM and the HSC, are outlined, alongside their respective qubit encodings for DQS. The simulation framework is then detailed, including the choice of observable used as an accuracy metric, the parameters of the emulated NISQ hardware, the circuit execution strategy, and the algorithm hyperparameters employed. Finally, the estimation of resource requirements is discussed, focusing specifically on the overhead introduced by ZNE and ISL over standard Trotterization.

\subsection{Trotterized Time Evolution for DQS}
\label{sec:trotterized-time-evolution}

Given a Hamiltonian $H$ of interest acting on a system of qubits and an initial state $\ket{\varphi_0}$, the goal of DQS is to find the state of the system at a later time $t>0$, using a DQC. 

For a time-independent Hamiltonian, the unitary evolution of the state is given as $\ket{\varphi(t)} = U(t) \ket{\varphi_0}$, where $U(t) = e^{-iHt}$ is the time evolution operator\footnote{Natural units are used, where $\hbar = 1$.}. To simulate this evolution on a DQC, the operator $U(t)$ must be decomposed into a sequence of elementary quantum gates that can be directly implemented on the chosen quantum hardware. However, this decomposition is generally intractable for Hamiltonians of large dimensionality due to the exponentially scaling number of terms. Therefore, approximate methods are often required~\cite{nielsen_quantum_2010}.

The Trotter product formula (also known as Suzuki-Trotter decomposition) provides a framework for approximating exponentials of operator sums~\cite{ Kluber_2025}. Following the formulation of~\cite{nielsen_quantum_2010}, it states that for any two Hermitian operators $A$ and $B$, and any $t\in\mathbb{R}$, the following identity holds: \begin{equation}\label{eq:trotter-product}
    \lim_{n\rightarrow\infty}{(e^{iAt/n}e^{iBt/n})^n}=e^{i(A+B)t}.
\end{equation} By defining a step size $\Delta t \equiv t / N_T$ for some $N_T \in \mathbb{Z}_+$, one obtains the first-order Trotter approximation:
\begin{equation}
    e^{i(A+B)\Delta t}=e^{iA\Delta t}e^{iB\Delta t}+\mathcal{O}(\Delta t^2).
\end{equation} This result naturally generalizes to sums of more than two terms. In particular, for a time-independent Hamiltonian $H = \sum_{m=1}^M H_m$, the time evolution operator can be approximated as:\begin{equation}\label{eq:trotter-evolution}
    U(t)= \left( \prod_{m=1}^M e^{-iH_m\Delta t} \right)^{N_T} + \mathcal{O}(\Delta t^2).
\end{equation} Defining the Trotter step as $T = \prod_m e^{-iH_m \Delta t}$, the time evolution operator can then be approximated by repeated applications of $T$, i.e., $U(n \Delta t) \approx T^n$ for $n = 0, 1, \ldots, N_T$. In DQS, $T$ is represented as a sequence of quantum gates that must be applied repeatedly for each time step, leading to deeper circuits with increasing time step, which can be challenging to evaluate on NISQ devices due to noise accumulation.

This approximation is particularly efficient for local Hamiltonians, where each term $H_m$ acts on at most a constant number of qubits. In such cases, the individual exponentials $e^{-iH_m \Delta t}$ are easier to compute or approximate on DQC hardware. The Hamiltonian terms $\{H_m\}$ are typically mapped to Pauli strings (i.e., tensor products of single-qubit Pauli operators, including the single-qubit identity operator) \cite{steudtner_fermion--qubit_2018}, which admit well-studied decompositions into quantum gate sequences~\cite{sarkar_2024, Dion_2024}.

Higher-order Trotter-Suzuki decompositions can also be employed to achieve more accurate approximations, at the cost of additional terms and increased circuit depth~\cite{nielsen_quantum_2010, Yang_2022}. 

It is worth noting that if all terms in the Hamiltonian commute, meaning $[H_m, H_l] = 0$ for all $m$ and $l$, then the time evolution operator factorizes exactly as $U(t)=\prod_{m=1}^M{e^{-iH_mt}}$, and no approximation is required. However, this condition rarely holds in practice~\cite{nielsen_quantum_2010}.

\subsection{QEM with ZNE}
For an observable $O$ of interest, QEM aims to reduce the deviation between the expectation value $\langle O \rangle$ estimated on noisy hardware and the corresponding ideal, noiseless value $\bar{O}$. Specifically, QEM methods seek to minimize the mean squared error $(\langle O\rangle - \bar{O})^2$~\cite{Cai_2023}. These techniques involve post-processing the outputs from ensembles of circuit runs to improve aggregate results, rather than reducing noise on individual runs. QEM encompasses a broad range of approaches, including probabilistic error cancellation~\cite{Zhang_2020,van_den_berg_probabilistic_2023}, readout error mitigation~\cite{Maciejewski_2020, Bravyi_2021}, Clifford data regression~\cite{Czarnik_2021, Lowe_2021, Perez_2024}, and tensor network methods \cite{filippov2023}, among others. QEM is particularly relevant for near-term devices, with several studies demonstrating measurable improvements over unmitigated execution on NISQ hardware~\cite{endo_practical_2018, Song_2019, Larose_2022, Russo_2023, Moradi_2023}.

In this work, ZNE~\cite{li_efficient_2017, temme_error_2017} is implemented to mitigate errors in circuits for Trotterized time evolution. ZNE operates by executing the same quantum circuit under systematically scaled noise levels and fitting the resulting measurement outcomes to an analytical model, which is then extrapolated to the zero-noise limit. The local gate folding technique is adopted~\cite{giurgica-tiron_digital_2020}, in which individual gates $U$ in the input quantum circuit are replaced by repeated applications and their inverses, $U \mapsto U(U^\dagger U)^l$, for some positive integer $l$. This amplifies the effect of noise by increasing the number of physical operations, without altering the underlying computation. As applying folding to the entire circuit scales the effective depth by odd integers, final folding can be applied to a subset of $s$ gates in the circuit to obtain a finer-grained noise scaling factor. Thus, for an input circuit of depth $d$, the total number of layers in the folded circuit increase by a factor of $\lambda =2(l+s/d)+1$. 

In this investigation, the resulting noisy expectation values $\langle O \rangle(\lambda_i)$ are fitted using an exponential ansatz,
\begin{equation}
   \langle O \rangle(\lambda) = a+b e^{-c\lambda},
\end{equation} where $c>0$, and $a \in \mathbb{R}$ represents the noisy asymptote value of the observable for $\lambda\rightarrow\infty$. This model is motivated by the assumption of exponential decay in fidelity with respect to circuit depth under weak stochastic noise~\cite{endo_practical_2018}. The fit is performed via linear regression on the logarithmically transformed values: \begin{equation}
    \log \left( \langle O \rangle(\lambda) - a \right) = \log b - c \lambda,
\end{equation}
with $a$ treated as a hyperparameter defined a priori based on knowledge of the observable. This approach is favored over polynomial fits such as Richardson extrapolation~\cite{temme_error_2017, li_efficient_2017}, as the assumptions underlying a low-order Taylor expansion may break down in deeper circuits or for small noise levels~\cite{endo_hybrid_2021}.

\subsection{ISL Protocol for VQS}\label{subsec:ISL-for-VQS}
ISL recompilation is a protocol designed to recursively construct hardware-efficient variational circuits that approximate the action of a target sequence of deep (or otherwise hardware-inefficient) circuits~\cite{jaderberg_minimum_2020}. Originally developed as a VQS method, ISL is particularly well-suited for Trotterized DQS.

In its standard VQS formulation, ISL constructs a sequence of variational circuits $\{V_n\}_{n=1}^{N_T}$ to approximate the outputs of Trotterized circuits $\{T^n\}_{n=1}^{N_T}$, with each $V_{n+1}$ trained to match the action of $T$ on the preceding variational state. Assuming the circuit acts on the all-zeros state $\ket{\bm{0}} = \ket{0}^{\otimes N_q}$, a successful approximation is expressed as $V_{n+1}\ket{\bm{0}} \approx T V_n \ket{\bm{0}}$, where $V_0 \equiv U_{\mathrm{st}}$ prepares the chosen initial state acting on $\ket{\bm{0}}$. This recursive approach exploits the local structure of Trotterization by constructing each $V_n\equiv V_n(\bm\theta)$ as a sequence of layers of 2-local subcircuits (called thinly-dressed CNOTs~\cite{jaderberg_minimum_2020}), consisting of a CNOT gate surrounded by single-qubit rotation gates, with both placement and parameters optimized using learning-based techniques. 

In particular, the parameters $\bm \theta$ are optimized at each step $n=1,\ldots,N_T$ to minimize the cost function \begin{equation}
C(\btheta) = 1 - |\braket{\mathbf{0}| V_n(\btheta)^\dagger T V_{n-1} | \mathbf{0}}|^2.
\end{equation}  In this study, ISL is implemented under realistic conditions, where both the Trotterized target circuit $T$ and the variational ansätze $V_n$ are executed on simulated noisy hardware. Consequently, the cost function is estimated from noisy measurements. The impact of gate errors and shot noise\footnote{Shot noise refers to the statistical fluctuations arising from the finite number of circuit executions (shots) used to estimate expectation values.} on the accuracy of the optimization is therefore central to the investigation. This implementation deviates from the standard formulation of VQS protocols and enables a realistic assessment of algorithmic robustness in the NISQ setting.

A crucial step in ISL optimization is the selection of qubit pairs for constructing the parameterized circuit, known as the ansatz. At each layer, reduced density matrices for physically connected qubit pairs are estimated via quantum state tomography (QST)~\cite{altepeter_4_2004}, allowing identification of the pair with the highest local entanglement measure. This pair is then selected for the next thinly-dressed CNOT layer; if all entanglement measures are zero, heuristic rules based on deviations in local expectation values are applied. While QST enables more effective circuit design and can accelerate convergence, it also introduces a non-negligible measurement overhead (see Section~\ref{sec:resource-estimation} for a quantitative analysis).

ISL recompilation proceeds as follows:
\begin{enumerate}
    \item \textbf{Initialization:}
    Define the initial state preparation circuit $U_\mathrm{st}$, and the Trotter step circuit $T$ governing time evolution. 
 
    \item \textbf{Recursive recompilation:} For each time step $n = 1, \ldots, N_T$:
    \begin{enumerate}
        \item Append a thinly-dressed CNOT layer to a qubit pair in the parameterized circuit $V_n(\bm\theta)$.
        \item Optimize the parameters in the new layer to minimize the cost function $C(\btheta)$ using the Rotoselect algorithm~\cite{ostaszewski_structure_2021} until convergence. Then, employ the following transpilation strategies\footnote{In this work, this subprotocol is referred to as ISL transpilation, to avoid confusion with (backend) transpilation of quantum circuits to a specific hardware.} to simplify the circuit: perform the Rotosolve\footnote{Rotosolve minimizes the angle for each rotation gate around a fixed axis ($x$, $y$, or $z$), whereas Rotoselect performs all three and chooses the optimal configuration.} algorithm~\cite{ostaszewski_structure_2021} once on all parameters,
        merge adjacent rotation gates acting around the same axis, remove pairs of consecutive CNOTs, and remove rotation gates with angles below a threshold $\theta_\mathrm{th}$ (fixed to $10^{-3}$ in this investigation). 
        \item Assess convergence: If the cost drops below a user-specified cost threshold $C_\mathrm{suff}$, or if improvement saturates or resource constraints are reached, finalize $V_n$. Otherwise, repeat steps (a) and (b).
    \end{enumerate}
    \item \textbf{Output:} After $N_T$ steps, obtain a sequence of shallow circuits $\{V_n\}$ that collectively approximate the Trotterized evolution.
\end{enumerate}
Further technical details, including a graphical illustration of the ISL workflow, are provided in Appendix~\ref{app:isl}.

In the literature, ISL has been applied to VQS for a variety of systems, including the Fermi--Hubbard model via dynamical mean field theory~\cite{jaderberg_minimum_2020}, the TCM and Jaynes--Cummings model~\cite{fitzpatrick_evaluating_2021}, and electron-phonon dynamics~\cite{jaderberg_recompilation-enhanced_2022}. Outside of VQS, ISL has also been employed to quantum neural networks in order to optimize the number of two-qubit gates needed~\cite{jaderberg_quantum_2022}. While ISL can substantially reduce circuit depth and thereby improve robustness to noise, this benefit comes at the cost of introducing classical optimization overhead and a greater number of quantum circuit executions required for optimizing circuit parameters and gate placement~\cite{fitzpatrick_evaluating_2021}.

\subsection{Tavis--Cummings model}
\label{sec:tcm-model}

Various spin-boson models have been used as model Hamiltonians to demonstrate and investigate specific instances of DQS~\cite{di_paolo_variational_2020, fitzpatrick_evaluating_2021, miessen_quantum_2021, li_efficient_2023}. Among these, the TCM is considered here as the $N$-atom generalization of the Jaynes--Cummings model~\cite{jaynes_comparison_1963}. The TCM describes a system of $N$ two-level atoms coupled equally to a quantized electromagnetic field. Its Hamiltonian is given by~\cite{larson_jaynescummings_2021}
\begin{equation} \label{eq:tavis-cummings1}
    \mathcal{H}_{\text{TC}} = \omega a^{\dag}a + \Omega \sum_{i=1}^N\frac{\sigma_z^{(i)}}{2} + g \sum_{i=1}^N \left( a\sigma_+^{(i)} + a^{\dag}\sigma_-^{(i)} \right).
\end{equation}
Here, $\omega$ is the frequency of the field mode, $\Omega$ is the separation between the atom energy levels, and $g$ is the coupling constant. The operator $a^{(\dag)}$ denotes the annihilation (creation) operator for the bosonic mode, and the Pauli operators $\{ \sigma^{(i)}_{\alpha} \mid \alpha \in \{x,y,z\} \}$ act on the $i$th atom. The spin raising ($+$) and lowering ($-$) operators are defined as $\sigma^{(i)}_{\pm} \equiv \left( \sigma^{(i)}_x \pm i\sigma^{(i)}_y  \right)/2$. Since each term in Eq.~\eqref{eq:tavis-cummings1} contains an equal number of creation/spin-raising and annihilation/spin-lowering operators, the total number of excitations is conserved throughout time evolution~\cite{Jorge_2022}. 

Throughout this work, all simulations are performed under the resonance condition $\Omega=\omega$. The initial state (before mapping to the qubit basis) is defined as $\ket{\psi_0} \equiv \ket{1} \bigotimes_{i=1}^N \ket{g}_i$, where the field mode contains a single photon and each atom is in its ground state.

\subsubsection{Tavis--Cummings qubit encoding}
\label{subsec:tcm-qubit-encoding}
To simulate the TCM on a DQC, both the Hamiltonian and initial state must be mapped to the qubit formalism. 

Each two-level atom in the system is represented by a qubit: the ground state $\ket{g}_i$ and excited state $\ket{e}_i$ of the $i$th atom are mapped to the computational basis states $\ket{1}_i$ and $\ket{0}_i$ of the $i$th qubit, respectively. In this encoding, the spin operators $\sigma_\alpha^{(i)}$ (for $\alpha\in\{x,y,z\}$) correspond directly to the standard Pauli operators acting on the $i$th qubit.

The bosonic field mode is described by Fock states $\ket{n}$, where $n$ is the photon number. As the Hilbert space of the field is infinite-dimensional, it must be truncated to $D$ states, retaining only $\ket{n}$ with $0\le n < D$~\cite{sawaya_resource-efficient_2020}. In this study, because the initial state contains exactly one photon, the dynamics are restricted to the single-excitation subspace spanned by the Fock states $\ket{0}$ and $\ket{1}$. Consequently, the field mode can be efficiently encoded in a single qubit, with $\ket{0}$ (zero photons) mapped to the qubit state $\ket{0}_0$ and $\ket{1}$ (one photon) mapped to $\ket{1}_0$.

The resulting qubit system consists of $N_q = N + 1$ qubits: qubit $0$ encodes the field mode, and qubits $1$ to $N$ represent the atoms. The qubit-encoded TCM Hamiltonian takes the form
\begin{equation} \label{eq:qubit-tcm}
    H_{\text{TC}} = \frac{\omega}{2}\left( \id - \sigma_z^{(0)} \right) + \frac{\omega}{2}\sum_{i=1}^N \sigma_z^{(i)} \\
    + \frac{g}{2}\sum_{i=1}^N \left( \sigma_x^{(0)}\sigma_x^{(i)} - \sigma_y^{(0)}\sigma_y^{(i)} \right).
\end{equation}

The corresponding encoded initial state is $\ket{\varphi_0} = \ket{\bm1}\equiv \ket{1}^{\otimes N_q}$, which can be prepared using the state preparation circuit $U_{\mathrm{st}} \equiv \bigotimes_{i=0}^{N_q-1} \sigma_x^{(i)}$ acting on the all-zeros state $\ket{\bm{0}} = \ket{0}^{\otimes N_q}$.

To perform Trotterized time evolution under this Hamiltonian on quantum hardware, the evolution operator must be decomposed into implementable quantum gates. For the TCM, a single Trotter step can be expressed as
\begin{equation} \label{eq:trotter-tcm-step}
    T_{\text{TC}} = e^{-i \omega \Delta t/2}
    \left( R_z^{(0)}(-\omega \Delta t) \prod_{j=1}^N R_z^{(j)}(\omega \Delta t) \right)
    \left (\prod_{j=1}^N R_{xx}^{(0,j)}(g \Delta t) R_{yy}^{(0,j)}(-g \Delta t) \right),
\end{equation}
where $R_{\alpha \alpha}^{(j,k)} (\theta) \equiv \exp(-i\frac{\theta}{2}\sigma_{\alpha}^{(j)}\otimes\sigma_{\alpha}^{(k)})$. The global phase factor $e^{-i\omega \Delta t/2}$ does not affect measurement outcomes and can be omitted in practical circuit implementations.

\subsection{Heisenberg Spin Chain}
\label{sec:heisenberg-model}
The Quantum Heisenberg Model (QHM) describes nearest-neighbor spin-to-spin interactions in lattices of spin-$\frac{1}{2}$ systems~\cite{heisenberg_model_1928}. The one-dimensional version of the QHM, often referred to as the Heisenberg Spin Chain (HSC)~\cite{candu2013spin_chains}, consists of a lattice of $L$ sites arranged linearly, with each site hosting a spin-$\frac{1}{2}$ particle that interacts with its two nearest neighbors. The system is governed by the Hamiltonian
\begin{equation} \label{eq:hsc-Hamiltonian}
\mathcal{H}_{\text{HSC}}=-\frac{1}{2}\sum_{j=1}^{L}{(J_x\sigma_x^{(j)}\sigma_x^{(j+1)}+J_y\sigma_y^{(j)}\sigma_y^{(j+1)}+J_z\sigma_z^{(j)}\sigma_z^{(j+1)}+\bm{h}\cdot\bm{\sigma}^{(j)})}.
\end{equation} 
Based on this formulation, $\bm{J}\equiv(J_x,J_y,J_z)\in \mathbb{R}^3$ is defined as the vector coupling constant between adjacent lattice sites~\cite{jiang2022}. The individual components specify the interaction strengths along the three canonical axes; each term $J_\alpha \sigma_\alpha^{(j)}\sigma_\alpha^{(j+1)}$ couples the $\alpha$-components of neighboring spins.

The vector $\bm{h}\equiv(h_x, h_y, h_z)\in\mathbb{R}^3$ represents a uniform external magnetic field applied to all spins. Each component $h_\alpha$ specifies the field strength along the corresponding axis, and the term $\bm{h}\cdot\bm{\sigma}^{(j)}$ describes the interaction of the $j$th spin with the field. The notation $\bm{\sigma} ^{(j)}\equiv{(\sigma_x^{(j)}, \sigma_y^{(j)}, \sigma_z^{(j)})}$ is used for brevity.

To properly define the term $\sigma_\alpha^{(L+1)}$ in the Hamiltonian, boundary conditions must be specified. Here, periodic boundary conditions are employed, such that $\sigma_\alpha^{(L+1)} \equiv \sigma_\alpha^{(1)}$ for all $\alpha\in\{x,y,z\}$.

In order to perform DQS, the system is initialized in the state $|\psi_0\rangle\equiv\ket{\downarrow_z}^{\otimes L}$, where $\ket{\downarrow_z}_i$ represents the spin-down eigenstate along the $z$-axis for the $i$th spin.

\subsubsection{Heisenberg Spin Chain Qubit Encoding}
Each site in the Heisenberg spin chain hosts a spin-$\frac{1}{2}$ particle, which can be naturally mapped to a qubit. The mapping chosen here assigns the spin-down state along the $z$-axis, $\ket{\downarrow_z}_i$, to the computational basis state $\ket{1}_i$, and the spin-up state, $\ket{\uparrow_z}_i$, to $\ket{0}_i$. Under this encoding, the spin operators $\sigma_\alpha^{(i)}$ correspond directly to the standard Pauli gates acting on the $i$th qubit.

With this mapping, the Hamiltonian $\mathcal{H}_\mathrm{HSC}$, which is already expressed in terms of Pauli operators, translates directly to a qubit Hamiltonian without additional transformation required. Therefore, the resulting qubit system consists of $N_q=L$ qubits. This direct correspondence allows for efficient quantum circuit implementations, particularly when applying Trotterization.

The Trotterized time evolution operator for the HSC reads
\begin{equation} \label{eq:trotter-hm-step}
T_{\text{HSC}} = \Biggl(\prod^L_{j=1}{\prod_{\alpha \in P}{R_{\alpha\alpha}^{(j,j+1)}\Bigl(-\frac{1}{2}J_\alpha \Delta t\Bigr)} \Biggr) \Biggl( \prod_{i=1}^L{\prod_{\beta \in P}{R^{(i)}_\beta \Bigl(-\frac{1}{2} h_\beta \Delta t \Bigr)}} \Biggr)},
\end{equation} where $P\equiv\{x,y,z\}$ is used as a shorthand for the canonical axes. 

The initial state for all simulations is taken to be fully spin-down along the $z$-axis, which corresponds in the qubit encoding to $\ket{\varphi_0}=\ket{\bm1}$, analogous to the TCM encoding.

\subsection{Simulations}
\label{sec:simulations}
The numerical experiments in this work are designed to benchmark the performance of standard Trotterization, as well Trotterization enhanced with ZNE and ISL, under realistic noise conditions. Each simulation begins with a unitary state preparation circuit $U_\mathrm{st}$ (as defined in Section~\ref{subsec:tcm-qubit-encoding}), which maps the vacuum state to the desired initial state $\ket{\varphi_0}$.

The time evolution is performed in $N_T=40$ discrete steps of size $\Delta t=0.01$. After $n$ time steps, the system state is represented by the density operator $\rho_n$, corresponding to the noisy execution of $T^n U_\mathrm{st}\ket{\bm0}$, or its ISL-approximate version. 

The main observable tracked throughout the simulations is the overlap between the evolved state and the noiseless initial state, with an expectation value given by
\begin{equation}
    \mathbb{P}\left( \ket{\varphi_0} \right) \equiv \bra{\varphi_0}\rho_n\ket{\varphi_0}.
\end{equation} This observable is accessible for all methods and enables direct comparison, as ZNE mitigates errors in expectation values rather than correcting state vectors directly. The time evolution under noiseless Trotterization is set as the baseline for optimal performance. Thus, Trotter error is not considered, and any observed deviations from the noiseless evolution arise from a combination of hardware noise, algorithmic variability (ZNE and ISL), and approximation errors introduced during ISL recompilation.

\begin{figure}[htb]
    \centering
    \includegraphics[width=0.65\linewidth]{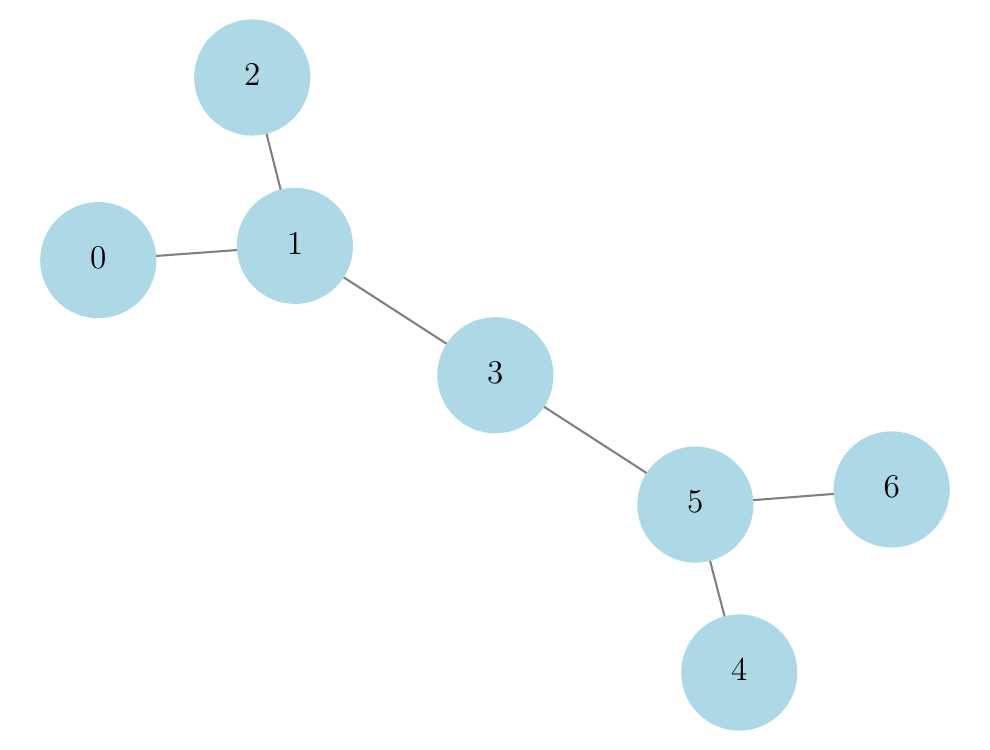}
    \caption{Qubit connectivity graph for the \texttt{ibm\_nairobi} backend. Nodes represent qubits and edges indicate pairs that can perform native $\mathrm{CNOT}$ gates without $\mathrm{SWAP}$ routing.}
    \label{fig:backend-coupling-map}
\end{figure}

All quantum simulations are performed classically with Qiskit~\cite{Qiskit}, with the seven-qubit \texttt{ibm\_nairobi} FakeBackend to emulate the noise model and characteristics of a contemporary NISQ device\footnote{While the IBM Nairobi processor was publicly available at the time of selecting this noise model, it was retired in November 2023~\cite{ibm_documentation_2023}.}~\cite{ibm_quantum_2023}. The available basis gates are $\{ \id, R_z, \sqrt{\sigma_x}, \sigma_x, \mathrm{CNOT} \}$, and the qubit connectivity graph is shown in Fig.~\ref{fig:backend-coupling-map}. This topology determines which qubits can perform native two-qubit gates without SWAP routing, directly influencing circuit depth and gate counts. Backend transpilation routines were run with maximal optimization to minimize single- and two-qubit gate counts, thus reducing depth and mitigating cumulative noise ~\cite{ibm_quantum_transpiler}.

ZNE is implemented using the Mitiq library~\cite{mitiq}, and the code for ISL is available on GitHub~\cite{islcode}, as an adaptation of the original implementation~\cite{jaderberg_minimum_2020}. All code for the simulations presented, including TCM and HSC implementations, the noise model, and applying quantum simulation libraries, is publicly available (see~\cite{ahaukis}). Simulations were executed on the Triton computing cluster provided by the Aalto University School of Science ``Science-IT" project~\cite{triton_cluster}.

Each protocol is tested on instances of both the TCM and HSC for a range of parameters. System sizes  with $N_q=2,3,4$ qubits are considered. The number of shots per circuit evaluation, $k$, is varied over $k=1\,024, 4\,096, 8\,192,$ and $16\,384$ shots, for all methods. This allows an assessment of the influence of shot noise on protocol performance, under comparable circuit evaluation resolution across all methods. However, the total number of required circuit executions differs between methods, due to the distinct number of circuits evaluated per time step (see Section~\ref{subsubsec:circuit-depth-shot-cost}). 

For ZNE, noise mitigation is carried out using three different noise amplification rates, $\lambda=1, 2, 3$. For each $\lambda$, the observable is estimated by averaging results from $n_\mathrm{avg}=10$ independent circuit evaluations, to reduce statistical fluctuations in the zero-noise predictions. Extrapolation is performed by fitting an exponential model, with the asymptote parameter $a$ fixed at $1/N_q^2$. This value reflects the probability of measuring the initial state in a fully randomized outcome, providing a physically motivated constraint for the fit. 

Some additional considerations apply to ISL. An inadequately chosen cost threshold $C_\mathrm{suff}$ can result in insufficient convergence if set too high, or excessive computational effort if too low. To address this, two thresholds are considered: $C_\mathrm{suff}=10^{-2}$ (as considered in~\cite{jaderberg_recompilation-enhanced_2022, jaderberg_minimum_2020}) and $C_\mathrm{suff}=10^{-4}$ for stricter optimization. Furthermore, due to the recursive structure of ISL and variability in the optimization (especially under high shot noise), the approximate states produced at later time steps can vary significantly across independent runs. To assess robustness, six independent ISL runs are performed under noisy conditions, for each combination of system instance, cost threshold, and shot count. Unless otherwise specified, all reported ISL results correspond to the median performance across these independent runs.

To evaluate protocol accuracy, three complementary metrics are used. First, the median absolute error $\tilde{\varepsilon}$ in $\mathbb{P}(\ket{\varphi_0})$ for a given protocol, with respect to noiseless Trotterized evolution. Second, the fidelity of the simulated state with respect to the noiseless Trotterized state is considered. Here, the density matrix is obtained directly from the underlying Qiskit simulation at each time step. For a density operator $\rho_n$ resulting from noisy simulation at time step $t=n\Delta t$, and the corresponding noiseless state $\ket{\varphi_n}=T^nU_\mathrm{st}\ket{\bm0}$, the fidelity is defined as \begin{equation}
    F(\rho_n, \ket{\varphi_n})=\bra{\varphi_n}\rho_n\ket{\varphi_n}.
\end{equation}
This provides a direct measure of how closely the simulated state follows the target noiseless trajectory. Importantly, because the density matrix is computed directly from the simulator rather than reconstructed through QST, this metric cannot be meaningfully extended to ZNE. QEM methods reduce errors only at the level of aggregated measurement outcomes, not across individual circuit executions. Consequently, this metric is applied only to plain Trotterization and ISL.

 Finally, dephasing in ISL is quantified by extracting the phase shift $\tau_\mathrm{lag}$ between the first major peak of the noiseless probability curve and the corresponding feature in the ISL evolution. Negative dephasing indicates the ISL curve is lagging behind the noiseless evolution, while positive curves indicate a lead. Peaks in the noiseless evolution are detected using the \texttt{find\_peaks} function from SciPy~\cite{Virtanen_2020}, with a height threshold of 0.3. The initial time step (a trivial local maximum for all systems) is excluded. Within a search window of five steps from the noiseless peak, the ISL peak is identified using the same procedure; if no clear peak exists, the maximum within the window is taken instead. This approach ensures consistent peak identification across parameter regimes. Only the first oscillation is considered, since weakly coupled system ($g$ or $\|\bm J\|$ small) may exhibit limited oscillations within the simulated time window, and later peaks are more strongly degraded by accumulated noise.

\subsection{Estimating Required Circuit Executions} \label{sec:resource-estimation}

In the NISQ era, quantum circuits must remain shallow enough to mitigate decoherence and gate errors, as increasing circuit depth rapidly degrades result fidelity and limits the potential for quantum advantage~\cite{De_Palma_2023}. However, even circuits with low depth can become impractical if they require a prohibitively large number of executions. This issue arises prominently in both VQAs~\cite{cerezo_variational_2021} and QEM techniques~\cite{Takagi_2022}, which rely on repeated circuit executions per input circuit. If the total number of circuit executions scales unfavorably, such as exponentially with the number of qubits, any theoretical quantum speedup may be negated by excessive runtime. 

Consequently, the total number of circuit executions required to perform time evolution over a fixed period, $c_\mathrm{tot}$---here referring to a fixed number of Trotter steps---is presented for each algorithm. This metric is critical for assessing the practical feasibility of quantum algorithms, alongside their accuracy.

\subsubsection{Trotterization and ZNE Circuit Execution Counts}
For plain Trotterization, circuit executions are only required to estimate the target expectation value for each Trotter circuit. Thus, the total number of executions over $N_T$ Trotter steps is given by
\begin{equation}
    c_{\text{Trotter}} = N_T k, 
\end{equation}  where $k$ denotes the total number of measurement shots employed per circuit evaluation.

When ZNE is applied, each Trotterized circuit must be evaluated at the three chosen noise amplification rates. For each noise rate, $n_\mathrm{avg}$ independent circuit evaluations are performed, each consisting of $k$ measurement shots, to obtain a robust average. This leads to a total execution count of
\begin{equation}
c_{\mathrm{ZNE}} = 3n_{\mathrm{avg}} c_{\mathrm{Trotter}}.
\end{equation} With $n_\mathrm{avg}=10$, as used throughout this work, ZNE-enhanced Trotterization therefore requires 30 times as many executions as the plain case.

\subsubsection{ISL Circuit Execution Counts}
The total number of circuit executions needed in ISL-approximated Trotterization is determined by three main factors: (i) the number of cost function evaluations during variational optimization, (ii) the number of measurements required for qubit pair selection, and (iii) the evaluation of the target observable.

Qubit pair selection is primarily performed via QST, a protocol for reconstructing the reduced density matrix of a subset of qubits by measuring them in different bases~\cite{Cramer_2010}. In ISL, pairwise QST is carried out at each layer to determine the local entanglement structure, which guides the construction of the ansatz. For each qubit pair, QST requires measurements in nine different basis settings~\cite{altepeter_4_2004, Garcia_2020}, adding a fixed overhead of measurement circuits per pair, per layer. 

Thus, for an ansatz constructed on hardware with $N_{qp}$ connected qubit pairs and $N_l$ layers, the total QST resource cost is at most $9 N_l N_{qp} k$ circuit executions per Trotter step. This value is only an approximation, as practical heuristics are employed to reduce the number of qubit pairs measured when adding a new layer. For instance, the algorithm keeps track of recently selected layers that did not yield significant cost improvements and temporarily avoids their reselection. Moreover, an exact resource cost would also need to account for any additional measurements required for qubit pair selection routines when all measured pairs exhibit zero local entanglement. Since ISL behavior is difficult to predict in advance, the exact QST resource cost cannot be determined prior to execution.

Including the contributions from variational optimization and observable evaluation, the total number of circuit executions required for ISL is hence approximated by:
\begin{equation}
    c_{\mathrm{ISL}} \approx \left( 1+ N_{ce} + 9 N_{l} N_{qp}\right)c_{\mathrm{Trotter}},
\end{equation}
where $N_{ce}$ is the total number of cost function evaluations across all Trotter steps. Both $N_{ce}$ and $N_l$ depend on system size, optimization dynamics, and problem instance, and thus cannot be determined a priori.

Shot noise, governed by $k$, is another key consideration in determining $c_\mathrm{ISL}$. Increasing $k$ reduces statistical fluctuations in measured expectation values and can improve optimization accuracy, possibly accelerating convergence, yielding smaller $(N_\mathrm{ce},N_l)$. However, increasing $k$ also linearly raises the total number of circuit executions. Balancing $k$ is therefore a primary challenge: systems with higher $N_q$ are typically associated with higher $c_\mathrm{ISL}$ due to a larger number of qubit pairs and increased Hamiltonian terms, which may require greater $(N_\mathrm{ce},N_l)$ to obtain sufficient ansatz expressibility. For these larger systems, increasing $k$ may be necessary to address greater statistical uncertainty in expectation values; but such increases can render ISL computationally infeasible if not properly tuned.

The scaling of $c_\mathrm{ISL}$ with respect to $k$, $N_\mathrm{ce}$, and $N_l$ is complex and often problem-dependent. For practical applications, careful empirical benchmarking is required to determine realistic resource requirements and to ensure that ISL-approximate quantum simulations remain feasible on current hardware.

\section{Results} \label{sec:results}
This section presents a comprehensive analysis of the performance, accuracy, and resource requirements of the quantum simulation protocols studied in this work.

The results are organized to progressively build an understanding of each protocol's performance across different quantum systems, simulation parameters and evaluation metrics. First, simulations of the TCM are analyzed for representative parameter sets, with varying coupling strengths characterizing the speed of state evolution. Second, analogous analyses are performed for instances of the HSC, including both TCM-inspired systems and higher-dimensional interactions with subsequently larger Hamiltonian terms. This is followed by a cross-protocol and cross-system comparison, including relevant statistics on resource requirements, accuracy measures, and a focused discussion on the direct impact of noise during recompilation in ISL.

Within each subsection, simulation results are presented in detail, with key findings supported by figures and tables as appropriate. Where relevant, comparisons are made with respect to the number of qubits ($N_q$), the number of measurement shots per circuit evaluation\footnote{Recall that for ZNE and ISL, $k$ characterizes measurement resolution, but is not equal to the total circuit shots per time step.} ($k$), and other algorithmic parameters.

Throughout this section, \textbf{plain Trotterization} refers to the evaluation of Trotter circuits directly under noise, \textbf{ZNE} refers to plain Trotterization enhanced by zero-noise extrapolation, and \textbf{ISL} refers to Trotterized evolution approximated by incremental structural learning under noisy conditions. Similarly, \textbf{noiseless evolution} refers to the evaluation of Trotter circuits in an ideal, noiseless backend, which is used as the baseline for ideal performance. Furthermore, $k_\mathrm{max}\equiv16\,384$ denotes the maximum number of shots per circuit evaluation studied. In addition, the two tested ISL cost thresholds will be indicated by $C_{-2}\equiv10^{-2}$ and $C_{-4}\equiv10^{-4}$.

\subsection{Tavis--Cummings Simulations} \label{subsec:tcm-simulations}
The qubit-encoded TCM Hamiltonian (Eq.~\eqref{eq:qubit-tcm}) is simulated using a coupling strength of $g=10$ and a system transition frequency of $\omega=1$. The frequency $\omega$ introduces only a relative phase in the Trotter step and thus does not affect the chosen observable. The coupling $g$ was chosen to ensure that the resulting time evolution displays clear and non-trivial oscillatory dynamics within the simulated time window.

From this section onward, the \textbf{main TCM parameterization} refers to the physical system defined by this parameter choice, for a varying number of two-level atoms $N$. 

\subsubsection{Evolution of probabilities}\label{subsubsec:tcm-main-data}
\begin{figure}[htb]
    \centering
    \includegraphics[width=0.9\textwidth]{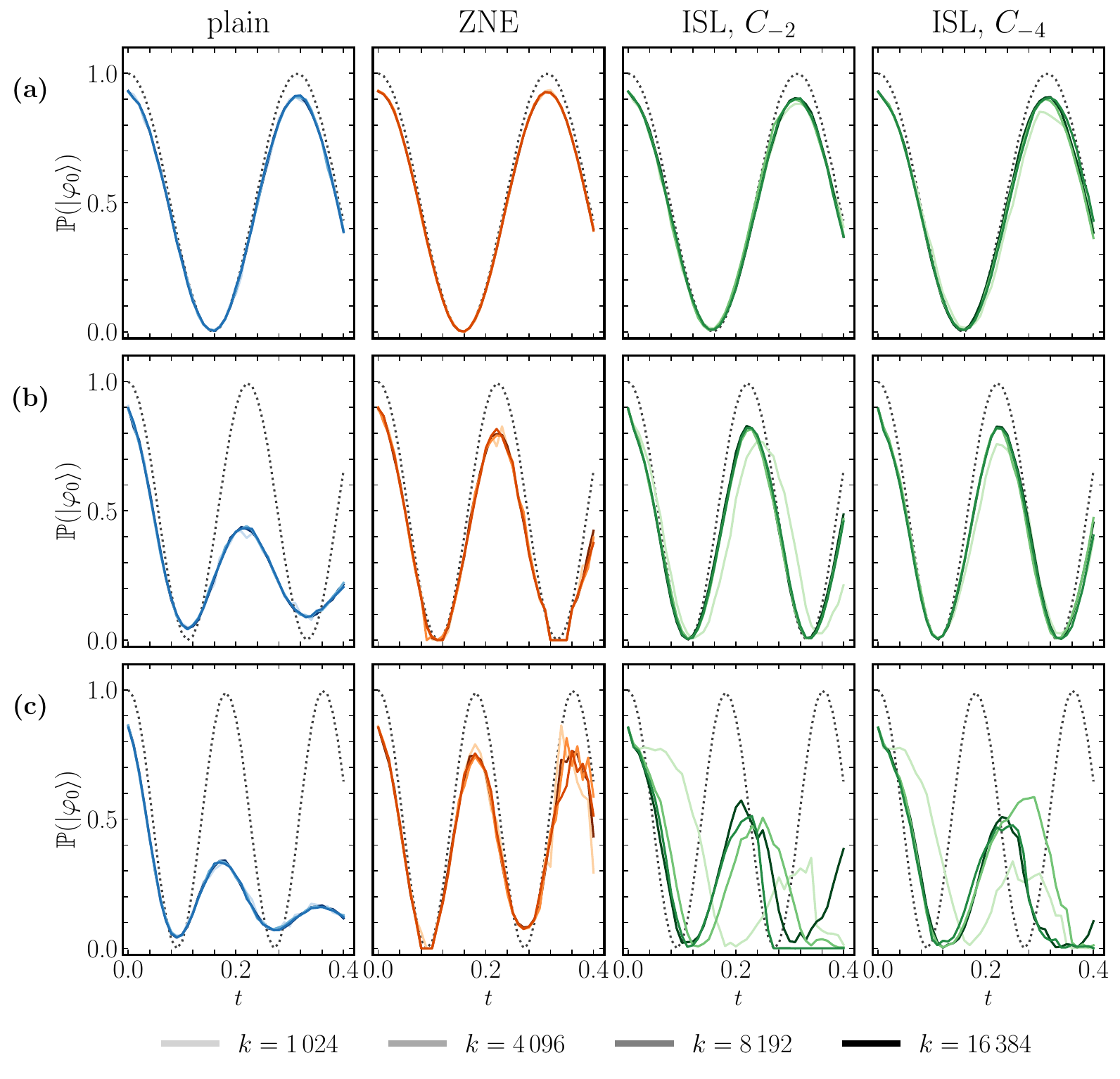}
    \caption{Time evolution of $\mathbb{P}(\ket{\varphi_0})$ for (a) $N = 1$, (b) $N = 2$, and (c) $N = 3$. \emph{Plain} denotes plain Trotterization; other columns show results with noise reduction protocols. Each method uses $k$ shots per circuit evaluation (darker lines: higher $k$). The dotted line is the noiseless reference curve (evaluated with $k=k_\mathrm{max}$ shots). 
    }
    \label{fig:time-evolution}
\end{figure}

The time evolution of the probability of measuring the initial state, $\mathbb{P}(\ket{\varphi_0})$, is presented in Fig.~ \ref{fig:time-evolution}. For ISL, the plotted probability curves correspond to the median value at each time step across all independent recompilations; the absolute error at each time step is computed with respect to this median value.

\begin{figure}[htb]
    \centering
    \includegraphics[width=\textwidth]{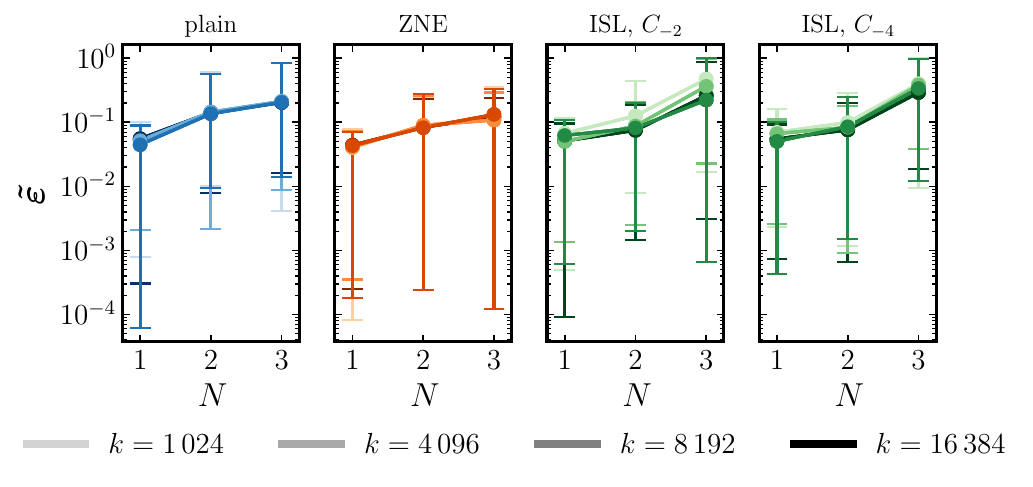}
    \caption{Median of the absolute error $\tilde{\varepsilon}$ over all time steps as a function of system size $N$, for the studied methods and experimental setup as in Fig.~\ref{fig:time-evolution}. Vertical bars show the range (minimum to maximum) of absolute error observed across all time steps for each $k$.}
    \label{fig:error-scaling}
\end{figure}

To provide a concise summary of algorithmic accuracy, the median of the absolute error over all time steps is reported in Fig.~ \ref{fig:error-scaling}. The use of the median, rather than the mean, reduces the influence of outlier values that can arise from rare but large deviations at individual time steps, yielding a more robust measure of typical performance.

% Absolute error across time evolution, removed for conciseness (previously not referenced in the analysis)

For this parameterization, plain Trotterization accurately captures the noiseless time evolution for $N=1$ (see Fig.~\ref{fig:time-evolution}(a)), showing minimal degradation from noise over increasing time steps. The application of ZNE provides marginal improvements, while ISL exhibits slight under-performance, showing minor deviations from the noiseless evolution, especially for lower shot counts with both $C_\mathrm{suff}$. These algorithmic trends are reflected in the median absolute errors $\tilde{\varepsilon}$ (see Fig.~\ref{fig:error-scaling}).

However, the accuracy of plain Trotterization begins to degrade for $N=2$ (see Fig.~\ref{fig:time-evolution}(b)), particularly in reproducing the amplitude of the second peak. While the ZNE method improves probability estimates, it introduces instability near extrema, suggesting that gate noise results in more frequent fluctuations at the maxima and minima of the noiseless expectation value. ISL outperforms ZNE, achieving a median absolute error of $0.075$ ($0.076$) with $C_{-2}$  ($C_{-4}$) compared to $0.084$ for ZNE at $k_\mathrm{max}$ shots (see Fig.~\ref{fig:error-scaling}), and visibly improved agreement with the reference curve at the second trough, indicating increased robustness over extended time steps.

The performance of plain Trotterization degrades further for $N=3$ (see Fig.~\ref{fig:time-evolution}(c)), consistent with the expected increase in noise accumulation for wider circuits (i.e., higher qubit count). Although noise bias is not directly measured here, it is well established that both decoherence and gate errors compound as circuit width and depth increase~\cite{preskill_quantum_2018, Ayral_2021}, leading to greater deviations from the noiseless reference. ZNE remains sensitive to increasing noise and displays growing fluctuations in later time steps. 

However, ZNE still outperforms ISL, which struggles with the larger system size: ISL fails to reconstruct the final peak and instead plateaus to near-zero probabilities in the final time steps for most recompilations. This effect is compounded by phase-lag errors, which visibly affect all ISL recompilations, and cause the ISL time evolution to increasingly lag behind the noiseless reference. As illustrated in Fig.~\ref{fig:error-scaling}, these effects lead to a significant scaling in errors for ISL recompilation at $N=3$. While ISL recompilation with larger shot counts generally improves accuracy, as seen in the figure, large errors still persist at $k_\mathrm{max}$, with phase lag attributed as a significant source of error. This suggests that a further increase in $k$ may be necessary to improve the viability of ISL for TCM systems with $N\ge3$. However, this requirement poses a challenge for the scalability of the method. In contrast, for plain Trotterization and ZNE, increasing the shot count beyond a certain threshold yields minimal changes in accuracy. This observation indicates that, for these approaches, factors other than shot noise are the main limiting sources of error under the tested conditions, while also demonstrating the viability of shot counts lower than $k_\mathrm{max}$ for these protocols.

Notably, both tested cost thresholds yield broadly similar performance, as seen by the close agreement of their probability and error curves. However, a more detailed inspection shows that the looser threshold $C_{-2}$ yields slightly lower errors than $C_{-4}$, despite the latter requiring more demanding optimization. For instance, at $k_\mathrm{max}$, the median curve for $C_{-4}$ yields higher $\tilde{\varepsilon}$ (0.292) than $C_{-2}$ (0.261); with similar trends also present for $N=1$ and $N=2$.

\begin{figure}[htb]
    \centering
    \includegraphics[width=\textwidth]{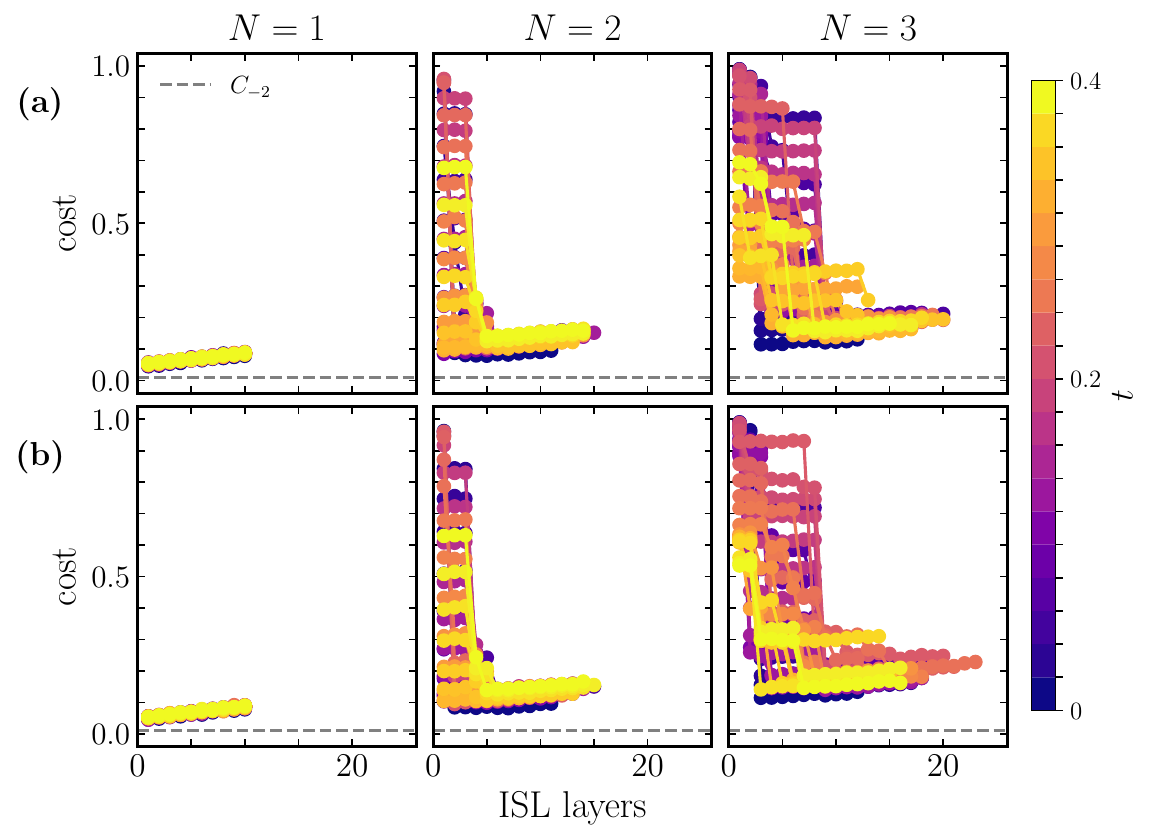}
    \caption{ISL recompilation cost at the end of each layer, for the main TCM parameterization, evaluated at $k_\mathrm{max}$. The rows correspond to sampled individual recompilations performed with the cost thresholds:
    (a) $C_{-2}$ and (b) $C_{-4}$. The color scale represents the corresponding time step. The dashed horizontal line marks the cost $C_{-2}$.}
    \label{fig:cost-progress}
\end{figure}

This somewhat counterintuitive result motivates the inspection of the recompilation cost  history for ISL in the studied systems. The evolution of the recompilation cost per ansatz layer\footnote{Here, ansatz layer is used interchangeably with thinned-CNOT layer (see Section~\ref{subsec:ISL-for-VQS}).} is presented for a single generic run in Fig.~\ref{fig:cost-progress}. The median cost across runs is not shown here, as varying number of layers between independent recompilations would complicate the interpretation. Notably, Fig.~\ref{fig:cost-progress} demonstrates that the cost function does not reach either of the $C_\mathrm{suff}$ thresholds tested. Instead, the ISL procedure most often halts the addition of new layers when further optimization fails to yield appreciable reductions in the cost function. In fact, all tested curves exhibit a gradual upward drift in the cost with additional layers during later time steps. This behavior likely reflects the accumulation of noise and approximation errors as circuit depth increases, reinforcing the need for careful control over the ansatz depth in noisy environments. 

This result explains why both $C_{-2}$ and $C_{-4}$ lead to similar performance: the chosen threshold has not affected the final Rotosolve optimization step of any layer, meaning any difference must be marginal, as it can only arise in intermediate optimization steps. As a result, the overall pattern in the cost curves is visibly similar for both $C_\mathrm{suff}$ values.

\subsubsection{Further TCM parameters}

To further assess the robustness of the observed trends, additional simulations are performed for alternative coupling strengths, specifically $g=1, 5, 20$, while $\omega$ is kept constant. All simulations are performed using $k_\mathrm{max}$ shots to minimize shot noise, and ISL circuits are recompiled for both studied cost tolerances $\{C_{-2},C_{-4}\}$. In this study, these systems will be collectively referred to as the \textbf{alternative TCM parameterizations}. 

\begin{figure}[htb]
    \centering
    \includegraphics[width=0.9\linewidth]{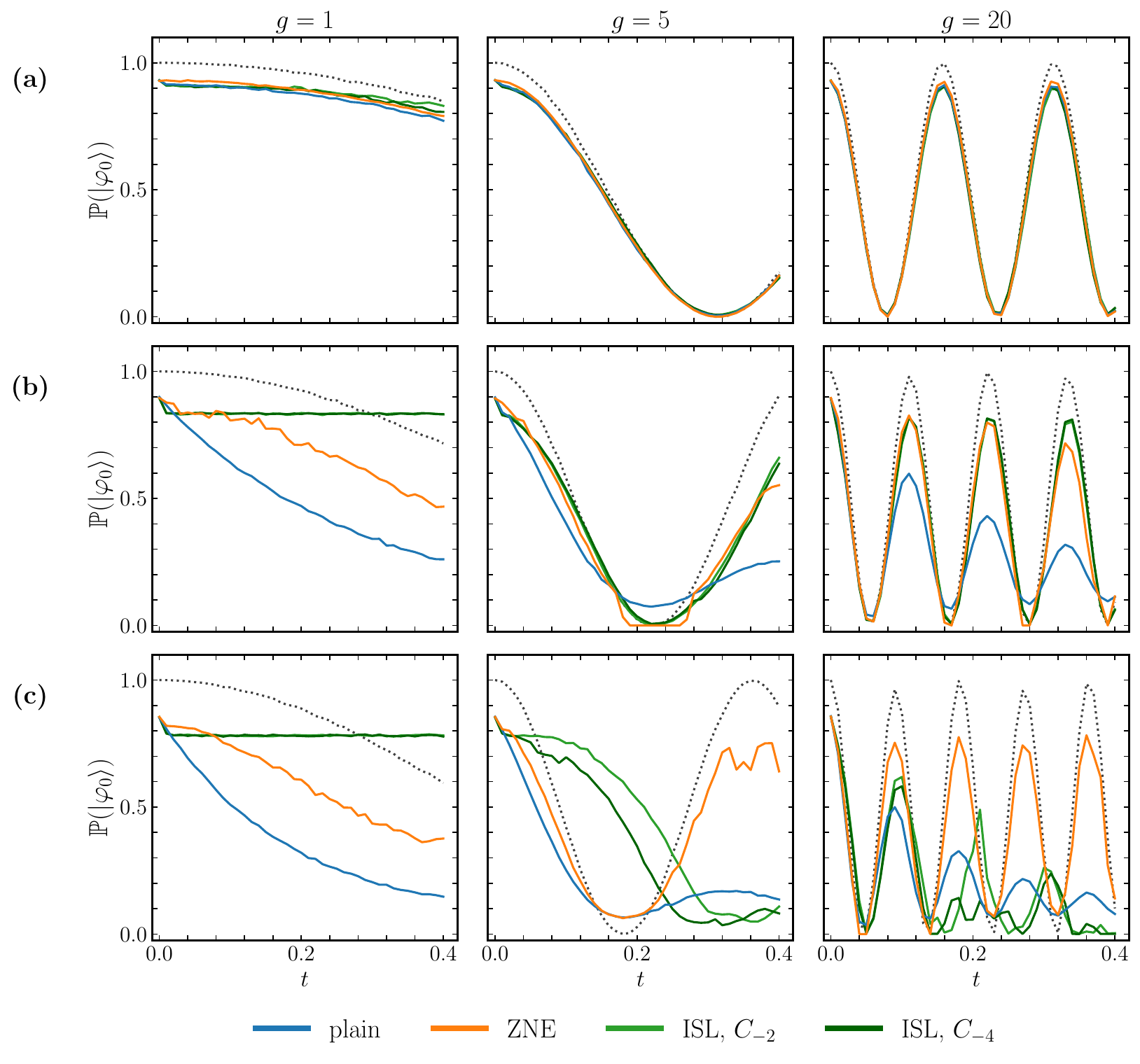}
    \caption{Time evolution of $\mathbb{P}(|\varphi_0\rangle)$ for plain Trotterization, ZNE and ISL, for different coupling strengths and system sizes. Columns (left to right) correspond to $g=1$, $g=5$ and $g=20$, and the system sizes are (a) $N=1$ (b) $N=2$ and (c) $N=3$. The dotted line denotes the noiseless time curve. All algorithms are evaluated with $k=k_\mathrm{max}$. For ZNE reference curves, points outside the interval $[0, 1]$ have been clipped.}
    \label{fig:tcm-g-time-evolution}
\end{figure}

The results for the probability evolution of the initial state in the alternative TCM parameterizations are presented in Fig.~ \ref{fig:tcm-g-time-evolution}.

As in the main TCM parameterization, results for $N=1$ across all $g$ values indicate that plain Trotterization remains viable in this regime. This protocol continues to provide an accurate time evolution that is robust to increasing time steps, with ISL and ZNE only providing marginal differences in the results, despite their associated overhead compared to the baseline method.

For $N\ge 2$ and the smallest coupling factor tested ($g=1$), plain Trotterization and ZNE both outperform ISL, as the latter produces nearly constant probability amplitudes after the initial time step. This outcome illustrates that noisy ISL recompilation can fail to approximate the target unitary when interaction strengths are small. In this regime, ISL is expected to exhibit low sensitivity to parameter updates, since each Trotter step induces only minimal state changes that may not be fully captured by the recompilation protocol. Indeed, ISL transpilation discards rotation gates in the ansatz for sufficiently small angles, which can limit the output circuit's ability to encode weak dynamical effects. Additionally, ansatz growth may terminate prematurely due to noise accumulation with added layers, as the stopping criterion is then reached before sufficient depth is achieved. Consequently, the resulting ansatz may stabilize at too shallow depths, restricting expressivity, resulting in noise effects dominating dynamics over the weak evolution in the cost function being optimized. 

For the remaining systems studied, with coupling strengths $g=5$ and $g=20$, the observed trends are similar to those in the main TCM parameterization. For $N=2$, ISL outperforms the other two protocols, with Trotterization suffering from the expected decay due to noise over increasing time steps, and ZNE being particularly unstable near the troughs of the evolution curves, often resulting in points outside the valid domain $[0, 1]$ that had to be clipped in postprocessing. As such, ISL with $C_{-2}$ ($C_{-4}$) generally obtains a relative improvement in $\tilde{\varepsilon}$ over ZNE, of 16.4\% (23.4\%) for $g=5$, and of 9.1\% (11.7\%) for $g=20$.

For $N=3$, however, ZNE achieves the best performance, effectively mitigating the increasing noise-induced decay in plain Trotterization, while ISL exhibits great phase lag errors in both cases. Notably, in the $g=20$ case, there is substantial variation in the dephasing across independent ISL runs, resulting in median values that fail to capture oscillatory dynamics beyond the initial time steps, as explored in more detail in Section~\ref{subsubsec:ISL-variation}. As a result, ZNE is able to achieve substantial relative improvements in $\tilde{\varepsilon}$ over ISL with $C_{-2}$ ($C_{-4}$) of 74.3\% (68.6\%) for $g=5$, and of 38.5\% (49.6\%) for $g=20$; demonstrating a consistent advantage in these systems.

\subsubsection{Error Summary Across TCM Systems}\label{subsubsec:error-summary-TCM}
\begin{figure}[htb]
    \centering
    \includegraphics[width=0.95\linewidth]{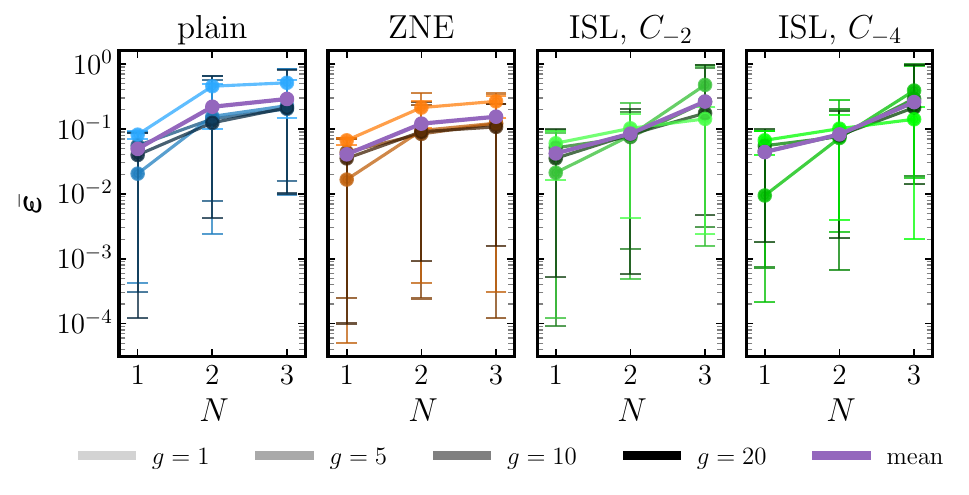}
    \caption{Median absolute error $\tilde{\varepsilon}$ in the time evolution of $\mathbb{P}(|\varphi_0\rangle)$ for TCM systems, shown for each protocol executed at $k_\mathrm{max}$, as a function of system size $N$ and coupling strength $g$. Error bars denote the range observed (minimum to maximum) across all time steps. The purple curve shows the mean across all $g$.}
    \label{fig:tcm-g-error-amplitude}
\end{figure}

A comprehensive comparison of amplitude errors in all tested TCM parameterizations is presented in Fig.~\ref{fig:tcm-g-error-amplitude}, which illustrates the median error over all simulation steps, per algorithm, for all considered TCM systems.

Across all protocols, the median amplitude error $\tilde{\varepsilon}$ increases substantially from $N=1$ to $N=2$, likely reflecting the effect of increased noise associated with deeper Trotter circuits and higher qubit count. Median error is also sensitive to the coupling strength $g$: for Trotterization and ZNE, the highest errors occur at $g=1$ for all $N$; illustrating how different time evolution curves may lead to significant variations in baseline accuracy errors. Although this coupling strength achieves the lowest median error for ISL with both cost thresholds at $N=3$, this does not correspond to improved algorithmic performance; given the constant probability produced by ISL in this case.

Furthermore, the advantage of ISL for $N=2$ can be confirmed, with $\tilde{\varepsilon}$ achieving 0.085 (0.082) for ISL with $C_{-2}$ ($C_{-4}$) averaged over all parameters, compared to 0.219 and 0.121 for plain Trotterization and ZNE, respectively. In contrast, at $N=3$, ZNE is able to provide a relative improvement over ISL of 41.6\% for $C_{-2}$, and of 40.4\% for $C_{-4}$.

\begin{figure}[htb]
    \centering
    \includegraphics[width=0.9\linewidth]{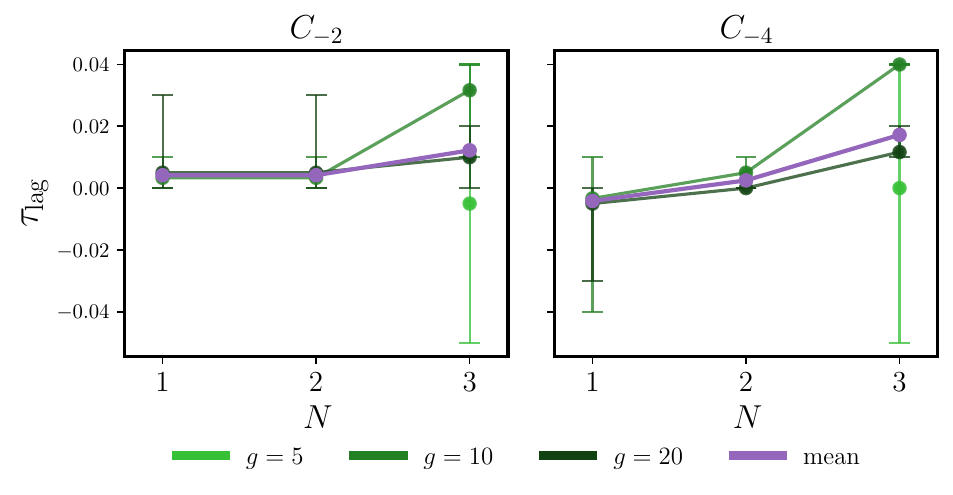}
    \caption{Dephasing between the first major peak of the noiseless evolution of TCM systems and the corresponding peak in the ISL evolution, for $k=k_\mathrm{max}$, with varying system size $N$ and coupling strength $g$. Scatter points represent the mean phase shift across independent ISL recompilations, while error bars denote the range. The purple curve shows the mean across all $g$. Positive (negative) values indicate a delayed (advanced) peak relative to the noiseless case.}
    \label{fig:tcm-g-phase-error}
\end{figure}

Phase shift (dephasing) errors in ISL recompilations across the studied TCM parameterizations are presented in Fig.~\ref{fig:tcm-g-phase-error}. The phase error analysis contrasts trends between dephasing errors and $C_\mathrm{suff}$. Notably, for $N=1$, the mean dephasing across $g=10$ and $g=20$ is negligible---0.004 (-0.004) for $C_{-2}$ ($C_{-4}$)---and is driven primarily by outlier runs, with most recompilations showing zero dephasing. For $N=2$, $C_{-2}$ yields a larger dephasing error on average, but it remains negligible ($|\tau_\mathrm{lag}|<0.01$). For $N=3$, average dephasing increases by a factor of 2.93 (6.89) for $C_{-2}$ ($C_{-4}$). Therefore, average dephasing in ISL tends towards leads (positive shifts) for $N=3$, although partially offset by lags. Furthermore, at this system size, variability in dephasing is somewhat greater for $C_{-2}$, with the distance between error bars having an average length of 0.047, compared to 0.033 for $C_{-4}$. 

Despite the previously discussed bias in $\tau_\mathrm{lag}$ toward larger coupling strengths, the mean magnitude of dephasing errors is smallest in magnitude for $g=5$ at $N=3$, reaching $|\tau_\mathrm{lag}|=0.005$ (0.000), whereas the maximum magnitude is 0.032 (0.040), reached by $g=10$ for $C_{-2}$ ($C_{-4})$. However, this does not reflect a genuine reduction in dephasing for $g=5$, but instead arises from large variability in dephasing, resulting in significant lags and leads which partially offset each other. This is reflected in the error bars having a length of 0.090 for both $C_{\mathrm{suff}}$ values in this system. Visual inspection of the time evolution curves (Fig.~\ref{fig:qhm-params-time-evolution}) confirms that the median ISL traces at $g=5$, $N=3$ exhibit significant phase lag, suggesting that many individual runs fail to correctly identify the peak structure. In contrast, ISL runs at $g=20$ achieve error bars of distance 0.020 (0.010) for $C_{-2}$ ($C_{-4}$), suggesting more consistent alignment with the noiseless evolution for the initial identified peak and resulting in lower and more reliable per-run dephasing errors. 

Taken together, these results show that accounting for dephasing variability yields agreement with expected trends based on the oscillation pattern. Since this variability arises from the systematic procedure for dephasing, no strong evidence is found for a link between Hamiltonian structure and dephasing in  the considered systems. Instead, the observed dephasing arises from the natural accumulation of error during ISL recompilation, stemming from approximate ansatz optimization (see Section~\ref{subsec:ISL-for-VQS}), and the algorithm's tendency to favor shallower circuits due to the high noise observed with increasing layer counts (Fig.~\ref{fig:cost-progress}).

Table~\ref{tab:tcm-table} summarizes the general trends observed for each studied configuration of the TCM. Overall, protocol performance across $(N,g)$ has been classified into four different cases. Trotterization remains reliable throughout all two-qubit systems, ISL generally performs best for $N=2$ systems, but ZNE achieves the greatest improvements for increased system size and weak coupling strengths.

\begin{table}[htb]
\centering
\begin{tabularx}{\linewidth}{|c|c|X|X|X|}
\hline
$N$ & $g$ & plain & ZNE & ISL ($C_{-2}$/$C_{-4}$) \\[0.5em]  \hline
1 & all & Close to noiseless evolution & Marginal improvement & Slightly worse than Trotter \\[2em] \hline
2, 3  & 1     & Large errors at all $t$ & Substantial improvement  & Nearly constant after $t=1$  \\[2em]  \hline
2 & 5, 10, 20  & Increasing errors with $t$  & Substantial improvement; outlier points need clipping  & Outperforms Trotter and ZNE  \\[2em]  \hline
3 & 5, 10, 20     & Errors slightly higher than $N=2$ & Substantial improvement; small fluctuations, no outliers & Significant phase lags; higher amplitude errors \\[2em]  \hline
\end{tabularx} \caption{Summary of qualitative performance in accuracy for all tested TCM parameterizations, for each protocol.}
\label{tab:tcm-table}
\end{table}

\subsection{Heisenberg Spin Chain Simulations}\label{subsec:qhm-simulations}

To assess the performance of the considered methods in a spin-lattice model, simulations are conducted for the one-dimensional HSC model. The chosen parameterization maintains the same single- and two-qubit Pauli operator magnitudes and signs as in the main TCM parameterization, but applies them to the cyclic interactions of the spin chain. This enables a direct comparison of how the coupling pattern impacts the performance of all tested algorithms.

The coupling vector is set to 
\begin{equation}
    \bm{J}_\mathrm{main} \equiv (-10, 10, 0),
\end{equation}
and the external magnetic field is chosen as \begin{equation}
    \bm{h}_\mathrm{main} \equiv (0, 0, -2) 
\end{equation} 

As in the TCM, the term \textbf{main HSC parameterization} refers to the physical system described by this choice of parameters, for a varying number of spin sites $L$. 

\subsubsection{Evolution of probabilities}

\begin{figure}[htb]
    \centering
    \includegraphics[width=0.9\linewidth]{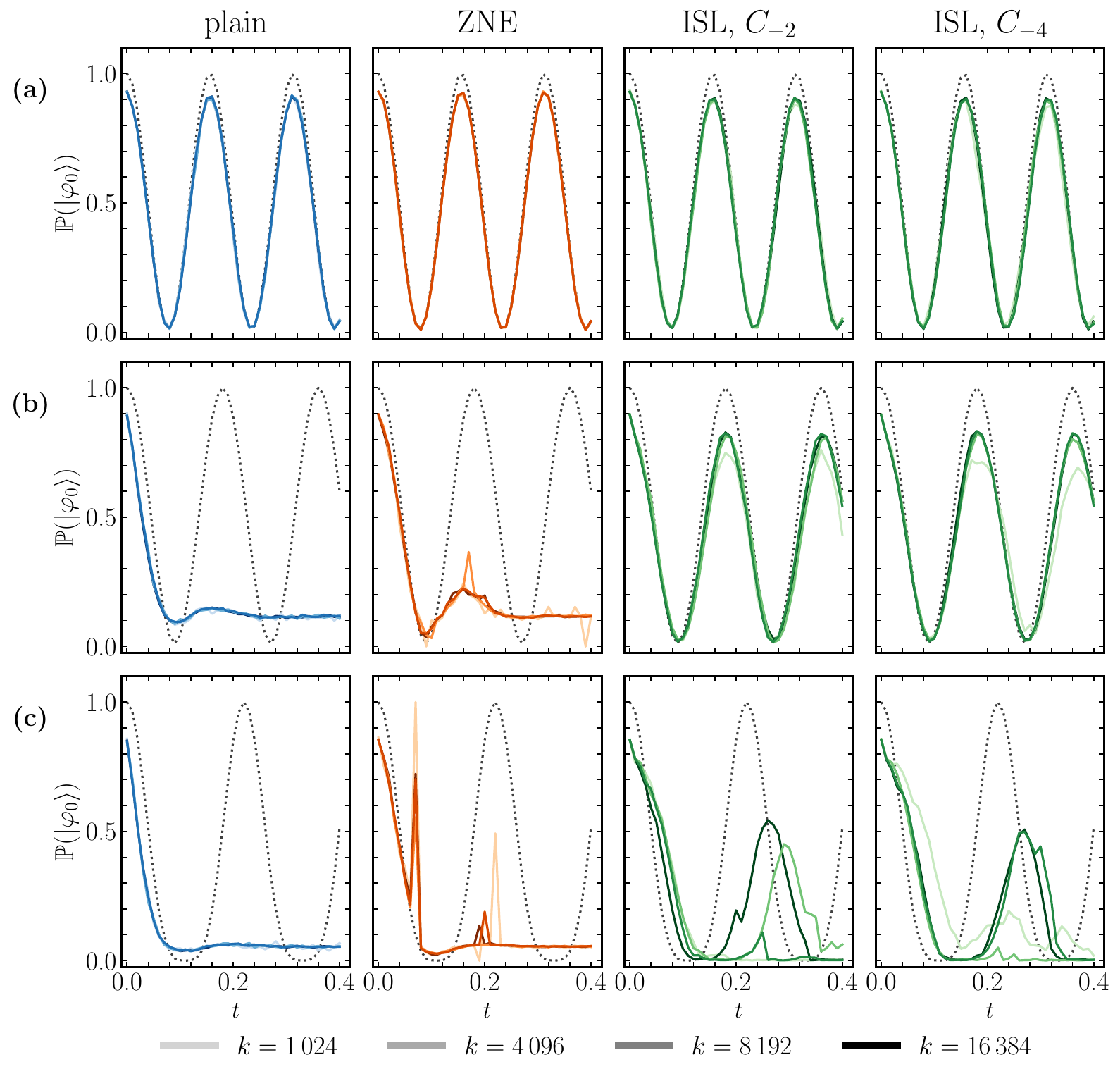}
    \caption{Time evolution of $\mathbb{P}(|\varphi_0\rangle)$ for the studied algorithms, with system sizes (a) $L=2$, (b) $L=3$, and (c) $L=4$. Darker shades correspond to an increased number of shots per circuit evaluation, and the dotted line denotes the noiseless time curve, with $k=k_\mathrm{max}$.}
    \label{fig:qhm-time-evolution}
\end{figure}

\begin{figure}[htb]
    \centering
    \includegraphics[width=\linewidth]{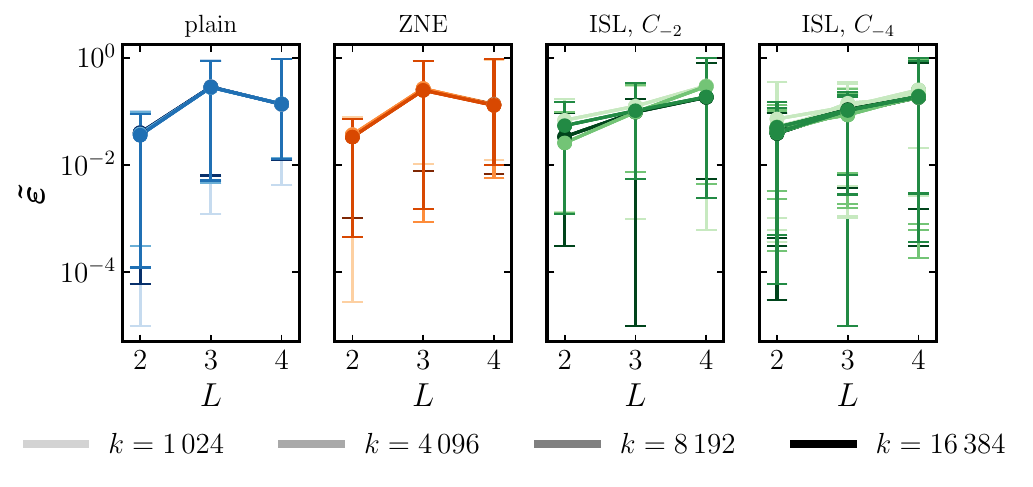}
    \caption{Median of the absolute error $\tilde{
    \varepsilon}$ over all time steps as a function of system size $L$ for the examined methods. As in Fig.~ \ref{fig:error-scaling}, the vertical bars indicate the range between the minimum and maximum error observed across all time steps. For clarity, minima below $10^{-5}$ are displayed at $10^{-5}$}
    \label{fig:HSC-error-scaling}
\end{figure}

For the main HSC parameterization and varying site counts $L=2,3,4$, the time evolution of $\mathbb{P}(|\varphi_0\rangle)$ is presented in Fig.~ \ref{fig:qhm-time-evolution}, while the corresponding median absolute errors are presented in Fig.~ \ref{fig:HSC-error-scaling}. As in TCM systems, all ISL time-evolution probabilities are obtained as the median of multiple independent runs, for each time step.

The results for $L=2$ in the HSC model closely mirror those discussed for $N_q=2$ (i.e., $N=1$) in the TCM simulations, and are therefore not analyzed in further detail here to avoid redundancy. In the two-qubit system, the cyclical spin interactions in the chosen HSC parameterization lead to a Hamiltonian structure that is effectively equivalent to the main TCM parameterization, but with doubled coupling strength due to the inclusion of both $\sigma_{x,y}^{(1)}\sigma_{x,y}^{(2)}$ and $\sigma_{x,y}^{(2)}\sigma_{x,y}^{(1)}$ terms. As these operators contribute additively, the resulting dynamics correspond exactly to the TCM evolution with $g=20$, as can be verified by comparing the time evolution in both cases (see Fig.~\ref{fig:tcm-g-time-evolution}). This equivalence results in comparable performance across all methods, with ZNE and ISL only providing marginal differences relative to plain Trotterization.

However, key differences in algorithmic performance emerge between the two models at $L=3$, where the Hamiltonian equivalence no longer holds. In this case, plain Trotterization rapidly deteriorates, as evidenced by the convergence of the probability curve toward the noisy asymptote $1/N_q^2$, leading to large accuracy errors near amplitude peaks. ZNE provides a small relative improvement of $10.1\%$. Nevertheless, it still exhibits convergence towards the same asymptote, resulting in similarly large deviations from the ideal evolution. In contrast, ISL maintains a stable reconstruction of the underlying time evolution and visibly outperforms the other methods, achieving a relative improvement in $\tilde{\varepsilon}$ by 27.2\% and 26.4\% over ZNE for $C_{-2}$ and $C_{-4}$, respectively, further illustrated in Fig.~\ref{fig:HSC-error-scaling}, showing lower errors for ISL at both $C_\mathrm{suff}$.

At $L=4$, performance degrades across all methods due to the increased noise per time step, as expected from the greater depth and width of the Trotter circuits. The convergence of plain Trotterization towards the noisy asymptote is further amplified, visibly flattening the evolution curve after the first trough. Correspondingly, ZNE offers only minimal improvement over plain Trotterization ($2.2\%$), and displays increased instability at early time steps ($t\le0.2$), even at $k_\mathrm{max}$. Interestingly, both methods achieve a lower $\tilde{\varepsilon}$ compared to $L=3$; however, this appears to be an artifact of the reduced oscillation frequency in the noiseless evolution, which better aligns the troughs with the asymptotic value. Visually, ISL is able to provide with the closest reconstruction of the oscillatory pattern in the time evolution. However, it still suffers from significant phase lags, large amplitude errors, and eventual plateauing near $t=0.36$. As a result, ISL yields the largest errors among all methods for this system size, as shown in Fig.~\ref{fig:HSC-error-scaling}, of 0.184 (0.182) for $C_{-2}$ ($C_{-4}$), compared to 0.137 and 0.134 for plain Trotterization and ZNE, respectively.

Reducing the instability of ZNE in these systems would likely require more extensive sampling (i.e., larger $n_\mathrm{avg}$) at each time step, which could help suppress outlier behavior in the extrapolated expectation values by enabling more stable statistical averages. However, such an approach would linearly increase the number of required circuit executions, and it may still be unable to provide significant improvements over plain Trotterization in the presence of substantial circuit noise---as appears to be the case for circuits considered here. Ultimately, the large circuit depths associated with Hamiltonians with large term counts and longer time ranges imposes a fundamental limitation on ZNE for DQS, as evidenced here, since the method mitigates measurement error in expectation values but does not address the underlying accumulation of circuit-level noise. Indeed, increasing shot counts for plain Trotterization and ZNE in these systems provides minimal improvement against noisy plateauing; only reducing instability and fluctuations.

As with the TCM, the evolution of the cost function for ISL in this system provides insight into the performance limitations of the method. However, the observed trends closely resemble those in the main TCM parameterization (see Fig.~\ref{fig:cost-progress}), including costs not reaching below the set $C_\mathrm{suff}$ thresholds and gradually increasing with added layers. The cost evolution plots for a generic run of the main HSC system is presented in Appendix~\ref{app:HSC-cost-progression} for completeness.

\subsubsection{Further HSC parameters}

To further assess the performance of the tested methods in HSC systems, additional simulations were performed for systems with variable coupling strengths $\bm J$, with a focus on interaction structures with additional Hamiltonian terms. These alternative configurations were selected to introduce a range of anisotropic and fully coupled regimes, while maintaining coupling magnitudes comparable to the main parameterization $(\bm{J}_\mathrm{main})$. The goal is to explore the robustness and generalizability of the algorithms beyond a single interaction profile, under more entanglement-rich and diverse coupling conditions.

Specifically, four alternative coupling vectors were selected, denoted $\bm J_1$ through $\bm J_4$, which span a range of physically relevant regimes: \begin{itemize}
    \item $\bm J_1\equiv-(20, 20, 2)$, a weakly perturbed XXZ extension of XY coupling systems;
    \item $\bm J_2 \equiv -(2, 2, 20)$, a XXZ system with dominant $z$-axis coupling;
    \item $\bm J_3 \equiv -(20, 20, 20)$, fully balanced interactions (termed XXX spin chains~\cite{candu2013spin_chains});
    \item $\bm J_4 \equiv -(2, 20, 4)$, asymmetric XYZ coupling with strong $y$- and $z$-components.
\end{itemize} 

These $\bm J$ vectors were chosen heuristically, but they enable consistent comparison across qualitatively distinct coupling structures. The negative signs in each case ensure that, after accounting for the $-\frac{1}{2}$ prefactor in the HSC Hamiltonian (cf. Eq.~\ref{eq:hsc-Hamiltonian}), the resulting 2-local terms yield positive interaction strengths with magnitudes comparable to those used in the presented TCM simulations.

To complement the varied interaction terms, the following three-dimensional magnetic field was applied across all configurations:
\begin{equation}
    \bm{h}_\mathrm{alt}\equiv(-20, -20, -20).
\end{equation} 

This choice introduces isotropic 1-local terms along all spatial directions, increasing the Hamiltonian terms per qubit, and subsequently the gate overhead in the Trotter decomposition. The magnitude of magnetic field components was selected to match the scale of the $\bm J$ components, ensuring that single-qubit effects remain significant in the dynamics---unlike previously analyzed systems, where only relative phase evolution from local terms was observed. As in the TCM, these systems are collectively referred to as the \textbf{alternative HSC parameterizations}. 

The explored parameter space remains necessarily limited. While a broader scan over $(\bm J, \bm h)$ configurations would strengthen the empirical validation of observed patterns, it would also incur substantial computational cost; particularly due to ZNE and ISL time evolution, and extending the scope of the analysis beyond the intended focus. Later sections analyze the selected systems in greater detail to highlight relevant performance trends.

\begin{figure}[hbt]
    \centering
    \includegraphics[width=0.9\linewidth]{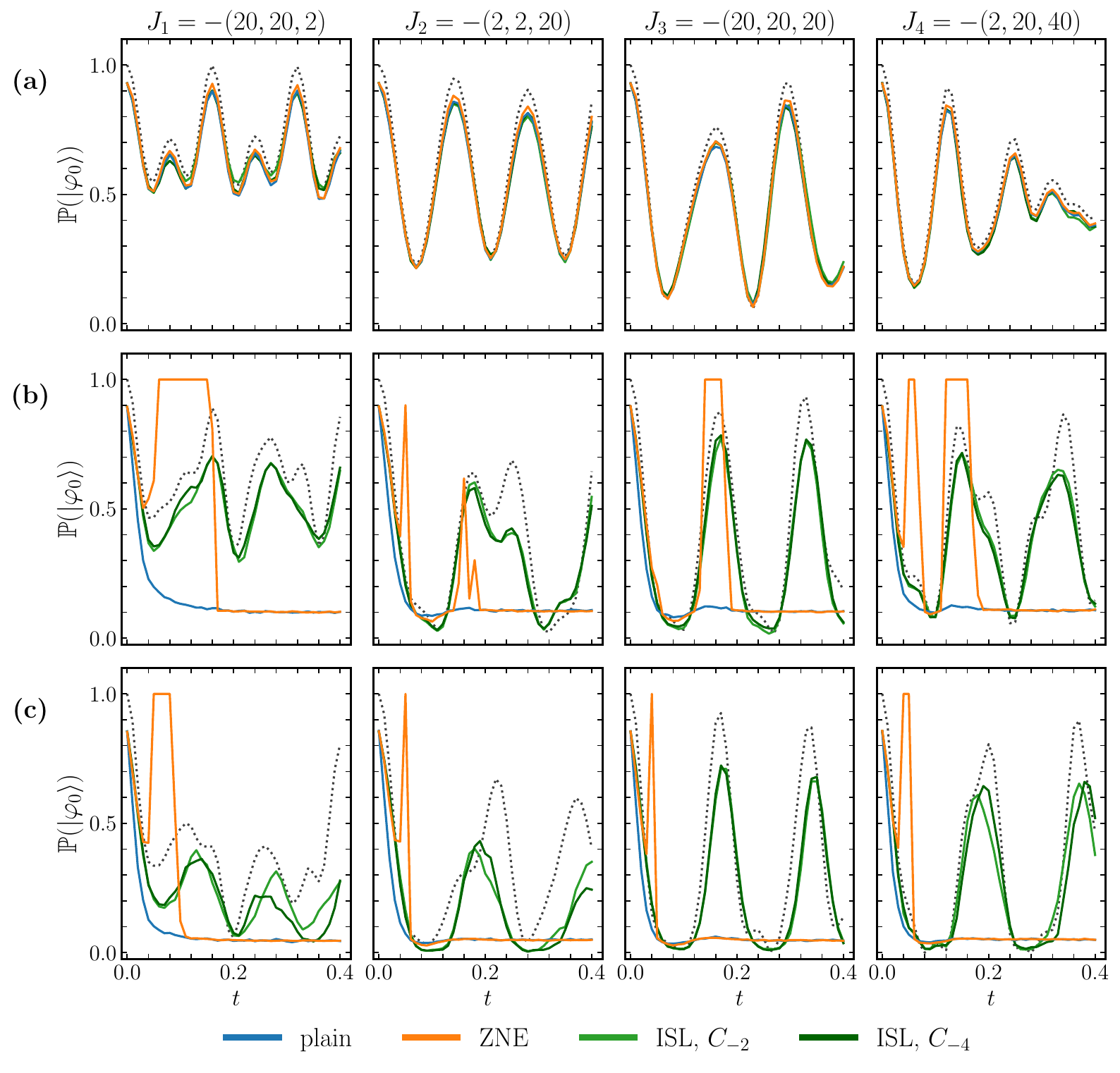}
    \caption{Time evolution of $\mathbb{P}(|\varphi_0\rangle)$ for plain Trotterization, ZNE, and ISL, under different HSC parameterizations. Columns (left to right) correspond to different coupling vector $\bm  J$ values , while system sizes are (a) $L=2$, (b) $L=3$, and (c) $L=4$. The dotted line denotes the noiseless evolution curve. All simulations are evaluated with $k=k_\mathrm{max}$. For ZNE reference curves, points outside the interval $[0, 1]$ have been clipped.}
     \label{fig:qhm-params-time-evolution}
\end{figure}

The results of the probability evolution of $\mathbb{P}(\ket{\varphi_0})$ under the chosen coupling strengths are presented in Fig.~ \ref{fig:qhm-params-time-evolution}. 

With the increased dimensionality of parameters, the $L=2$ systems studied no longer exhibit direct equivalence to TCM systems with $N=1$, as the coupling between qubits is no longer anti-symmetric\footnote{That is, of equal magnitude but opposite sign.} along the $x$- and $y$-axes. Additional differences arise from the inclusion of $z$-axis coupling and isotropic single-qubit rotations, as previously described. Despite these structural differences, the relative performance among the three algorithms remains consistent at this system scale, as shown in Fig.~\ref{fig:qhm-params-time-evolution}(a). This is primarily due to plain Trotterization maintaining high simulation accuracy, similar to earlier two-qubit systems. These results reinforce the robustness of plain Trotterization in yielding low errors in small-scale HSC systems, consistent with observations from the TCM.

For $L=3$, pronounced differences between the protocols emerge across all four parameterizations (see Fig.~\ref{fig:qhm-params-time-evolution}(b)), consistent with trends observed in the main HSC case. As before, plain Trotterization rapidly converges toward the noisy asymptote, resulting in significant accuracy loss beyond early time steps. ZNE suffers similarly, with large accuracy errors, persistent fluctuations, and eventual collapse toward near-constant asymptotes---limiting its gains. Relative improvements over plain Trotterization vary substantially: ZNE achieves its largest relative improvement in $\tilde{\varepsilon}$ for $\bm J_2$ (48.7\%), and the smallest for $\bm J_4$ ($-0.4$\%).

In contrast, ISL maintains accurate reconstructions of the noiseless time evolution, avoiding plateauing and showing minimal dephasing over time. Nonetheless,
non-negligible errors remain, with visible deviations in certain $\bm J$, potentially due to the increased Hamiltonian term count associated with the larger system, thus increasing the depth of Trotter circuits. The largest relative improvement over plain Trotterization occurs for $C_{-2}$ in $\bm J_4$ (81.6\%), with the smallest improvement is for $C_{-2}$ in $\bm J_3$ (54.0\%), indicating a consistent advantage over plain Trotterization across all tested systems, in contrast to ZNE.

For $L=4$ (see Fig.~\ref{fig:qhm-params-time-evolution}(c)), despite the larger system size, the performance of plain Trotterization remains relatively stable, likely due to saturation in noise impact per Trotter step. In some cases, changing noiseless trends even reduce $\tilde{\varepsilon}$ relative to $L=3$ for plain Trotterization, despite plateauing behavior. A similar trend is observed for ZNE, with fewer outlier points than at $L=3$ but persistent variability across all four systems. For instance, in $\bm J_1$, errors decrease for plain Trotterization (ZNE) from 0.496 (0.439) at $L=3$ to 0.290 (0.288) at $L=4$.

ISL performance degrades with increasing system size, though the extent varies across parameterizations. Some cases exhibit significant phase lags and distortions in oscillatory patterns. For $\bm J_1$, ISL errors with $C_{-2}$ ($C_{-4}$) increase by 22.5\% (55.9\%) relative to $L=3$. For $\bm J_2$, relative error increases are more severe: 163\% (116\%). Even in $\bm J_3$, with fully isotropic interactions, errors increase by 7.25\% (38.5\%), though visual inspection shows that oscillatory structure is still captured. In contrast, despite highly asymmetric couplings, $\bm J_4$ shows minimal dephasing and high robustness, with a slight improvement of 4.26\% for $C_{-2}$, and of 1.69\% for $C_{-4}$. Notably, this case corresponds to coupling vector with  highest Euclidean norm of, providing additional evidence that stronger overall interactions may be more effectively captured by ISL in noisy larger-scale systems.

These findings further illustrate the sensitivity of ISL to the specific structure of the target Hamiltonian, across the different HSC parameterizations studied here. In particular, ISL recompilation appears more robust in the alternative HSC configurations, most notably for $\bm J_3$ and $\bm J_4$. Despite the increased number of Hamiltonian terms compared to the TCM and the main HSC systems, and thus deeper Trotter circuits, the results suggest that ISL benefits from isotropic 1- and 2-local terms, as well as stronger interactions; characterized by larger norms in $\bm J$ for the HSC.

\subsubsection{Error Summary Across HSC Systems}

\begin{figure}[htb]
    \centering
    \includegraphics[width=\linewidth]{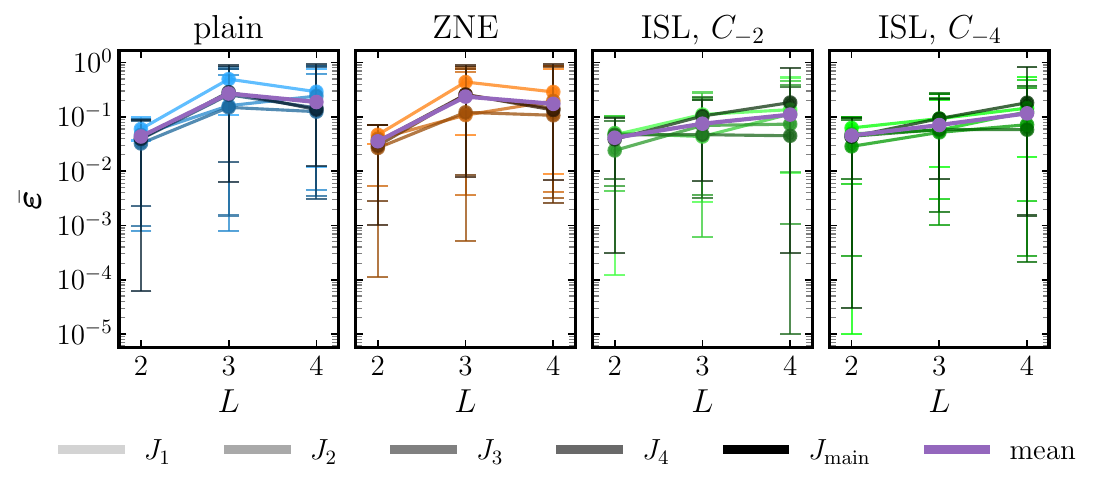}
    \caption{Median absolute error $\tilde{\varepsilon}$ in the time evolution of $\mathbb{P}(\ket{\varphi_0})$ for HSC systems, shown for each protocol executed at $k=k_\mathrm{max}$, as a function of system size $L$ and coupling vector $\bm J$. Error bars denote the range across all time steps, and the purple curve shows the mean across all $\bm J$. For clarity, minima below $10^{-5}$ are displayed at $10^{-5}$.}
    \label{fig:qhm-J-error-amplitude}
\end{figure}

As was done for the TCM, the median error over all time steps across the considered HSC systems is plotted for each algorithm in Fig.~\ref{fig:qhm-J-error-amplitude}.

Analyzing trends across system size reveals that, as in the main HSC parameterization, increasing system size from $L=3$ to $L=4$ leads to relatively smaller errors $\tilde{\varepsilon}$ in plain Trotterization by 30.4\% , and similarly for ZNE by 26.5\%.  As before, this appears to be an artifact of the specific evolution patterns that emerge with increasing $L$. As visually observed across time evolution curves, these lower error values do not necessarily indicate improved performance for these two protocols over different time ranges.

In contrast, ISL exhibits a consistent increase in average error with system size, unaffected by curve plateauing; increasing by 46.7\% and 63.4\% for $C_{-2}$ and $C_{-4}$, respectively. Despite this, ISL maintains the lowest average error of all three methods for $L\ge3$. In addition, results are consistent across both $C_\mathrm{suff}$ values, with $\bm J_\mathrm{main}$ and $\bm J_4$ corresponding to the systems with the largest and smallest $\tilde{\varepsilon}$, respectively, on the two larger system sizes. 

\begin{figure}[htb]
    \centering
    \includegraphics[width=\linewidth]{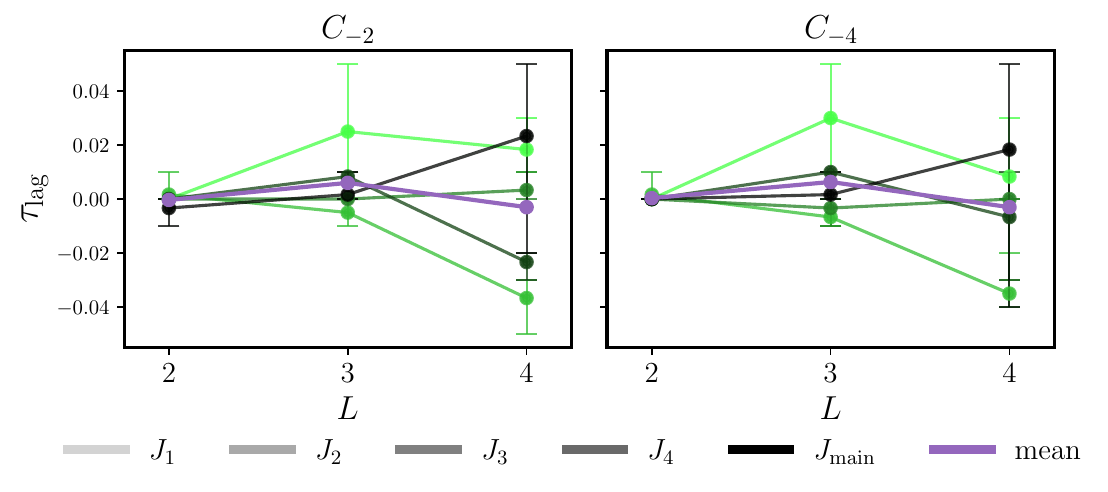}
    \caption{Dephasing between the first major peak of the noiseless evolution of HSC systems and the corresponding peak in the ISL evolution, for $k=k_\mathrm{max}$, with varying system size $L$ and coupling vector $\bm J$. Scatter points represent the mean phase shift across independent ISL recompilations, while error bars denote the range. The purple curve shows the mean across studied $\bm J$. Positive (negative) values indicate a delayed (advanced) peak relative to the noiseless case.}
    \label{fig:qhm-J-phase-error}
\end{figure}

To complement the discussion of amplitude error, a dephasing analysis was conducted for all HSC parameterizations, as presented in Fig.~\ref{fig:qhm-J-phase-error}.

All studied HSC systems contain suitable peaks across all considered parameterizations, with peaks obtained at earlier time steps relative to certain TCM systems. This enables a larger sample for the dephasing analysis under more comparable noise conditions across systems. However, in some systems with more complex dynamics, peaks identified in the noiseless evolution become less visually prominent. For example, the \verb|find_peaks| function identifies a peak at $t=0.11$ for $\bm J_1$, $L=3$ (Fig.~\ref{fig:qhm-time-evolution}(b)), which may be more prone to errors in reconstruction across individual ISL runs, due to its lower height relative to subsequent peaks. Nonetheless, to maintain consistency and methodological rigor, initial peaks are defined as identified by this automated procedure. 

With this nuance in mind, Fig.~\ref{fig:qhm-J-phase-error} illustrates key trends across varying system size. As expected, for $L=2$, dephasing remains negligible across all runs, yielding mean dephasing values smaller than 0.001 for both cost thresholds. Conversely, for $L=4$, the mean dephasing magnitude is at most 0.003, for both cost thresholds. While this  indicates that average dephasing remains relatively small across larger system scales, this trend is partly caused by lags being offset by leads of comparable magnitude, as can be observed in the error bars. This is reflected in the mean error bar length, going from 0.020 (0.004) at $L=2$ to 0.047 (0.033) at $L=4$ for $C_{-2}$ ($C_{-4}$).  In contrast to the TCM, mean dephasing in the studied HSC systems tends towards negative values, indicating a greater prevalence of lags across individual ISL runs; potentially indicating model-specific behavior. However, the parameter size considered remains limited, and this trend does not hold for all considered $\bm J$ systems. For instance, $\bm J_1$ and $\bm J_\mathrm{main}$ show positive average dephasing across $L\ge3$.

Nevertheless, certain observations can be made regarding performance across $\bm J$. For $L=4$, $\bm J_3$, yields the smallest average dephasing magnitude, $|\tau_\mathrm{lag}|=0.002$, averaged across both $C_\mathrm{suff}$. This is followed by $\bm J_1$ and $\bm J_4$, with 0.013 and 0.015, respectively. Finally, $\bm J_\mathrm{main}$ and $\bm J_2$, produce the largest average dephasing magnitudes: 0.036 and 0.021, respectively. This result complements previous observations on ISL accuracy being smallest for $J_\mathrm{main}$. This is further supported by large variation across runs, with error bars extending to 0.07 (0.09) for $C_{-2}$ ($C_{-4}$). 

Overall, these trends indicate reduced dephasing and variability in systems with increasing interaction strength, as well as showing optimal results for isotropic coupling. Since greater interaction strengths did not always yield earlier peaks in the noiseless curve, the bias observed in TCM systems is avoided; providing more robust evidence. Moreover, the observation dephasing of ISL in the HSC is in line with expectations, indicating that this is a general property of ISL approximation error, rather than model-specific, as previously discussed.

Table \ref{tab:qhm-table} summarizes the general trends in protocol performance across all considered $(\bm J, \bm h)$ HSC configurations. Overall, the performance of plain Trotterization and ZNE degrades significantly for larger systems ($L\ge3$) due to convergence toward noisy asymptotes, indicating high noise levels in the underlying circuits. In contrast, ISL recompilation circumvents this issue and provides more robust time evolution reconstruction, although still demonstrating approximation errors that  vary in scale across $\bm J$ parameterizations.

\begin{table}[htb]
\centering
% \footnotesize
\begin{tabularx}{\linewidth}{|c|c|X|X|X|}
\hline
$L$ & $\bm J$ & plain & ZNE & ISL ($C_{-2}$/$C_{-4}$) \\[0.5em]  \hline
2 & All & Close to noiseless evolution & Marginal improvement & Slightly worse than Trotter \\[2em] \hline
3  & All  & Rapid convergence to $1/N_q^2$ & Minor improvements, unbounded fluctuations  & Visually stable reconstruction, lowest $\tilde{\varepsilon}$ \\[2em]  \hline
4 & $\bm J_{3-4}$ & Faster convergence to asymptote  & Minimal improvement, some fluctuations persist & Slight worsening relative to $L=3$ \\[2em]  \hline
4 & Rest & Same as for $\bm J_{3-4}$ & Same as for $\bm J_{3-4}$ & Greater errors relative to $L=3$ \\[2em]  \hline
\end{tabularx} \caption{Summary of qualitative performance in accuracy for all tested HSC parameterizations, for each protocol.}
\label{tab:qhm-table}
\end{table}

\subsection{Cross-System Comparison of Algorithmic Behavior}
Having examined the behavior of the studied protocols within individual instances of both the TCM and the HSC, additional cross-model analyses are considered to further investigate trends between the two models. In particular, this comparative analysis is organized as follows. 

Section~\ref{subsubsec:fidelity} presents the time evolution of fidelities for plain Trotterization and ISL relative to the noiseless evolution, across the two main parameterizations. Section~\ref{subsubsec:circuit-depth-shot-cost} examines the scaling of circuit depths across time steps and system sizes, alongside a discussion of the total circuit shot execution requirements, for all three time-evolution methods in the main parameterizations. Finally, Section~\ref{subsubsec:ISL-variation} presents statistics on the variation across independent ISL evaluations, as well as a comparison between noisy ISL time evolution trends and those obtained through noiseless circuit execution and/or noiseless ansatz optimization, for all parameterizations previously considered.

\subsubsection{Protocol fidelities}
\label{subsubsec:fidelity}

\begin{figure}[htb]
    \centering
    \includegraphics[width=0.9\textwidth]{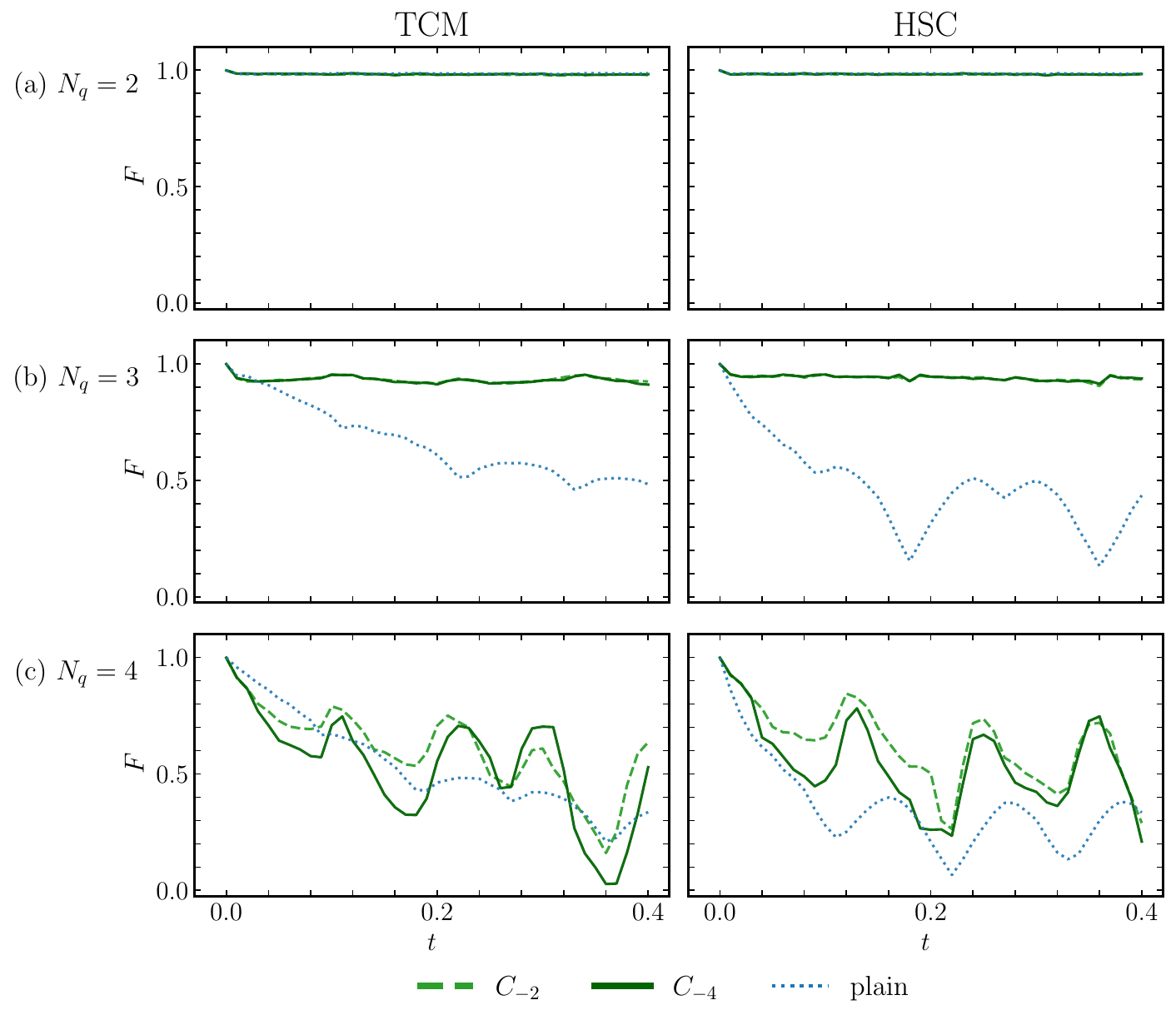}
    \caption{Fidelity $F$ between the noisy quantum state and the corresponding noiseless Trotterized state, for the main TCM parameterization (left) and the main HSC parameterization (right). Results are shown for qubit counts (a) $N_q=2$, (b) $N_q=3$, and (c) $N_q=4$, with $k=k_\mathrm{max}$. Blue dotted curves: plain Trotterization. Green dashed and solid curves: ISL median fidelity across all tested runs, for cost thresholds $C_{-2}$ and $C_{-4}$, respectively.}
    \label{fig:fidelity}
\end{figure}

As outlined in Section~\ref{sec:simulations}, a key metric for evaluating the performance of simulation protocols is the fidelity $F\equiv F(\rho, \ket{\varphi})$ of output states $\rho_n$ relative to the ideal (noiseless) Trotterized evolution $\ket{\varphi}$. The results for different system sizes are shown in Fig.~\ref{fig:fidelity}.

For $N_q=2$, both plain Trotterization and ISL achieve fidelities close to unity throughout the evolution, with no clear advantage for either method. For $N_q=3$, ISL shows a clear benefit, with both $C_\mathrm{suff}$ values exhibiting minimal fidelity decay, whereas plain Trotterization suffers a pronounced drop---particularly in the main HSC system. For $N_q=4$, ISL fidelities display stronger fluctuations but generally remain higher than those from plain Trotterization, which exhibits a more severe and monotonic fidelity decay. The exception is $C_{-4}$ in the TCM, where plain Trotterization achieves marginally larger mean fidelities (0.538) compared to this method (0.530).

Moreover, across both parameterizations, no clear fidelity advantage is observed from using a stricter $C_\mathrm{suff}$ threshold in ISL. For $N_q\le3$, the results for both thresholds are visually indistinguishable, while for $N_q=4$, $C_{-2}$ mean fidelities exceed those of $C_{-4}$ by 0.058 (0.090) in the TCM (HSC). This supports earlier observations that performance differences between thresholds are generally small, with the largest deviations occurring at $N_q=4$. In principle, a tighter threshold should improve accuracy and fidelity by enabling more extensive optimization of the circuit parameters. However, due to the non-negligible noise in the target circuits, optimizations rarely terminate by surpassing the threshold. Moreover, stricter optimization may actually amplify deviations from the noiseless evolution: because the optimization process minimizes a noisy approximation of the true cost function (which corresponds to fidelity), it can systematically steer the reconstructed state away from the ideal noiseless one. 

Overall, the results validate the applicability of ISL in improving circuit fidelities over plain Trotterization, extending beyond correcting expectation values, which is not possible with QEM methods. Consequently, a direct comparison with ZNE is not feasible here, as ZNE only mitigates errors through aggregate circuit measurements and therefore cannot be directly associated with a corrected state vector.

However, addressing the pronounced fidelity decay observed at $N_q=4$ is essential to ensure ISL scalability, to allow for sustained fidelities in larger system sizes and over longer time evolutions. As such, improving suboptimal parameter optimization and ansatz construction under noise remains a key consideration, for instance through QEM applied to ISL cost function evaluations, potentially enabling tighter cost thresholds to achieve finer-grained tuning.

\subsubsection{Circuit Evaluation Cost and Depth Analysis }\label{subsubsec:circuit-depth-shot-cost}
As discussed throughout the text, the depth of the resulting circuits obtained through Trotterization and ISL recompilation plays a key role in determining algorithmic performance. Here, circuit depth refers to the length of the longest path in the transpiled circuit, measured in \texttt{ibm\_nairobi} native gates. For ZNE, the reported depth corresponds to the longest folded circuit, which is three times the plain depth under the chosen noise amplification scheme. In NISQ hardware, large gate noise and limited native two-qubit connectivity mean that Trotterized Hamiltonians with a larger number of terms acting on qubits not connected in the hardware topology are more susceptible to noise (see Fig.~\ref{fig:backend-coupling-map}). In addition to introducing noise into Trotterization and ZNE, higher-depth Trotter steps can limit the accuracy of ISL optimization, as these correspond to deeper, and therefore noisier, target circuits to be recompiled at each time step. Furthermore, analyzing circuit depths in optimized ISL circuits can complement prior results on the cost evolution (see Fig.~\ref{fig:cost-progress}); the final number of ansatz layers in ISL does not always translate directly to circuit depth, as the latter is also influenced by ISL transpilation and posterior Qiskit transpilation to the QPU, potentially benefitting from more compact representations.

Another critical metric for assessing scalability is the total number of circuit executions $c_\mathrm{tot}$ required across all Trotter steps (see Section~\ref{sec:resource-estimation} for the definition). These two metrics capture complementary aspects of algorithmic practicality: circuit depth directly relates to noise accumulation per execution, while $c_\mathrm{tot}$ reflects the overall runtime burden on quantum hardware. The remainder of this subsection first focuses on circuit depth trends, followed by a combined analysis of depth alongside total circuit executions. 

\begin{figure}[hbt]
    \centering
    \includegraphics[width=0.95\linewidth]{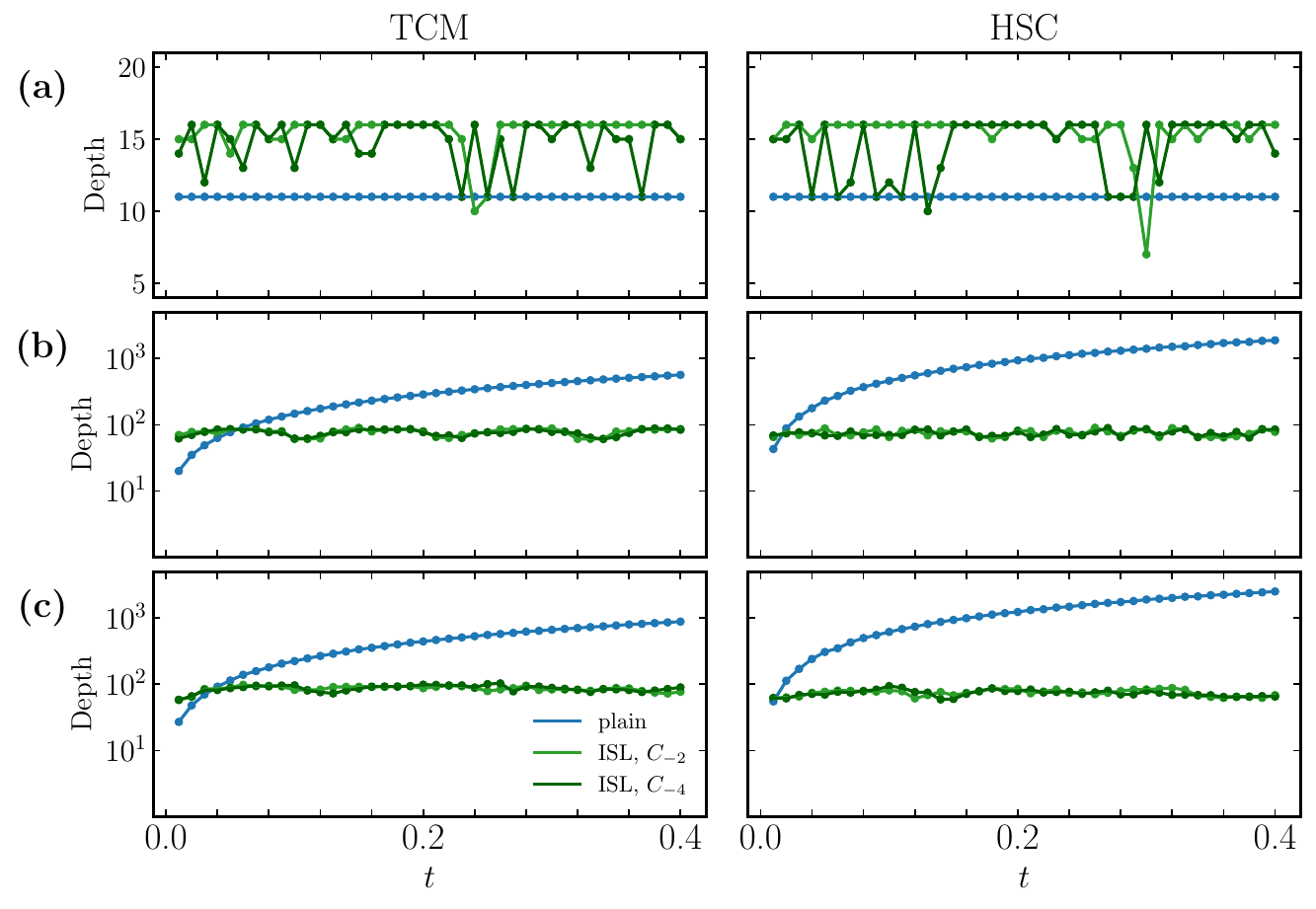}
    \caption{Depth of the time-evolution circuits for plain Trotterization and ISL, as a function of the time step, $t$, measured in the \texttt{ibm\_nairobi} native gates, for qubit counts (a) $N_q=2$, (b) $N_q=3$, and (c) $N_q=4$. For visual clarity, the y-axis scale of the $N_q=2$ results has been adjusted from the other rows. }
    \label{fig:TCM-vs-HSC-depth}
\end{figure}

\begin{figure}[htb]
    \centering
    \includegraphics[width=0.85\textwidth]{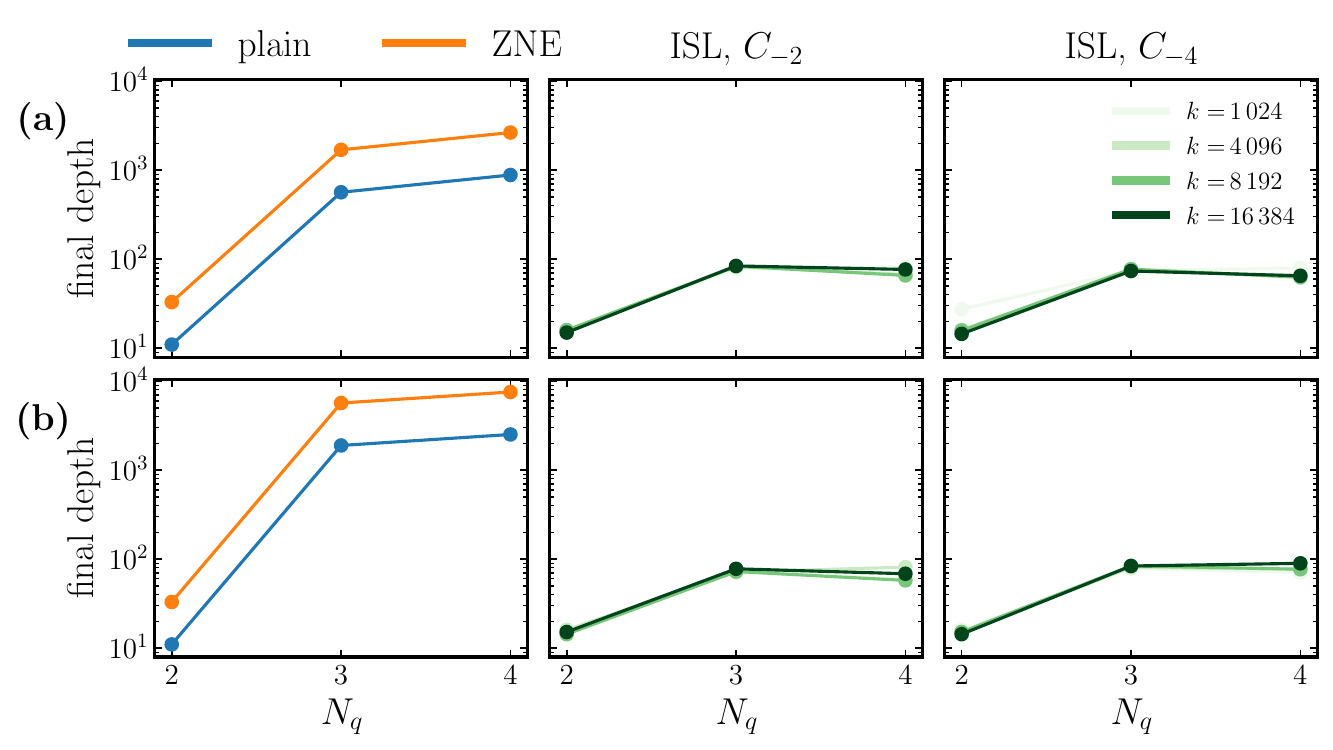}
    \caption{Depth of the final time-evolution circuits achieved by the different methods as a function of the qubit-size $N_q$ for the main parameterizations studied in (a) the TCM and (b) the HSC.}
    \label{fig:final-depth-scaling}
\end{figure}

The circuit depths for plain Trotterization and ISL for each time step after transpiling the circuits with the Qiskit transpiler are presented in Fig.~\ref{fig:TCM-vs-HSC-depth}, while final depths as a function of system size $N_q$ are shown in Fig.~\ref{fig:final-depth-scaling}. For ISL, mean depths obtained across six independent runs are reported for all systems. Although this analysis is limited to the main parameterizations of both models, structural similarities in the Hamiltonians across the systems studied---and consequently in the corresponding circuit implementations---suggest that these results are indicative of broader trends, including those in the alternative parameterizations.

Figure~\ref{fig:TCM-vs-HSC-depth}(a) shows that efficient backend transpilation allows plain Trotterization to yield constant circuit depths ($=11$) for $N_q=2$ across both models. The shallow depths obtained explain the near-noiseless performance of plain Trotterization in two-qubit systems. Consequently, ZNE offers only marginal error mitigation, while the rigid ansatz structure in ISL generally produces larger depths and introduces approximation errors. For all two-qubit TCM and HSC parameterizations studied, and likely broader domains of both models, fixed-depth Trotterized evolution appears attainable with the employed methodology, explaining the comparable performance of these algorithms in this domain.

For $N_q\ge3$ (see Fig.~\ref{fig:TCM-vs-HSC-depth}(b)--(c)), plain Trotterization circuit depths grow approximately linearly with the number of time steps. This results in final depths roughly two orders of magnitudes larger when moving from $N_q=2$ to $N_q=3$, and increasing further for $N_q=4$ (see Fig.~\ref{fig:final-depth-scaling}). In the HSC case, individual Trotter steps are deeper still, due to the larger number of Hamiltonian terms and the additional qubit-swapping operations needed to implement interactions consistent with the system's cyclical coupling pattern, not directly allowed by the native QPU topology for $N_q\ge3$. This is in contrast to the TCM, where the qubit connectivity directly allows for circuits implementing the star-shaped interactions characterizing TCM systems of up to four qubits (see Fig.~\ref{fig:backend-coupling-map}), hence avoiding significant gate overhead associated with qubit-swapping. This relative depth difference explains the larger amplitude errors for plain Trotterization and for ZNE in the HSC time evolution (see Fig.~\ref{fig:qhm-time-evolution}). The same trend likely extends to the alternative parameterizations, where the higher dimensionality of $(\bm J, \bm h)$ would further increase depths. This is consistent with the stronger decay toward the noisy asymptote for Trotterization and ZNE, as observed in Fig.~\ref{fig:qhm-params-time-evolution}.

In contrast, while ISL-recompiled circuit depth also increases from $N_q=2$ to $N_q\ge 3$, it remains bounded at approximately 100 gates across all tested systems and models, including variations in $C_\mathrm{suff}$ and $N_q$. As a result, ISL depths are quickly surpassed by plain Trotterization for all systems, yielding final depths that scale more favorably for ISL. However, significant fluctuations in depth persist between time steps (on the order of 50 gates), likely correlating with variations in ansatz layers across time steps, as seen in the cost evolution trends in Fig.~\ref{fig:cost-progress} (and Fig.~\ref{fig:HSC-cost-progress} in the Appendix). Besides oscillatory-like deviations in ISL depths for $N_q=3$ in the TCM, no consistent systematic trends are observed across systems. 

The relatively constant ISL depth, even at $N_q=4$, may indicate insufficient ansatz expressibility, limiting reconstruction accuracy. This interpretation aligns with the dephasing and reduced probability amplitudes observed in the $N_q=4$ time evolution for both systems. However, it is questionable whether deeper ISL ansätze could improve accuracy in the tested noisy backend, given the increasing per-layer errors observed in the cost evolution trends.

\begin{figure}[htb]
     \centering
     \includegraphics[width=0.9\textwidth]{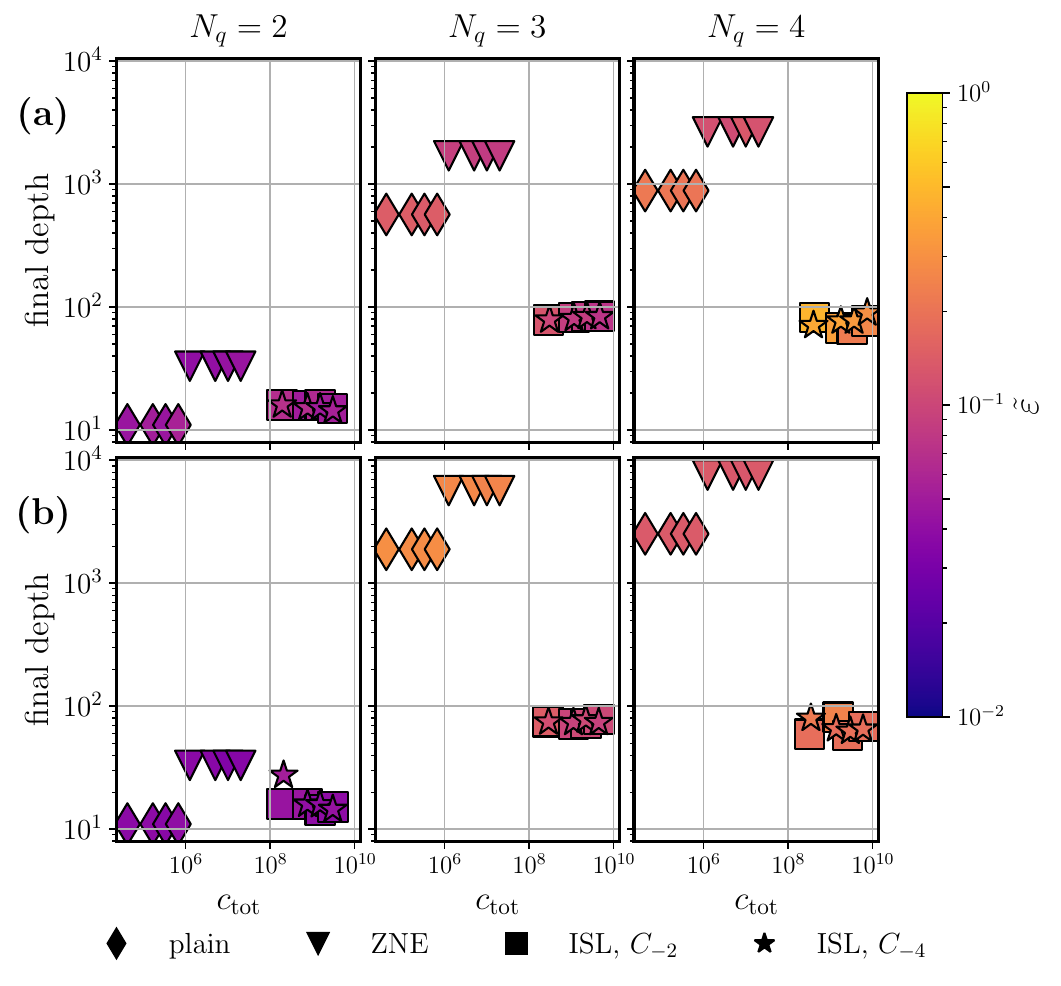}
     \caption{Final circuit depth, total circuit executions $c_\mathrm{tot}$, and the median absolute error $\tilde{\varepsilon}$ in the probability relative to noiseless Trotter evolution. All protocols are considered, for both the (a) main TCM and (b) main HSC parameterizations, evaluated across all four tested circuit shot counts $k$. For all subplots, an increase in $k$ always corresponds to a rightward shift in the figure for all methods, meaning a higher $c_\mathrm{tot}$.}
     \label{fig:colorbar}
\end{figure}

The relationship between final circuit depth and $c_\mathrm{tot}$ is summarized in Fig.~\ref{fig:colorbar}, along with the median absolute error $\tilde{\varepsilon}$ relative to the noiseless Trotter evolution. For plain Trotterization and ZNE, $c_\mathrm{tot}$ is computed using the analytical formulas discussed in Section~\ref{sec:resource-estimation} for all tested $k$ values. For ISL, final depths and $c_\mathrm{tot}$ are averaged across independent runs, with total cost approximated by Eq.~\eqref{eq:isl-cost} using average $N_\mathrm{ce}$ and $N_l$. For completeness, an alternative plot focusing solely on $c_\mathrm{tot}$ scaling across $N_q$ is provided in Section~\ref{app:scaling-executions} of the Appendix.

Figure~\ref{fig:colorbar} highlights the substantial $c_\mathrm{tot}$ overhead of ZNE and ISL compared to plain Trotterization. For ZNE, this overhead is exactly determined by the chosen noise amplification protocol, here producing a factor-of-30 increase, as discussed earlier. While this sampling cost could be reduced by adjusting amplification parameters (e.g., lowering $n_\mathrm{avg}$), the large errors and instability observed for the HSC for $N_q\ge3$ (see Fig.~\ref{fig:qhm-time-evolution}) suggests that such sampling overhead may be necessary to maintain stability in larger HSC instances. Indeed, for this parameterization, final ZNE depths reach almost $10^4$ for $N_q=4$, while providing no clear advantage in $\tilde{\varepsilon}$ relative to plain Trotterization. In contrast, in the main TCM systems, ZNE with $k=1\,024$ shots is already able to achieve low $\tilde{\varepsilon}$ across all system sizes, with the smallest $c_\mathrm{tot}$ overhead relative to plain Trotterization---even outperforming all ISL recompilations for $N_q=4$.  

In ISL, results across all systems illustrate significant $c_\mathrm{tot}$ overhead, relative to the other methods. Most of these trends remain stable across $C_\mathrm{suff}$ values, further highlighting ISL's inability to benefit from tighter optimization routines under noisy conditions, due to the nature of its cost history (see Fig.~\ref{fig:cost-progress}). Similar behavior appears across both models for fixed $N_q$, suggesting underlying ISL properties that are not system-specific. In particular, ISL with $k=1\,024$ shots already requires approximately two orders of magnitude more execution shots than plain Trotterization with $k_\mathrm{max}$ across all systems and both $C_\mathrm{suff}$ (equivalently, over three times the executions of ZNE with $k_\mathrm{max}$). For ISL with $k_\mathrm{max}$, $c_\mathrm{tot}$ approaches $\sim10^{10}$ for $N_q=4$. Indeed, $c_\mathrm{tot}$ scales approximately linearly with $k$ across all systems. Although limited sample sizes prevent firm conclusions, ISL data suggests that $N_\mathrm{ce}$ and $N_l$ remain roughly constant for fixed $N_q$ (and hence $N_\mathrm{qp})$, producing the observed linear scaling, as predicted by the analytical approximation. Nevertheless, non-negligible variability in $N_l$ is reflected in final depths across $k$. Averaging over both $C_\mathrm{suff}$, ISL with $k_\mathrm{max}$ yields $c_\mathrm{tot}\sim 3.0\times 10^9$ in both systems for $N_q=2$, increasing to $\sim7.5\times10^9$ ($\sim6.1\times10^9$) in the TCM (HSC) for $N_q=4$. Given the substantial shot noise in four-qubit systems, potentially requiring increased $k$, this scaling is unfavorable for larger $N_q$.

In summary, ISL and ZNE exhibit diverging resource profiles for DQS in the NISQ era: ISL achieves lower depths at the cost of dramatically higher $c_\mathrm{tot}$, while ZNE depth remains tied to plain Trotterization but incurs a fixed multiplicative execution overhead. The high $c_\mathrm{tot}$ overhead of ISL was also reported by Fitzpatrick et al.~\cite{fitzpatrick_evaluating_2021} in their comparison of Trotterization with ISL and alternative VQS approaches. Although increasing executions can improve ISL results for smaller systems, its scalability remains a concern. Similarly, ZNE faces inherent limitations in mitigating noise for deep Trotter circuits, particularly in systems with Hamiltonian with multiple non-local terms, and connectivity constraints that prevent optimal qubit mapping.

\subsubsection{Variation Across ISL Recompilations}\label{subsubsec:ISL-variation}
The purpose of this analysis is to assess the variability of ISL time evolution results across repeated recompilations, and to determine how this variability depends on system type, parameterization, and cost threshold. As discussed in previous sections, the recursive nature of ISL means that small stochastic differences in early steps can propagate through the time evolution, leading to variations in amplitude errors, dephasing, and qualitative behavior between independent runs. Evaluating this variability across multiple models and parameter sets provides insight into the robustness of ISL and its sensitivity to noise.

For clarity, three ISL variants are considered:
\begin{itemize}
    \item \textbf{Noisy ISL}: recompilation and evaluation carried out entirely on the noisy backend, as studied in previous sections.
    \item \textbf{Noiseless ISL}: recompilation and evaluation carried out on an ideal (noise-free) backend.
    \item \textbf{Mixed ISL}: recompilation carried out without noise\footnote{While maintaining the qubit-pair connectivity dictated by the QPU topology of \texttt{ibm\_nairobi}.}, followed by execution on the noisy backend.
\end{itemize}

This comparison separates the effects of noise during recompilation from those arising in the circuit execution, while also providing a reference for the ideal performance limit of ISL.

For each system and ISL variant, multiple independent time evolutions and aggregate statistics were performed. For noisy ISL, the six runs reported in earlier sections are considered,\footnote{The discrepancy arises from five configurations being performed for aggregate statistics, while an additional generic run was considered to provide a randomly selected cost history; this convention was extended to the other parameterizations for consistency in noisy ISL results.} whereas noiseless and mixed ISL were each evaluated over five independent runs per configurations. Both cost thresholds $\{C_{-2}, C_{-4}\}$ were considered for all ISL types. Given the high computational cost of ISL simulations, the sample size tested is limited. Tentative conclusions can only be drawn where trends among independent runs are consistent.

\begin{figure}[htb]
    \centering
    \includegraphics[width=\linewidth]{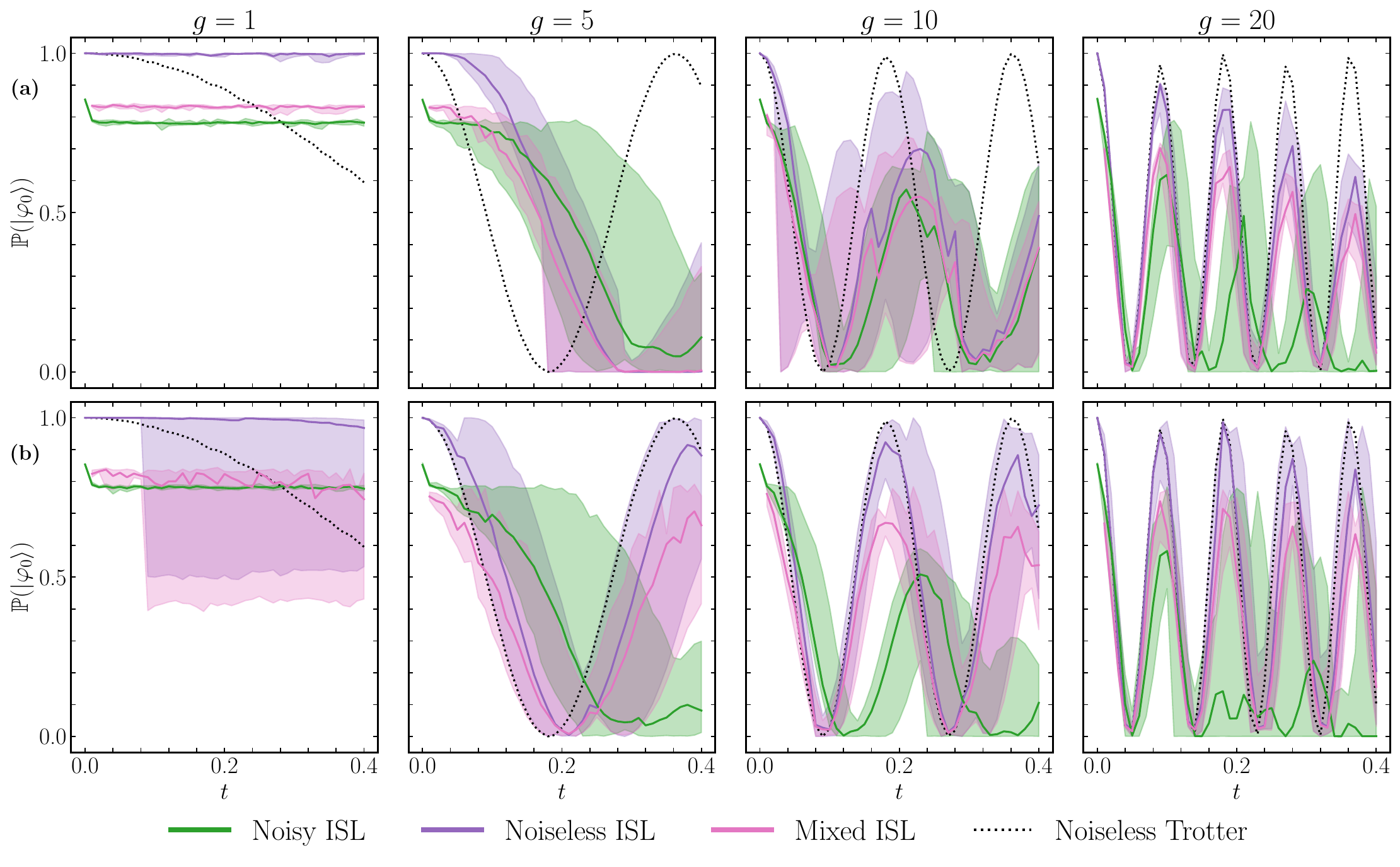}
    \caption{ISL time evolution for all tested TCM systems, for $N_q=4$. Methods: noisy ISL (green), noiseless ISL (purple), and mixed ISL (pink). Six independent runs are shown for noisy ISL and five for noiseless/mixed ISL. The bold line is the median expectation value; shading shows the min-max range at each time step. Results are for (a) $C_{-2}$ and (b) $C_{-4}$.}
    \label{fig:TCM-ISL-reps-4_qubits}
\end{figure}

\begin{figure}[htb]
    \centering
    \includegraphics[width=\linewidth]{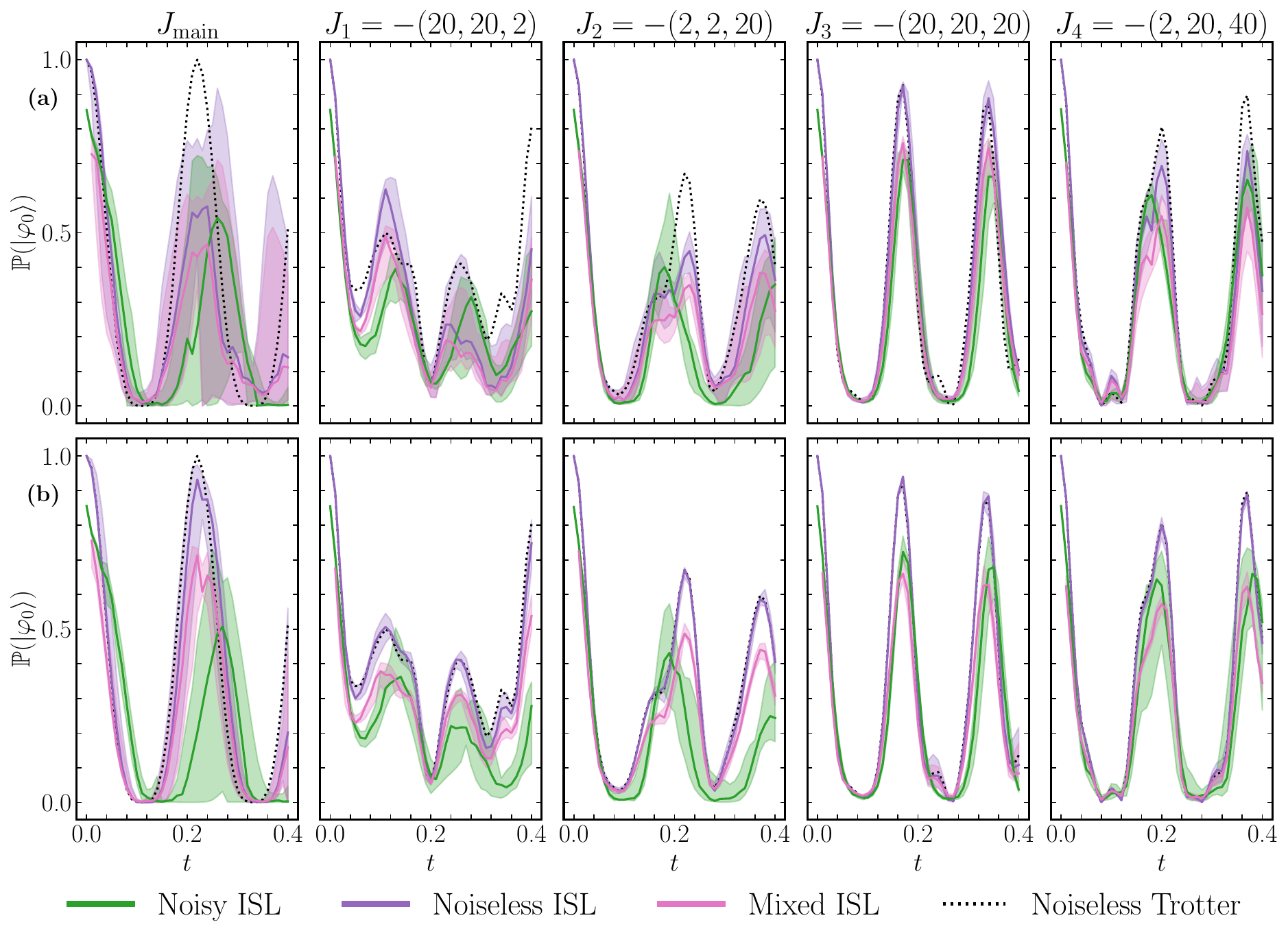}
    \caption{ISL time evolution for all tested HSC systems, for $N_q=4$. Methods: noisy ISL (green), noiseless ISL (purple), and mixed ISL (pink). Six independent runs are shown for noisy ISL and five for noiseless/mixed ISL. The bold line is the median expectation value; shading shows the min-max range at each time step. Results are for (a) $C_{-2}$ and (b) $C_{-4}$.}
    \label{fig:HSC-ISL-reps-4_qubits}
\end{figure}

Figures~\ref{fig:TCM-ISL-reps-4_qubits} and~\ref{fig:HSC-ISL-reps-4_qubits} show the results for the main parameterizations of the TCM and HSC, respectively, at $N_q=4$, with analogous plots for smaller $N_q$ given in Appendix~\ref{app:rep-ISL-smaller-Nq}.

To complement these results, overall ISL performance is summarized in Table~\ref{tab:isl-variation-summary}, which contains two key metrics for each configuration at $N_q=4$: (i) the mean amplitude spread,  $\overline{\Delta P}\equiv\frac{1}{N_T}\sum_t{\Delta P(t)}$, with $\Delta P(t) = \max_r P_r(t)-\min_r P_r(t)$, and (ii) the median absolute error $\tilde{\varepsilon}$.

\begin{table}[htb] \centering \scriptsize \setlength{\tabcolsep}{4pt} 
\label{tab:isl-variation-summary} \begin{tabular}{llcccc} \toprule Model & Param. & \multicolumn{2}{c}{$\overline{\Delta P}$} & \multicolumn{2}{c}{$\tilde{\varepsilon}$} \\ \cmidrule(lr){3-4} \cmidrule(lr){5-6} & & $C_{-2}$ & $C_{-4}$ & $C_{-2}$ & $C_{-4}$ \\ \midrule \multirow{3}{*}{TCM} & $g=1$ & 0.013, 0.014, \textbf{0.009} & \textbf{0.012}, 0.336, 0.382 & 0.143, 0.122, \textbf{0.111} & 0.141, 0.111, \textbf{0.106} \\ & $g=5$ & 0.389, \textbf{0.217}, 0.252 & 0.368, \textbf{0.364}, 0.469 & 0.477, \textbf{0.415}, 0.499 & 0.390, 0.182, \textbf{0.153} \\ & $g=10$ & \textbf{0.406}, 0.425, 0.521 & \textbf{0.290}, 0.294, 0.382 & \textbf{0.261}, 0.272, 0.270 & 0.292, 0.123, \textbf{0.050} \\ & $g=20$ & 0.349, \textbf{0.158}, 0.196 & 0.425, \textbf{0.252}, 0.323 & 0.175, 0.158, \textbf{0.045} & 0.213, 0.142, \textbf{0.033} \\[2pt]
& avg. & 0.289, \textbf{0.204}, 0.245 & \textbf{0.274}, 0.321, 0.389 & 0.264, 0.242, \textbf{0.231} & 0.259, 0.140, \textbf{0.086} \\

\midrule \multirow{3}{*}{HSC} & $\bm J_\mathrm{main}$ & \textbf{0.238}, 0.322, 0.398 & 0.208, \textbf{0.144}, 0.181 & 0.184, 0.067, \textbf{0.064} & 0.182, 0.074, \textbf{0.032} \\ & $\bm J_1$ & \textbf{0.121}, 0.124, 0.153 & 0.122, \textbf{0.053}, 0.062 & 0.136, 0.128, \textbf{0.087} & 0.145, 0.096, \textbf{0.013} \\ & $\bm J_2 $ & 0.133, \textbf{0.109}, 0.136 & 0.113, \textbf{0.032}, 0.036 & 0.113, 0.084, \textbf{0.035} & 0.124, 0.068, \textbf{0.005} \\ & $\bm J_3$ & 0.098, \textbf{0.058}, 0.069 & 0.065, \textbf{0.044}, \textbf{0.044} & 0.074, 0.043, \textbf{0.033} & 0.072, 0.039, \textbf{0.007} \\ & $\bm J_4$ & 0.082, 0.087, \textbf{0.011} & 0.157, 0.028, \textbf{0.027} & 0.045, 0.071, \textbf{0.036} & 0.058, 0.052, \textbf{0.005} \\[2pt] 
& avg. & \textbf{0.134}, 0.140, 0.153 & 0.133, \textbf{0.060}, 0.070 & 0.110, 0.079, \textbf{0.051} & 0.116, 0.066, \textbf{0.012} \\\bottomrule \end{tabular}\caption{Summary of ISL $\overline{\Delta P}$ and $\tilde{\varepsilon}$ metrics for \(N_q=4\), across all TCM and HSC parameterizations. Each cell is N, M, NL for noisy, mixed, and noiseless ISL, with the smallest value per cell highlighted in bold.} \end{table}

% Discussing variations on performance (amplitude errors, dephasing?) going from noisy -> mixed -> noiseless 
Across most parameterizations and both $C_\mathrm{suff}$ values, a consistent trend is observed: noiseless ISL does not always minimize $\overline{\Delta P}$. Counterintuitively, it often exhibits larger spreads than mixed ISL, and even noisy ISL. Mixed ISL achieves the smallest spread in 10/18 cases\footnote{The eighteen cases refer to four TCM systems plus five HSC systems, for each of which two cost thresholds are considered.}, compared to only 4/18 for noiseless ISL. A notable example is $g=1$, where noisy ISL yields nearly constant expectation values with minimal spread, while noiseless recompilation (with $C_{-4}$) generally restores oscillatory dynamics and therefore attains larger spread. Thus, larger $\Delta P$ does not necessarily indicate worse median accuracy, with this system as a clear example where noisy ISL is stable but fails to capture underlying dynamics. In most other cases, smaller spread appears to be the result of increased noise in execution, limiting the range of attainable probabilities, potentially due to biasing towards the noisy asymptote. In comparison, relative changes in $\tilde{\varepsilon}$ greatly illustrate the improvements of ideal recompilation and execution. Averaging across all systems, $\tilde{\varepsilon}$ decreases by 12.5\% (53.6\%) for TCM (HSC) at $C_{-2}$, and by 66.8\% (89.7\%) at $C_{-4}$ when going from noisy to noiseless ISL.

% Discuss differences across both C_suff values
Relative error changes reveal clear effects across cost thresholds: the tighter threshold provides substantial gains in noiseless ISL, but not in noisy ISL. For noiseless ISL, the average $\tilde{\varepsilon}$ with $C_{-4}$ is 62.8\% (76.5\%) smaller than with $C_{-2}$ for TCM (HSC) systems, and consistently provides the smallest error across all ISL types in all systems. In contrast, noisy ISL averages change by only 1.9\% in the TCM and $-5.5$\% in the HSC, when changing to the tighter threshold, consistent with earlier discussions (Section~\ref{subsubsec:fidelity}) that noisy optimization rarely falls below either cost threshold, and changes are only marginal. This further provides evidence that tighter thresholds may even introduce greater errors, by amplifying noise effects from cost function evaluations. Mixed ISL trends fall between these extremes, with relative improvements of 42.1\% (TCM) and 16.5\% (HSC). This indicates that optimization gains are primarily realized here when recompilation is noise-free, reinforcing the need for reduced hardware noise. Notably, the relative improvement for HSC systems in this case is smaller, which is likely a result of a greater proportion of error arising from noise in the execution for HSC systems. This results in larger relative improvement going from mixed to noiseless ISL for $C_{-4}$ in average HSC statistics (81.8\%), as compared to the TCM (38.5\%).

% Discussing model and parameterization differences
In addition to patterns across ISL parameters, key trends among both models can be established. In particular, HSC systems exhibit systematically smaller errors and spreads than TCM systems. For instance, going from TCM to HSC systems in noiseless ISL reduces $\tilde{\varepsilon}$ by 77.9\% ($C_{-2})$ and 86.0\% ($C_{-4}$) on average; with similar trends also observed for noisy and mixed ISL, which both have smaller average errors compared to their TCM counterparts. In addition to errors, mean variation is generally smaller in HSC systems, as reflected in average $\overline{\Delta P}$ being larger across TCM systems, for each ISL type. 

Within the TCM, the most challenging parameterizations are $g=1$ and $g=5$, while $\bm J_\mathrm{main}$ and $\bm J_1$ are the most challenging for the HSC, as these yield the largest $\tilde{\varepsilon}$ for noiseless ISL with $C_{-4}$ (here chosen as the reference for optimal performance). Although direct comparison between both models may be limited due to error and spread trends being strongly dependent on noiseless evolution trends (cf. periodic behavior in $N_q=3$ and $N_q=4$ in \ref{fig:qhm-time-evolution}) beyond ISL reconstruction capabilities, qualitative observation of time evolution trends confirms that observed metrics in the TCM are representative of larger errors in the corresponding ISL recompilations. 

In the TCM systems, significant dephasing is observed for $t\ge0.2$, along with insufficient probability amplitudes. While HSC systems also show large dephasing and amplitude errors for $C_{-2}$, these are significantly improved by the tighter threshold. In contrast, noiseless ISL with $C_{-4}$ also reduces phase errors for TCM systems, but non-negligible dephasing can still be seen. This appears to result from greater approximation errors in the ISL recompilation of Trotter circuits for TCM systems, arising from suboptimal ansatz optimization of the TCM Hamiltonian, regardless of hardware constraints. The greater accumulation of these errors therefore leads to the observed dephasing.

\begin{figure}[htb]
    \centering
    \includegraphics[width=0.9\linewidth]{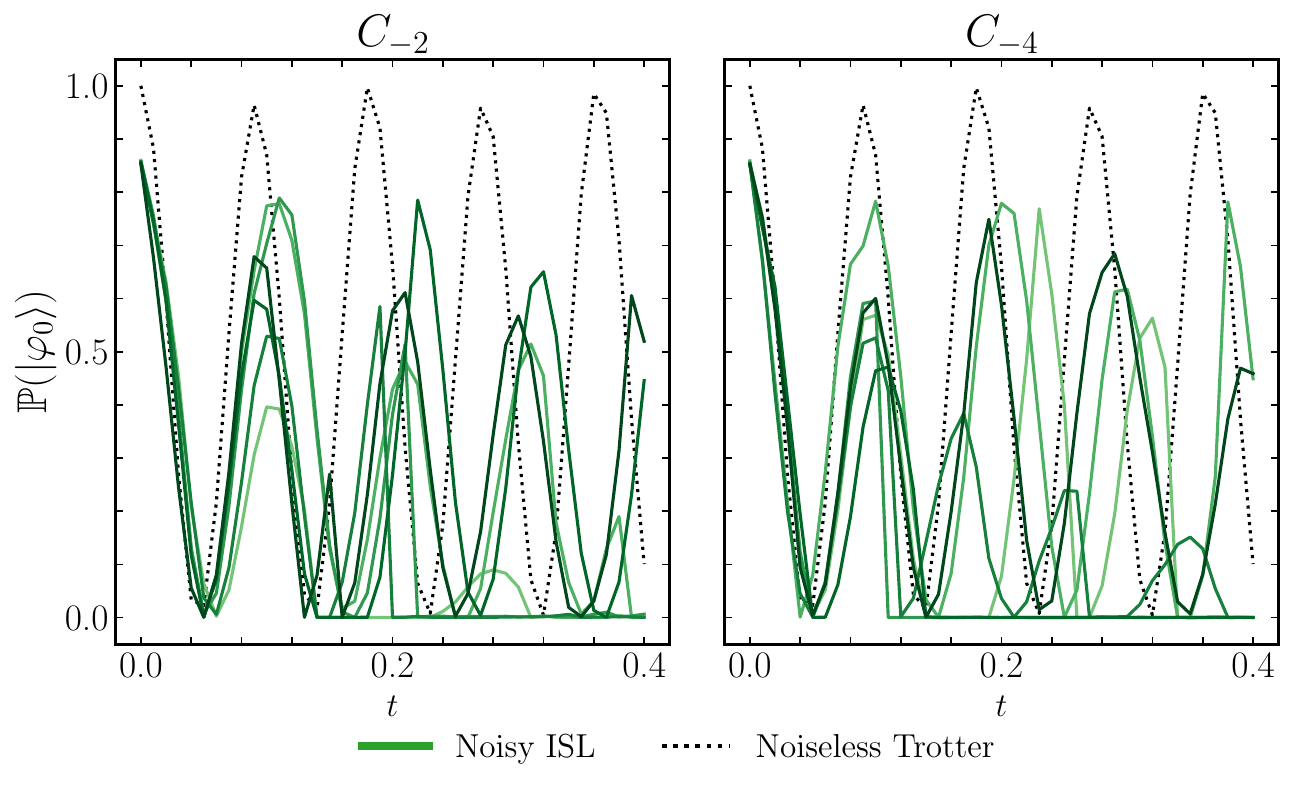}
    \caption{
    Time evolution for all noisy ISL runs for the TCM  with $g=20$, $N_q=4$, for both cost thresholds $\{C_{-2}, C_{-4}\}$, run at $k=k_\mathrm{max}$. Different shades indicate different runs for noisy ISL.}
    \label{fig:TCM-ISL-g-20_4_qubits}
\end{figure}

% Discussing g=20 case in relation to spread
On a related note, the TCM instance with $g=20$ illustrates the limits of median-based error metrics when dephasing varies significantly across runs. Here, noisy ISL runs exhibit positive phase shifts with large variability, and occasional plateauing near $P=0$ (Fig.~\ref{fig:TCM-ISL-g-20_4_qubits}). As such, the median ISL curve obtained smooths out oscillations and yields higher $\tilde{\varepsilon}$. This highlights that median trends, though representative in most cases, can obscure structured variability and dephasing.

\begin{figure}[htb]
    \centering
    \includegraphics[width=0.9\textwidth]{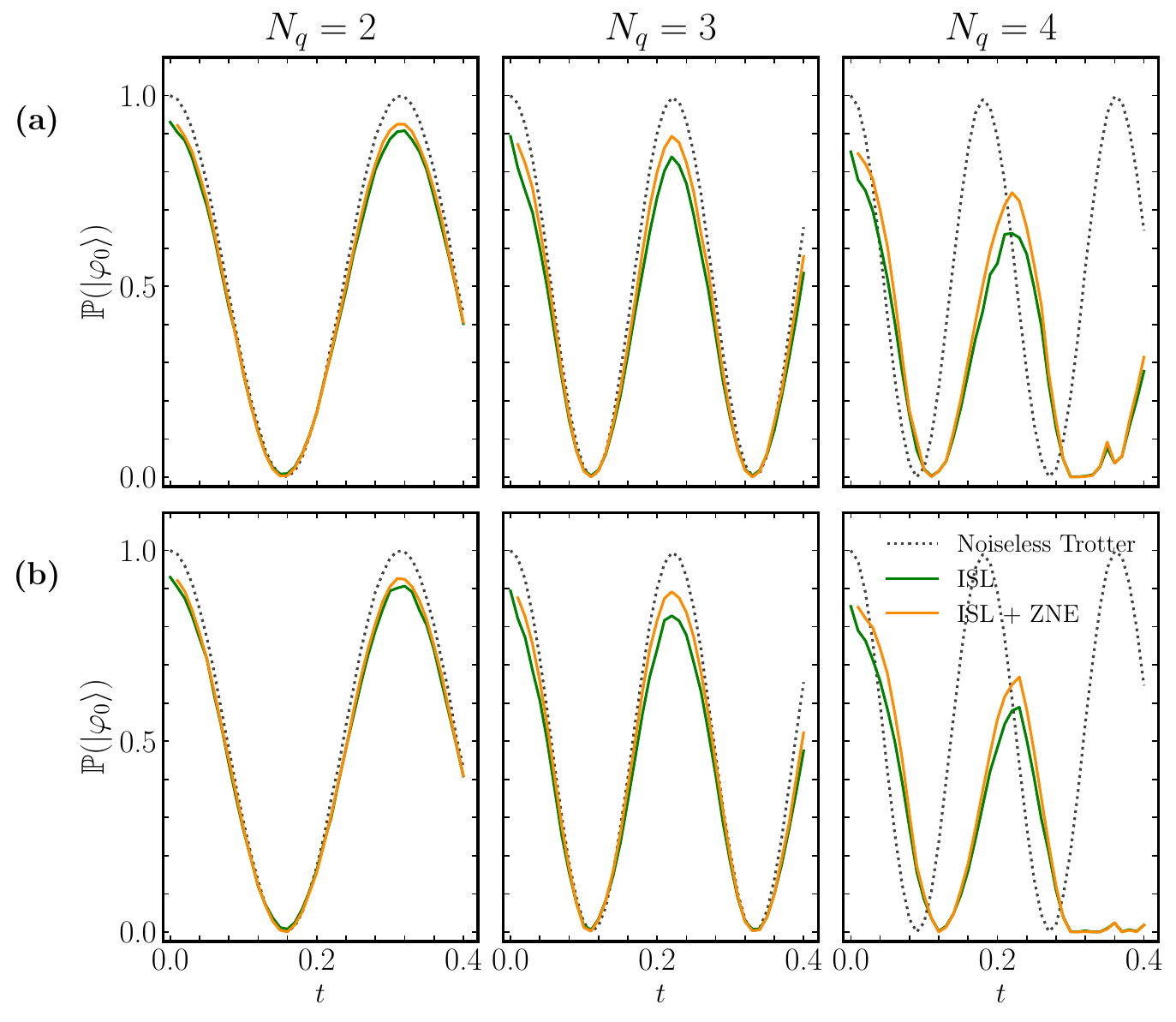}
    \caption{ZNE applied to noisy ISL-recompiled circuits for the main TCM parametrization, for a selected individual run, evaluated at $k = k_\mathrm{max}$, for cost thresholds (a) $C_{-2}$ and (b) $C_{-4}$.}
    \label{fig:mitigated-isl}
\end{figure}

Since ISL recompilation already produces circuits of reduced depth, an additional question is whether execution-level error mitigation can improve results. To test this, representative noisy ISL circuits (main TCM parameterization, $k_\mathrm{max}$) were re-executed with ZNE mitigation, as shown in Fig.~\ref{fig:mitigated-isl}. The improvements were marginal: ZNE reduced some gate-noise effects but did not correct the dominant errors in ISL time evolution for $N_q=4$, namely phase-lags and expectation value plateauing near $P=0$. As a result $\tilde{\varepsilon}$ decreases by $<0.1$\% for $C_{-2}$, and increases by $28.6\%$ for $C_{-4}$; likely due to reduced amplitude error (closer to $P=1.0$) amplifying net errors due to dephasing. This confirms that ISL accuracy is primarily limited by optimization variability and dephasing rather than by execution noise alone. Although applying ZNE within the recompilation loop could in principle address these limitations, its execution overhead would be substantially higher, and such approaches are left for future investigation.

Overall, these results illustrate the high sensitivity of ISL performance to noise in recompilation, with mixed and noiseless variants greatly outperforming noisy ISL for larger systems. Although noise in circuit execution persists, particularly for $N_q=4$, applying error mitigation techniques to execution alone provides marginal improvements, as tested here. Dependence to noise is particularly relevant when considering tighter cost thresholds, which mainly benefit noise-free recompilation, underscoring the need for hardware improvements. Model dependence is pronounced, with HSC parameterizations generally yielding smaller errors and spreads than TCM. These trends persist across mixed and noiseless ISL, indicating that system dependence extends beyond varying levels of noise in the underlying Trotterized circuits, but extends to additional factors, affecting the efficiency of the ansatz structure and parameter optimization. Indeed, noiseless ISL with $C_{-4}$ for TCM systems with weaker interactions (i.e., smaller $|g|$), and for HSC systems with coupling vectors closest to the TCM ($\bm J_\mathrm{main}$ and $\bm J_1$), yields the largest $\tilde{\varepsilon}$ out of all systems. Finally, strong run-to-run variability, particularly in conjunction with structured dephasing, may limit the reliability of single-run predictions, motivating more sophisticated ensemble approaches in ISL implementations.

\subsection{Optimal Methods}

The analysis across all tested systems shows that the optimal method depends strongly on the system type and qubit count. For $N_q=2$, plain Trotterization was consistently the most effective approach, as backend transpilation reduced circuits to minimal depth, yielding errors comparable to ZNE and ISL at negligible execution cost. For larger systems, however, Trotter circuits no longer benefited from such optimizations, and the balance between ISL and ZNE shifted depending on the Hamiltonian structure and system size.

For the TCM, ZNE was identified as the most viable approach. Although ISL outperformed this method for $N_q=3$, it did not present favorable scaling: suffering from significant phase-lags and large amplitude errors for $N_q=4$, leading to higher $\tilde{\varepsilon}$ values. For instance, for $N_q=4$ at $k_\mathrm{max}$, the average noisy ISL error across parameterizations was $0.216$ (across both $C_\mathrm{suff}$), whereas the ZNE average $\tilde{\varepsilon}$ was 28.2\% smaller, at $0.155$ (see Section~\ref{subsubsec:error-summary-TCM}). Moreover, the significantly higher $c_\mathrm{tot}$ overhead of ISL limits its practicality relative to ZNE, which consistently improved upon plain Trotterization while maintaining lower resource demands. While ZNE performance degraded with larger time horizons, it generally remained more robust than ISL under TCM dynamics for this system scale, suggesting a sustained advantage over the other tested methods at greater $N_q$.

In contrast, for the HSC, ISL clearly outperformed the other methods. Averaged across all five parameterizations at $N_q=4$, noisy ISL reduced errors relative to ZNE by 36.8\% (33.3\%), with average errors 0.110 (0.116) for $C_{-2}$ ($C_{-4}$). This advantage stems in large part from the fact that Trotter steps in the HSC system produced deep circuits in the considered backend, leading to significant noise effects. Most importantly, the limited backend topology influenced this outcome, with the \texttt{ibm\_nairobi} device requiring SWAP insertions in HSC Trotter circuits but not in TCM systems. This further amplified the depth difference between Trotterized and ISL circuits (see Fig.~\ref{fig:TCM-vs-HSC-depth}), favoring ISL performance. ISL circuits were able to address this difficulty by truncating Trotterized depths while retaining key dynamical features. Additionally, model parameterization played a role: while $\bm J_\mathrm{main}$ exhibited large errors across recompilations, alternatives such as $\bm J_2$, $\bm J_3$, and $\bm J_4$ yielded the smallest errors for Noisy ISL at $N_q=4$ across both cost thresholds (see Table~\ref{tab:isl-variation-summary}).

\section{Conclusion} 
\label{sec:conclusions}

This study evaluated the performance of Trotterized time evolution under noise for two spin-boson models (TCM and HSC), comparing plain Trotterization, ZNE with circuit folding, and ISL with noisy recompilation. The analysis shows that the optimal strategy depends strongly on both Hamiltonian structure and qubit count. For $N_q=2$, plain Trotterization remained most efficient, benefiting from aggressive backend transpilation optimizations. For larger systems, ZNE was optimal in the TCM, reducing errors by $\sim 28\%$ at $N_q=4$ relative to ISL, while ISL clearly outperformed ZNE in the HSC, achieving error reductions of $33$--$37\%$ across cost thresholds. This advantage, however, came at a steep execution overhead: ISL required around two orders of magnitude more circuit evaluations than ZNE at the highest circuit shot resolution tested ($k_\mathrm{max}$), consistent with earlier concerns about the resource scaling of VQAs~\cite{miessen_quantum_2021}.

From a practical perspective, these results indicate that the viability of quantum simulation on NISQ hardware depends greatly on the structure of the simulated Hamiltonian and device topology. ISL was particularly resilient in cases where Trotter steps produced prohibitively deep circuits (HSC), while ZNE remained the most viable in models with naturally shallow Trotter circuits (TCM). This highlights that no universal strategy exists, and that tailoring mitigation methods to both the physical system and the hardware connectivity is essential for accurate simulations. Moreover, in addition to favorable ISL performance across HSC systems, the results suggest more general applicability to other instances of physical models with stronger and isotropic interactions, as demonstrated  by smaller errors in the corresponding ISL simulations, when performed in both noisy and ideal hardware.

The present study is limited by its focus on qubit numbers $N_q\le4$, simulated noise rather than real hardware data, and a restricted set of Hamiltonian configurations. ISL simulations were limited to $N_q\leq 4$ due to the steep computational overhead. Notably, noisy ISL optimization histories were generally unable to reach the target cost thresholds $C_\mathrm{suff}$ and often showing worsening cost with added layers, following initial improvements. This suggests that ISL routines on NISQ hardware may terminate prematurely or produce under-expressive ansätze at scale. Furthermore, variability across independent ISL recompilations was only partially addressed; larger sample sizes would be needed to systematically quantify this effect. In addition, readout error mitigation was not applied, rendering all noisy results somewhat conservative. These constraints restrict scalability claims, but the observed trends nonetheless provide valuable benchmarks for the trade-offs between ISL and ZNE.

Future work should focus on refining ISL optimization routines. More advanced parameter optimization algorithms---such as Free-Axis selection~\cite{watanabe_2023}, extended use of Rotoselect in place of Rotosolve (the latter being computationally cheaper but potentially less optimal in cost), or state-of-the-art Rotoselect variants with freezing and initialization heuristics~\cite{pankkonen_2025_improving, pankkonen_2025_freezing}---could help overcome stagnation in the cost landscape. Alternatively, leveraging symmetries in the cost landscapes of parameterized quantum circuits could provide greater optimization, as explored in~\cite{fontana_non-trivial_2022}. Segmentation strategies in recursive recompilation, where ISL is applied to multiple Trotter steps at once or to partial steps, may also balance recursion errors with execution overhead. Hybrid approaches that incorporate QEM methods---such as probabilistic error cancellation~\cite{temme_error_2017, kim_evidence_2023} and Pauli twirling~\cite{li_efficient_2017, Cai_2020}---into ISL ansatz construction could further improve robustness against noise, demonstrating optimal performance on NISQ devices. Moreover, tailoring ansatz layers to Hamiltonian-specific symmetries\footnote{For example, exploiting the isotropic XX/YY couplings in the TCM or the cyclic structure of the HSC to minimize computational overhead in the ansatz construction.
} may yield more efficient recompilation. Finally, as simulated systems scale, selecting efficient qubit encodings~\cite{steudtner_fermion--qubit_2018, sawaya_resource-efficient_2020} will also be critical to reducing overhead. Such studies on larger qubit systems will be essential to empirically determine whether improvements in ISL routines can offset its steep execution overhead and establish its viability as a practical technique in addition to QEM in NISQ devices.

In summary, this work provides the first systematic assessment of ISL with noisy optimization steps for Trotterized dynamics in spin-boson models. The results demonstrate that while ISL can outperform traditional ZNE in specific regimes, particularly for HSC Hamiltonians with deep Trotter circuits, its high execution overhead and optimization challenges currently limit scalability. These findings underscore the need for improved optimization routines, ansatz designs, and hybrid strategies before ISL can be extended to simulating larger systems in NISQ devices. Ultimately, establishing whether ISL can evolve into a scalable method will be crucial for realizing practical quantum simulations in the near term.

\section{Acknowledgments}
% Triton acknowledgment
The authors acknowledge the computational resources provided by the Aalto University School of Science “Science-IT” project.

% Business Finland
The authors acknowledge funding from Business Finland for project 8726/31/2022 CICAQU.

%% file: appendix.tex
\section{Incremental structural learning (ISL)}
\label{app:isl}

\begin{figure}[htb]
    \centering
    $V_i(\btheta_i) =$
    \begin{quantikz}
    & \gate{R_{\alpha_{i,1}}(\theta_{i,1})} & \ctrl{1} & \gate{R_{\alpha_{i,2}}(\theta_{i,2})} & \qw \\
    & \gate{R_{\alpha_{i,3}}(\theta_{i,3})} & \targ{} & \gate{R_{\alpha_{i,4}}(\theta_{i,4})} & \qw
    \end{quantikz}
    \caption{A nested CNOT gate, which is a CNOT surrounded by four single-axis rotation gates. Here $\btheta_i = (\theta_{i,1}, \theta_{i,2}, \theta_{i,3}, \theta_{i,4})$ are the rotation angles and $\alpha_{i,k} \in \{x,y,z\}$ the rotation axes.}
    \label{fig:nested-CNOT}
\end{figure}

\noindent
This appendix describes the ISL algorithm in more detail, and is largely based on its original formulation~\cite{jaderberg_minimum_2020}. ISL is a method of recompiling circuits layer-by-layer, starting from a fixed initial state $\zeroket \equiv \ket{0}^{\otimes N_q}$, where $N_q$ is the number of qubits. To select another initial state, a state preparation circuit can be added to the beginning of the target circuit $U$. The target is compiled into a parameterized circuit
\begin{equation} \label{eq:isl-ansatz}
    V(\btheta) = V_1(\btheta_1) V_2(\btheta_2) \cdots V_l(\btheta_l),
\end{equation}
where each layer $V_i(\btheta_i)$ is a nested CNOT gate (see Fig.~\ref{fig:nested-CNOT}). New layers are added until a convergence criterion is reached, therefore the structure of the ansatz and the size of the parameter vector $\btheta$ change during the procedure.

The cost function that ISL seeks to minimize is\footnote{Note that the same considerations as in Section~\ref{subsec:ISL-for-VQS} regarding noisy recompilation hold.}
\begin{equation} \label{eq:isl-cost}
    C(\btheta) = 1 - \left| \braket{\mathbf{0} | V(\btheta)^{\dag} U | \mathbf{0}} \right|^2,
\end{equation}
or in other words, it seeks to maximize the fidelity
\begin{equation} \label{eq:isl-fidelity}
    F\left(V(\btheta)\zeroket,\, U\zeroket\right) = \left| \braket{\mathbf{0} | V(\btheta)^{\dag} U | \mathbf{0}} \right|^2.
\end{equation}
The ansatz is built as an adjoint $V^{\dag}$, with layers appended to the previous state of the cost circuit $V_l^{\dag} V_{l-1}^{\dag} \cdots V_1^{\dag} U \zeroket$ in order to bring it closer to the $\zeroket$ state. Considering an ansatz with $l$ layers (Eq.~\eqref{eq:isl-ansatz}), the ISL algorithm for layer $l+1$ proceeds as follows:

\begin{enumerate}
    \item A nested CNOT $V_{l+1}(\btheta_{l+1})$ is added between the qubit pair that achieves the highest value for a chosen entanglement measure, such as the entanglement of formation \cite{wootters_entanglement_1998}, yielding a new ansatz circuit $V(\btheta) \equiv\linebreak V_1(\btheta_1) \cdots V_l(\btheta_l) V_{l+1}(\btheta_{l+1})$. Evaluating the entanglement measure requires the use of quantum state tomography (QST) to estimate the density matrices between qubit pairs.

    As stated, the goal is to evolve the state $V_l^{\dag} V_{l-1}^{\dag} \cdots V_1^{\dag} U \zeroket$ back into $\zeroket$. Therefore adding $V_{l+1}^{\dag}$ between the most entangled pair of qubits is motivated by an attempt to "disentangle" them.
    
    \item The Rotoselect~\cite{ostaszewski_structure_2021} algorithm is applied to the rotation axes and angles of $V_{l+1}(\btheta_{l+1})$ to minimize Eq. \eqref{eq:isl-cost}. Note that all previous axes and angles belonging to $V_1, V_2, \ldots$ and $V_l$ are kept fixed. After this, Eq.~\eqref{eq:isl-cost} is minimized again, this time with the Rotosolve~\cite{ostaszewski_structure_2021} algorithm, such that all angles $\btheta = (\btheta_1, \btheta_2, \ldots, \btheta_{l+1})$ are allowed to change, but all axes are fixed.
    
    \item Finally, the circuit is simplified by removing duplicate gates and rotation gates whose angles have a magnitude below the user-specified threshold $\theta_\mathrm{th}$. A final evaluation of the cost function determines whether to terminate, or to add another layer and continue from the first step.
\end{enumerate}

An illustration of the construction of the ansatz can be seen in Fig. \ref{fig:isl-steps}.

\begin{figure}[hb]
    \centering
    \begin{adjustbox}{width=\textwidth}
    \tikzset{
        operator/.append style={rounded corners}
        }
    \large
    \begin{quantikz}
        \lstick[wires=4]{$\zeroket$} & \gate[4, style={fill=blue!20}][1.5cm]{U} & \gate[2, style={fill=orange!20}]{V_1(\btheta_1)^{\dag}} & \qw & \qw \midstick[4, brackets=none]{$\cdots$} & \qw & \qw & \qw & \meter{\text{\normalsize\bf\sffamily optimize } \btheta} \\
        & & \qw & \qw & \qw  & \gate[2, style={fill=orange!20}]{V_l(\btheta_l)^{\dag}}\slice[style=black]{{\normalsize\bf\sffamily QST on qubit pairs}} & \qw & \qw & \meter{} \\
        & & \qw & \gate[2, style={fill=orange!20}]{V_2(\btheta_2)^{\dag}} & \qw & & \qw & \gate[2, style={fill=orange!40}]{V_{l+1}(\btheta_{l+1})^{\dag}}\gategroup[2,steps=1,style={draw=none},background,label style={label
position=below,anchor=north,yshift=-0.2cm}]{{\normalsize\bf\sffamily add nested CNOT}} & \meter{} \\
        & & \qw & & \qw & \qw & \qw & \qw & \meter{}
    \end{quantikz}
    \end{adjustbox}
    \caption{Layer-by-layer construction of the ansatz circuit $V(\btheta)$ in ISL. The circuit is initially in the state $V_l^{\dag} V_{l-1}^{\dag} \cdots V_1^{\dag} U \zeroket$. First, QST is performed on all qubit pairs connected in the QPU to identify the most entangled one. A nested CNOT, $V_{l+1}(\btheta_{l+1})^{\dag}$, is appended after this pair, and the circuit parameters are optimized. The optimization may terminate or proceed to add another layer.}
    \label{fig:isl-steps}
\end{figure}

\clearpage
\section{Additional Statistics for ISL}
This appendix contains additional results from the performance of ISL in the considered systems, complementing the data presented in the study. Section~\ref{app:HSC-cost-progression} presents the cost history for an independent run of the main HSC parameterization, complementing the trends discussed for the TCM. Furthermore, Section~\ref{app:scaling-executions} provides an alternative figure illustrating the mean scaling in $c_\mathrm{tot}$ of ISL for the main systems, complementing the discussion in Section~\ref{subsubsec:circuit-depth-shot-cost}. Finally, Section~\ref{app:rep-ISL-smaller-Nq} presents the variations across independent runs for all ISL variants (noisy, mixed, noiseless) for all tested systems for $N_q=2$ and $N_q=3$, complementing the results presented for $N_q=4$ in Section~\ref{subsubsec:ISL-variation}.

\subsection{Cost History for HSC}\label{app:HSC-cost-progression}
\begin{figure}[htb]
    \centering
    \includegraphics[width=\linewidth]{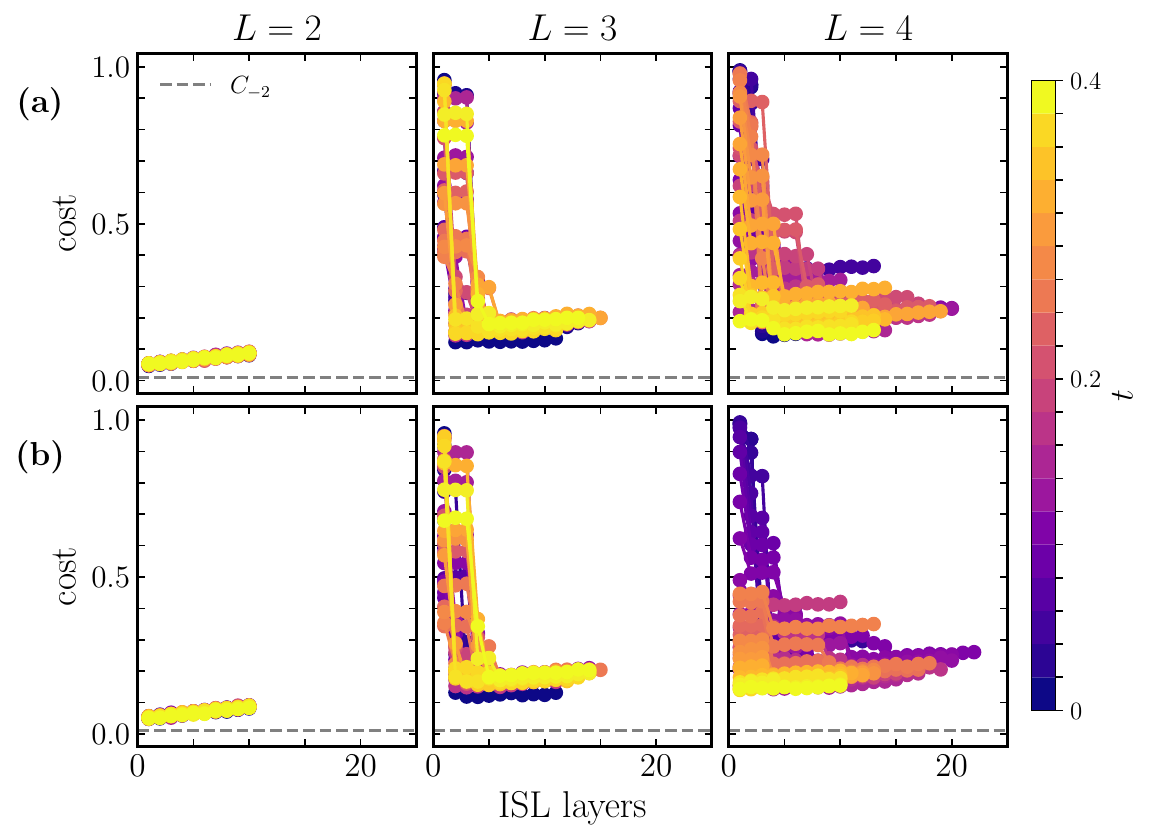}
    \caption{ISL recompilation cost at the end of each layer, for the main HSC parameterization, evaluated at $k_\mathrm{max}$. The rows correspond to an individual ISL recompilation for (a) $C_{-2}$ and (b) $C_{-4}$.}
    \label{fig:HSC-cost-progress}
\end{figure}

As is done in Sec. \ref{subsubsec:tcm-main-data} for the main TCM parameterization, the layer-by-layer cost evolution of ISL ansätze for $k=k_\mathrm{max}$ shots in the main HSC parameterization is illustrated in Fig. \ref{fig:HSC-cost-progress}, for chosen individual recompilations per tested $C_{\mathrm{suff}}$ value.

In this case, it is possible to see that the cost evolution of ISL follows similar patterns to that seen in the main TCM parameterization. Namely, this includes the final cost never falling below either $C_\mathrm{suff}$ threshold, the increasing cost with additional layers, and a comparable number of total layers.   

\subsection{Scaling of circuit executions}
\label{app:scaling-executions}
This appendix shows a different view of the total circuit executions data presented in Fig. \ref{fig:colorbar}, instead emphasizing the dependence on the qubit count $N_q$.

\begin{figure}[htb]
    \centering
    \includegraphics[width=\textwidth]{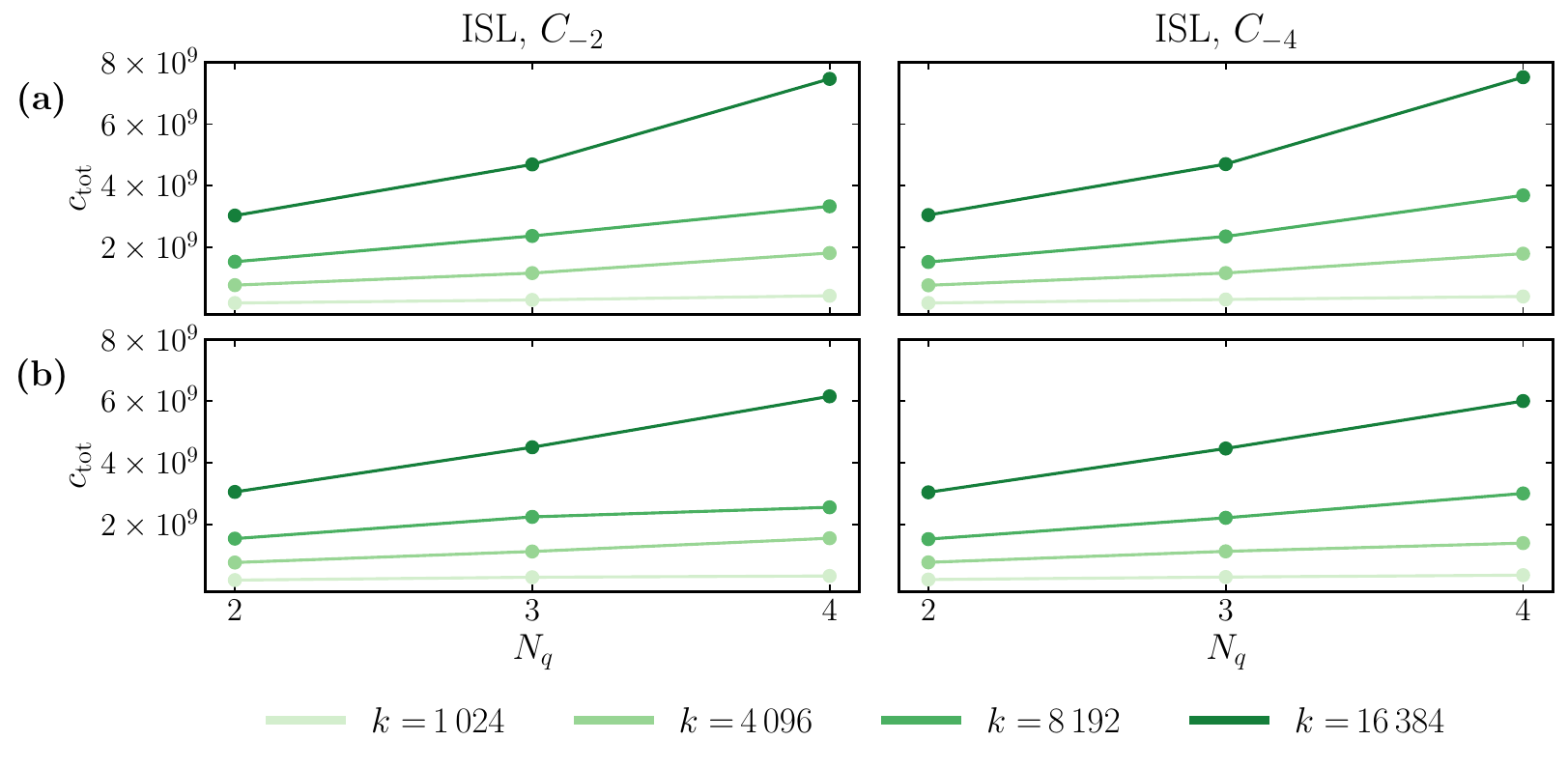}
    \caption{Total circuit executions, $c_\mathrm{tot}$, performed by ISL as a function of $N_q$ and $k$, averaged across independent runs, for the main considered parameterizations of the models: (a) TCM and (b) HSC.}
    \label{fig:total-ce-scaling}
\end{figure}

\subsection{Repeated ISL Evaluations for Smaller Qubit Counts}\label{app:rep-ISL-smaller-Nq}
\begin{figure}[htb]
    \centering
    \includegraphics[width=0.95\linewidth]{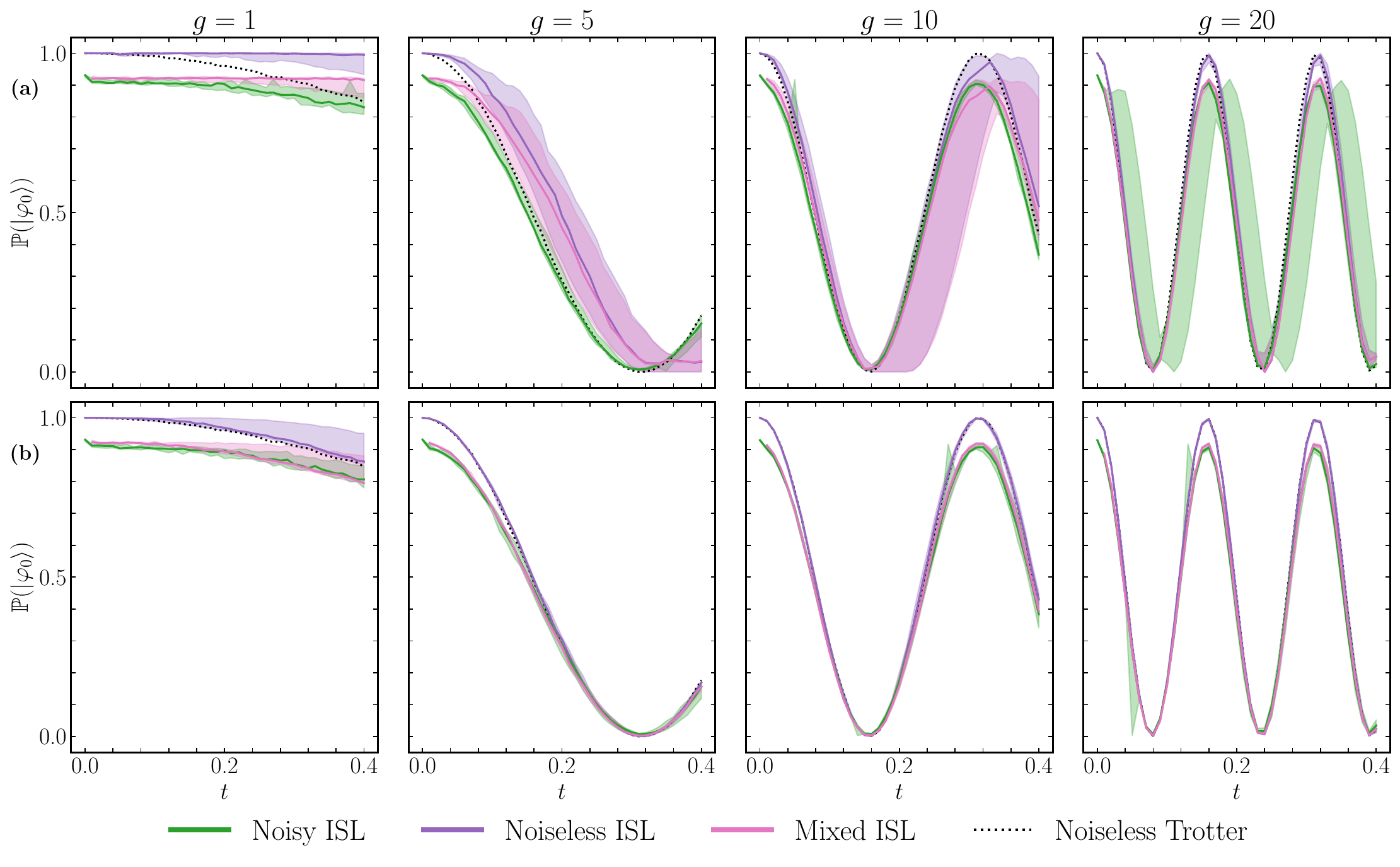}
    \caption{Repeated evaluations of ISL time evolution for all tested TCM systems, for $N_q=2$. The methodology remains the same as explained in Section~\ref{subsubsec:ISL-variation}, with noisy (green), mixed (pink), and noiseless (purple) ISL time evolution being considered for both cost thresholds (a) $C_{-2}$ and (b) $C_{-4}$.}
    \label{fig:TCM-ISL-reps-2_qubits}
\end{figure}

\begin{figure}[htb]
    \centering
    \includegraphics[width=0.95\linewidth]{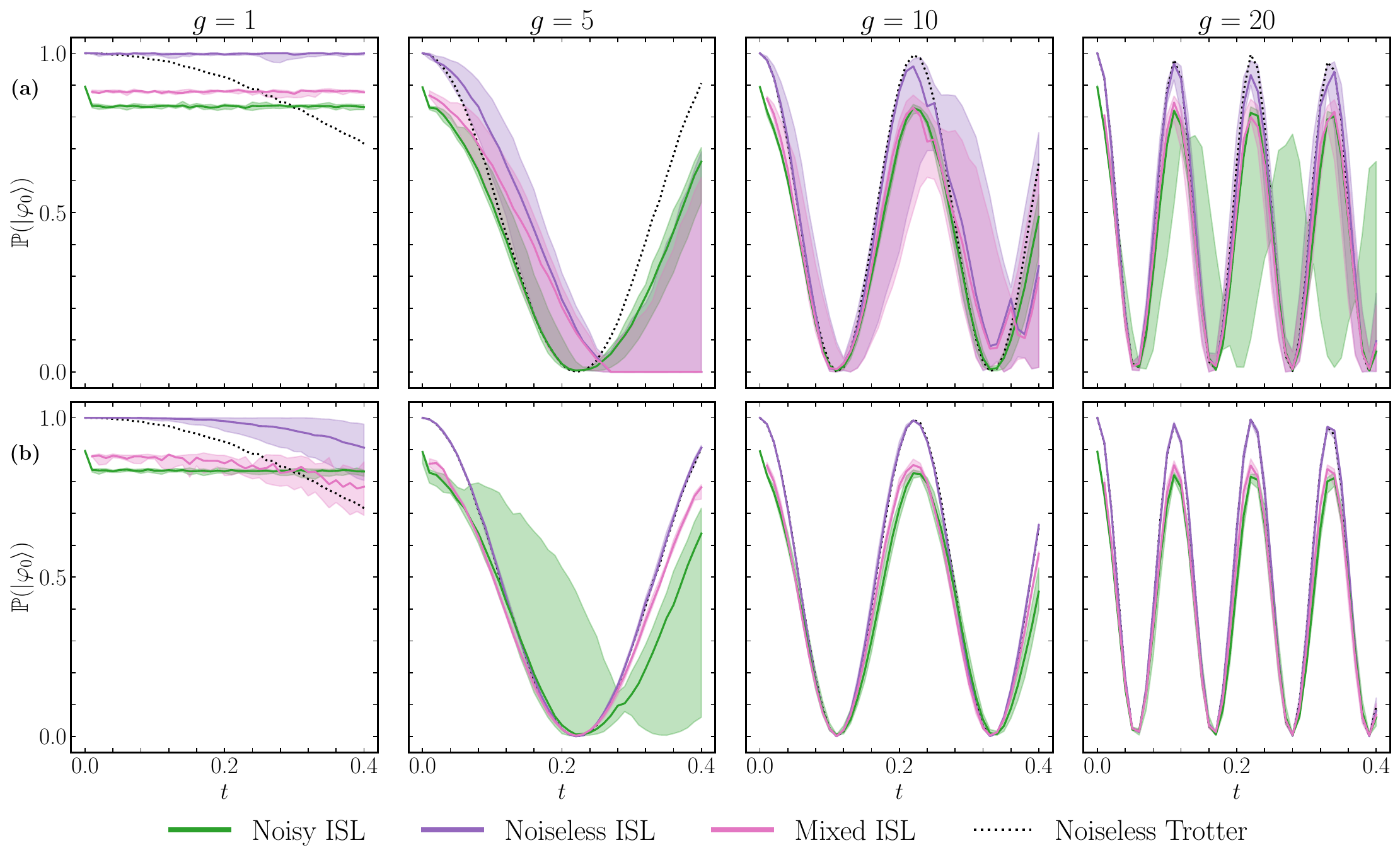}
    \caption{Repeated evaluations of ISL time evolution for all tested TCM systems, for $N_q=3$. The methodology remains the same as explained in Section~\ref{subsubsec:ISL-variation}, with noisy (green), mixed (pink), and noiseless (purple) ISL time evolution being considered for both cost thresholds (a) $C_{-2}$ and (b) $C_{-4}$.}
    \label{fig:TCM-ISL-reps-3_qubits}
\end{figure}

\begin{figure}[htb]
    \centering
    \includegraphics[width=0.95\linewidth]{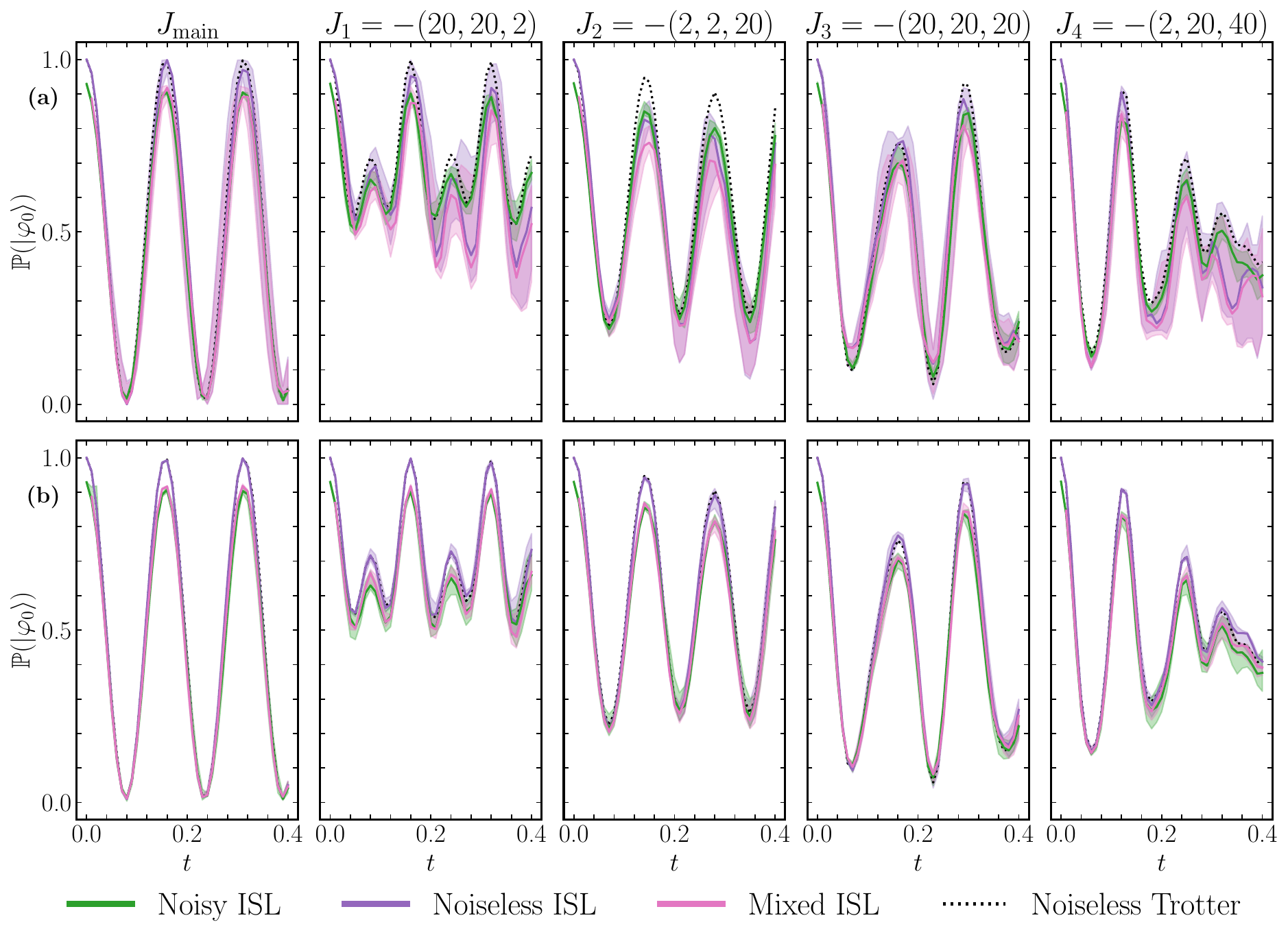}
    \caption{Repeated evaluations of ISL time evolution for all tested HSC systems, for $N_q=2$. The methodology remains the same as explained in Section~\ref{subsubsec:ISL-variation}, with noisy (green), mixed (pink), and noiseless (purple) ISL time evolution being considered for both cost thresholds (a) $C_{-2}$ and (b) $C_{-4}$.}
    \label{fig:HSC-ISL-reps-2_qubits}
\end{figure}

\begin{figure}[htb]
    \centering
    \includegraphics[width=0.95\linewidth]{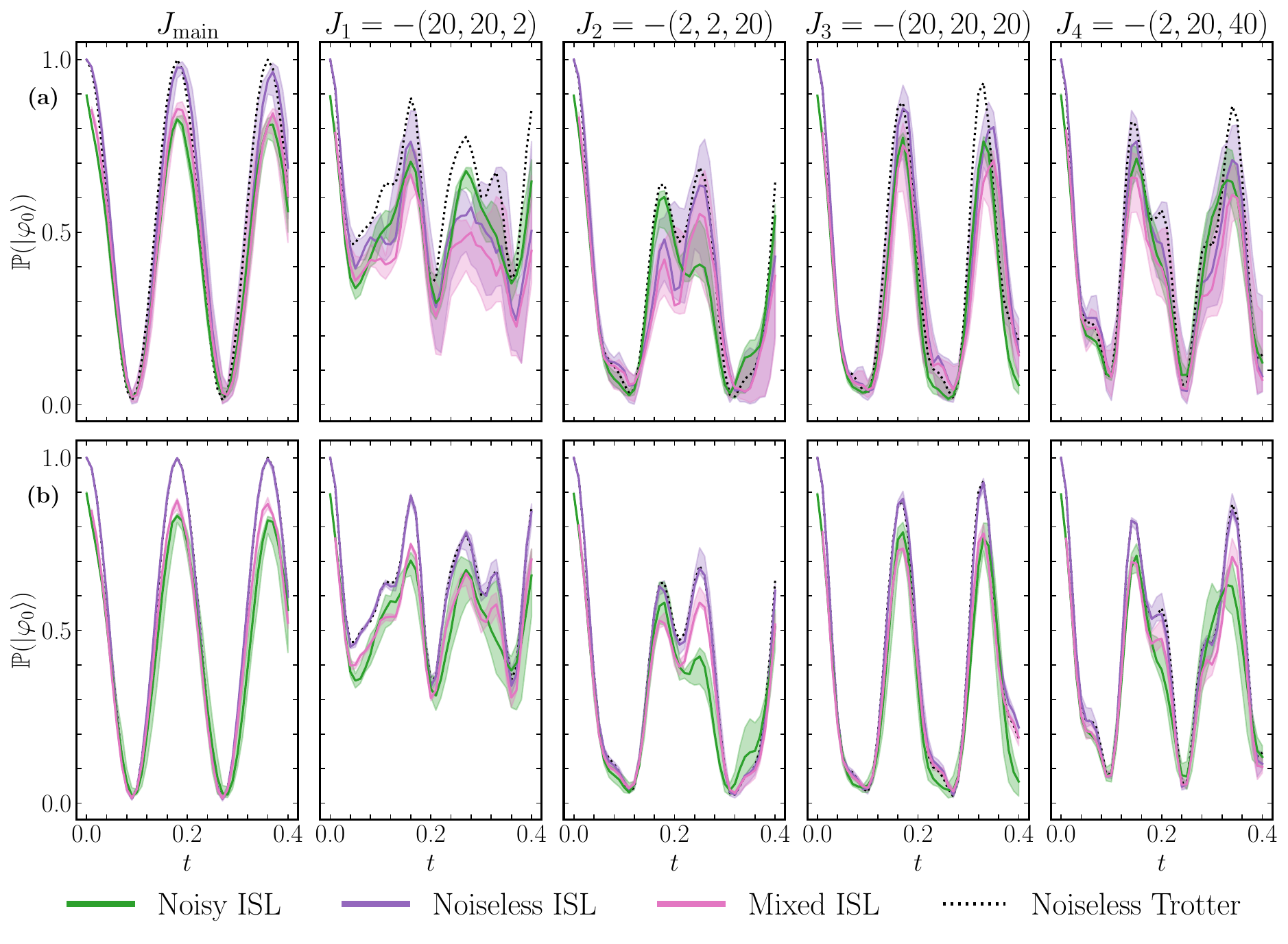}
    \caption{Repeated evaluations of ISL time evolution for all tested HSC systems, for $N_q=3$. The methodology remains the same as explained in Section~\ref{subsubsec:ISL-variation}, with noisy (green), mixed (pink), and noiseless (purple) ISL time evolution being considered for both cost thresholds (a) $C_{-2}$ and (b) $C_{-4}$.}
    \label{fig:HSC-ISL-reps-3_qubits}
\end{figure}

This appendix extends the analysis of  variations in ISL time evolution for the tested systems, as seen in Section~\ref{subsubsec:ISL-variation}. Particularly, Fig.~\ref{fig:TCM-ISL-reps-2_qubits} and Fig.~\ref{fig:TCM-ISL-reps-3_qubits} illustrate variations in the TCM for $N_q=1$ and $N_q=2$, respectively. Similarly, Fig.~\ref{fig:HSC-ISL-reps-2_qubits} and Fig.~\ref{fig:HSC-ISL-reps-3_qubits} illustrate variations in the HSC for $N_q=2$ and $N_q=3$, respectively.

In the case of the TCM, it is generally possible to observe greater stability in ISL time evolution, compared to the case $N_q=4$, as expected, although some variation is still present due to certain outliers. Furthermore, noiseless ISL time evolution still presents significant variations across its time evolutions, particularly for the looser $C_\mathrm{suff}$ threshold, even for the case $N_q=2$. Indeed, for the case $g=5$, noiseless ISL still presents cases where the time evolution plateaus near zero probability, whereas noisy ISL performs under visibly greater accuracy. This result suggests that backend noise may occasionally provide improvements in recompilation accuracy for weakly interacting systems, potentially preventing ansätze from terminating prematurely when shallow circuits already provide sufficiently low cost functions despite not capturing the evolved time dynamics. As such, enabling larger expressibility, which can allow the circuit to obtain closer overlap with the underlying time evolution, assuming proper optimization.

Similar to the TCM, in the case of the HSC, smaller system sizes show reduced variation across separate ISL evaluations. This is most apparent in the case of the main parameterization, where the highest visible variation is obtained for $L=4$. In most systems, all ISL methods provide comparable results, with marginal differences attributed to biases introduced by measurement error, or certain phase-lags under noiseless recompilation, which are corrected by the tighter $C_\mathrm{suff}$ threshold.

%% file: references.bib
@article{cerezo_variational_2021,
	title = {Variational quantum algorithms},
	volume = {3},
	url = {https://doi.org/10.1038/s42254-021-00348-9},
	number = {9},
	journal = {Nature Reviews Physics},
	author = {Cerezo, M. and Arrasmith, Andrew and Babbush, Ryan and Benjamin, Simon C. and Endo, Suguru and Fujii, Keisuke and McClean, Jarrod R. and Mitarai, Kosuke and Yuan, Xiao and Cincio, Lukasz and Coles, Patrick J.},
	year = {2021},
	pages = {625--644},
}

@article{endo_hybrid_2021,
	title = {Hybrid quantum-classical algorithms and quantum error mitigation},
	volume = {90},
	url = {https://journals.jps.jp/doi/10.7566/JPSJ.90.032001},
	number = {3},
	journal = {Journal of the Physical Society of Japan},
	author = {Endo, Suguru and Cai, Zhenyu and Benjamin, Simon C. and Yuan, Xiao},
	year = {2021},
	pages = {032001},
}

@article{preskill_quantum_2018,
	title = {Quantum computing in the {NISQ} era and beyond},
	volume = {2},
	url = {https://doi.org/10.22331/q-2018-08-06-79},
	journal = {Quantum},
	author = {Preskill, John},
	year = {2018},
	pages = {79},
}

@book{nielsen_quantum_2010,
	edition = {10th anniversary},
	title = {{Quantum Computation and Quantum Information}},
	isbn = {978-1-107-00217-3},
	publisher = {Cambridge University Press},
	author = {Nielsen, Michael A. and Chuang, Isaac L.},
	year = {2010},
}

@article{georgescu_quantum_2014,
	title = {Quantum simulation},
	volume = {86},
	url = {https://link.aps.org/doi/10.1103/RevModPhys.86.153},
	number = {1},
	journal = {Reviews of Modern Physics},
	author = {Georgescu, I. M. and Ashhab, S. and Nori, Franco},
	year = {2014},
	pages = {153--185},
}

@article{li_efficient_2017,
	title = {Efficient variational quantum simulator incorporating active error minimization},
	volume = {7},
	url = {https://link.aps.org/doi/10.1103/PhysRevX.7.021050},
	number = {2},
	journal = {Physical Review X},
	author = {Li, Ying and Benjamin, Simon C.},
	year = {2017},
	pages = {021050},
}

@article{di_paolo_variational_2020,
	title = {Variational quantum simulation of ultrastrong light-matter coupling},
	volume = {2},
	url = {https://link.aps.org/doi/10.1103/PhysRevResearch.2.033364},
	number = {3},
	journal = {Physical Review Research},
	author = {Di Paolo, Agustin and Barkoutsos, Panagiotis Kl. and Tavernelli, Ivano and Blais, Alexandre},
	year = {2020},
	pages = {033364},
}

@article{miessen_quantum_2021,
	title = {Quantum algorithms for quantum dynamics: A performance study on the spin-boson model},
	volume = {3},
	url = {https://link.aps.org/doi/10.1103/PhysRevResearch.3.043212},
	number = {4},
	journal = {Physical Review Research},
	author = {Miessen, Alexander and Ollitrault, Pauline J. and Tavernelli, Ivano},
	year = {2021},
	pages = {043212},
}

@misc{fitzpatrick_evaluating_2021,
      title={Evaluating low-depth quantum algorithms for time evolution on fermion-boson systems}, 
      author={Nathan Fitzpatrick and Harriet Apel and David Muñoz Ramo},
      year={2021},
      eprint={2106.03985},
      archivePrefix={arXiv},
      primaryClass={quant-ph},
      url={https://arxiv.org/abs/2106.03985}, 
}

@article{sawaya_resource-efficient_2020,
	title = {Resource-efficient digital quantum simulation of d-level systems for photonic, vibrational, and spin-s {Hamiltonians}},
	volume = {6},
	url = {https://www.nature.com/articles/s41534-020-0278-0},
	number = {1},
	journal = {npj Quantum Information},
	author = {Sawaya, Nicolas P. D. and Menke, Tim and Kyaw, Thi Ha and Johri, Sonika and Aspuru-Guzik, Alán and Guerreschi, Gian Giacomo},
	year = {2020},
	pages = {1--13},
}

@article{bassman_simulating_2021,
	title = {Simulating quantum materials with digital quantum computers},
	volume = {6},
	url = {https://dx.doi.org/10.1088/2058-9565/ac1ca6},
	number = {4},
	journal = {Quantum Science and Technology},
	author = {Bassman, Lindsay and Urbanek, Miroslav and Metcalf, Mekena and Carter, Jonathan and Kemper, Alexander F. and Jong, Wibe A. de},
	year = {2021},
	pages = {043002},
}

@article{bauer_quantum_2020,
	title = {Quantum algorithms for quantum chemistry and quantum materials science},
	volume = {120},
	url = {https://pubs.acs.org/doi/10.1021/acs.chemrev.9b00829},
	number = {22},
	journal = {Chemical Reviews},
	author = {Bauer, Bela and Bravyi, Sergey and Motta, Mario and Chan, Garnet Kin-Lic},
	year = {2020},
	pages = {12685--12717},
}

@article{steudtner_fermion--qubit_2018,
	title = {Fermion-to-qubit mappings with varying resource requirements for quantum simulation},
	volume = {20},
	url = {https://dx.doi.org/10.1088/1367-2630/aac54f},
	number = {6},
	journal = {New Journal of Physics},
	author = {Steudtner, Mark and Wehner, Stephanie},
	year = {2018},
	pages = {063010},
}

@article{yuan_theory_2019,
	title = {Theory of variational quantum simulation},
	volume = {3},
	url = {https://quantum-journal.org/papers/q-2019-10-07-191/},
	journal = {Quantum},
	author = {Yuan, Xiao and Endo, Suguru and Zhao, Qi and Li, Ying and Benjamin, Simon C.},
	year = {2019},
	pages = {191},
}

@book{larson_jaynescummings_2021,
author = {Larson, Jonas and Mavrogordatos, Themistoklis},
title = {{The Jaynes–Cummings Model and Its Descendants}},
publisher = {IOP Publishing},
year = {2021},
series = {2053-2563},
isbn = {978-0-7503-3447-1},
url = {https://doi.org/10.1088/978-0-7503-3447-1},
doi = {10.1088/978-0-7503-3447-1}
}

@article{li_efficient_2023,
	title = {Efficient quantum simulation of electron-phonon systems by variational basis state encoder},
	volume = {5},
	url = {https://doi.org/10.1103/PhysRevResearch.5.023046},
	number = {2},
	journal = {Physical Review Research},
	author = {Li, Weitang and Ren, Jiajun and Huai, Sainan and Cai, Tianqi and Shuai, Zhigang and Zhang, Shengyu},
	year = {2023},
	pages = {023046},
}

@article{bharti_noisy_2022,
	title = {Noisy intermediate-scale quantum ({NISQ}) algorithms},
	volume = {94},
	url = {https://doi.org/10.1103/RevModPhys.94.015004},
	number = {1},
	journal = {Reviews of Modern Physics},
	author = {Bharti, Kishor and Cervera-Lierta, Alba and Kyaw, Thi Ha and Haug, Tobias and Alperin-Lea, Sumner and Anand, Abhinav and Degroote, Matthias and Heimonen, Hermanni and Kottmann, Jakob S. and Menke, Tim and Mok, Wai-Keong and Sim, Sukin and Kwek, Leong-Chuan and Aspuru-Guzik, Alán},
	year = {2022},
	pages = {015004},
}

@article{jaderberg_recompilation-enhanced_2022,
	title = {Recompilation-enhanced simulation of electron-phonon dynamics on {IBM} quantum computers},
	volume = {24},
	doi = {10.1088/1367-2630/ac8a69},
	number = {9},
	journal = {New Journal of Physics},
	author = {Jaderberg, Ben and Eisfeld, Alexander and Jaksch, Dieter and Mostame, Sarah},
	year = {2022},
	pages = {093017},
}

@article{jaderberg_minimum_2020,
	title = {Minimum hardware requirements for hybrid quantum–classical {DMFT}},
	volume = {5},
	issn = {2058-9565},
	url = {https://dx.doi.org/10.1088/2058-9565/ab972b},
	doi = {10.1088/2058-9565/ab972b},
	abstract = {We numerically emulate noisy intermediate-scale quantum (NISQ) devices and determine the minimal hardware requirements for two-site hybrid quantum–classical dynamical mean-field theory (DMFT). We develop a circuit recompilation algorithm which significantly reduces the number of quantum gates of the DMFT algorithm and find that the quantum–classical algorithm converges if the two-qubit gate fidelities are larger than 99\%. The converged results agree with the exact solution within 10\%, and perfect agreement within noise-induced error margins can be obtained for two-qubit gate fidelities exceeding 99.9\%. By comparison, the quantum–classical algorithm without circuit recompilation requires a two-qubit gate fidelity of at least 99.999\% to achieve perfect agreement with the exact solution. We thus find quantum–classical DMFT calculations can be run on the next generation of NISQ devices if combined with the recompilation techniques developed in this work.},
	-language = {en},
	number = {3},
	urldate = {2023-07-14},
	journal = {Quantum Science and Technology},
       author={Jaderberg, B and Agarwal, A and Leonhardt, K and Kiffner, M and Jaksch, D},
	-month = jun,
	year = {2020},
	-note = {Publisher: IOP Publishing},
	keywords = {VQAs, ISL},
	pages = {034015},

}

@article{kim_evidence_2023,
	title = {Evidence for the utility of quantum computing before fault tolerance},
	volume = {618},
	copyright = {2023 The Author(s)},
	issn = {1476-4687},
	url = {https://www.nature.com/articles/s41586-023-06096-3},
	doi = {10.1038/s41586-023-06096-3},
	abstract = {Quantum computing promises to offer substantial speed-ups over its classical counterpart for certain problems. However, the greatest impediment to realizing its full potential is noise that is inherent to these systems. The widely accepted solution to this challenge is the implementation of fault-tolerant quantum circuits, which is out of reach for current processors. Here we report experiments on a noisy 127-qubit processor and demonstrate the measurement of accurate expectation values for circuit volumes at a scale beyond brute-force classical computation. We argue that this represents evidence for the utility of quantum computing in a pre-fault-tolerant era. These experimental results are enabled by advances in the coherence and calibration of a superconducting processor at this scale and the ability to characterize1 and controllably manipulate noise across such a large device. We establish the accuracy of the measured expectation values by comparing them with the output of exactly verifiable circuits. In the regime of strong entanglement, the quantum computer provides correct results for which leading classical approximations such as pure-state-based 1D (matrix product states, MPS) and 2D (isometric tensor network states, isoTNS) tensor network methods2,3 break down. These experiments demonstrate a foundational tool for the realization of near-term quantum applications4,5.},
	-language = {en},
	number = {7965},
	urldate = {2023-09-15},
	journal = {Nature},
	author = {Kim, Youngseok and Eddins, Andrew and Anand, Sajant and Wei, Ken Xuan and van den Berg, Ewout and Rosenblatt, Sami and Nayfeh, Hasan and Wu, Yantao and Zaletel, Michael and Temme, Kristan and Kandala, Abhinav},
	-month = jun,
	year = {2023},
	-note = {Number: 7965
Publisher: Nature Publishing Group},
	keywords = {error mitigation, Quantum information, Information technology, Quantum simulation},
	pages = {500--505},

}

@article{lloyd_universal_1996,
	title = {Universal Quantum Simulators},
	volume = {273},
	url = {https¨://www.science.org/doi/10.1126/science.273.5278.1073},
	number = {5278},
	journal = {Science},
	author = {Lloyd, Seth},
	year = {1996},
	pages = {1073--1078},
}

@article{feynman_simulating_1982,
	title = {Simulating physics with computers},
	volume = {21},
	issn = {1572-9575},
	url = {https://doi.org/10.1007/BF02650179},
	doi = {10.1007/BF02650179},
	-language = {en},
	number = {6},
	urldate = {2023-10-02},
	journal = {International Journal of Theoretical Physics},
	author = {Feynman, Richard P.},
	-month = jun,
	year = {1982},
	keywords = {Calcite, Cellular Automaton, Quantum Mechanic, Quantum System, Wigner Function},
	pages = {467--488},

}

@article{trotter_product_1959,
	title = {On the Product of Semi-Groups of Operators},
	volume = {10},
	url = {https://www.jstor.org/stable/2033649},
	number = {4},
	journal = {Proceedings of the American Mathematical Society},
	author = {Trotter, H. F.},
	year = {1959},
	pages = {545--551},
}

@article{cirstoiu_variational_2020,
	title = {Variational fast forwarding for quantum simulation beyond the coherence time},
	volume = {6},
	issn = {2056-6387},
	url = {https://www.nature.com/articles/s41534-020-00302-0},
	doi = {10.1038/s41534-020-00302-0},
	number = {1},
	journal = {npj Quantum Information},
	author = {Cîrstoiu, Cristina and Holmes, Zoë and Iosue, Joseph and Cincio, Lukasz and Coles, Patrick J. and Sornborger, Andrew},
	year = {2020},
	pages = {1--10},
}

@article{temme_error_2017,
	title = {Error mitigation for short-depth quantum circuits},
	volume = {119},
	url = {https://link.aps.org/doi/10.1103/PhysRevLett.119.180509},
	number = {18},
	journal = {Physical Review Letters},
	author = {Temme, Kristan and Bravyi, Sergey and Gambetta, Jay M.},
	year = {2017},
	pages = {180509},
}

@article{ostaszewski_structure_2021,
	title = {Structure optimization for parameterized quantum circuits},
	volume = {5},
	url = {https://quantum-journal.org/papers/q-2021-01-28-391/},
	doi = {10.22331/q-2021-01-28-391},
	abstract = {Mateusz Ostaszewski, Edward Grant, and Marcello Benedetti,
Quantum 5, 391 (2021).
We propose an efficient method for simultaneously optimizing both the structure and parameter values of quantum circuits with only a small computational overhead. Shallow circuits that use s…},
	-language = {en-GB},
	urldate = {2023-10-14},
	journal = {Quantum},
	author = {Ostaszewski, Mateusz and Grant, Edward and Benedetti, Marcello},
	-month = jan,
	year = {2021},
	-note = {Publisher: Verein zur Förderung des Open Access Publizierens in den Quantenwissenschaften},
	pages = {391},

}

@misc{Qiskit,
author = {{Qiskit Development Team}},
  title        = {Qiskit: An Open-Source Framework for Quantum Computing},
  year         = {2024},
  url          = {https://qiskit.org/},
  urldate         = {2025-03-01}
}

@article{wootters_entanglement_1998,
	title = {Entanglement of formation of an arbitrary state of two qubits},
	volume = {80},
	url = {https://link.aps.org/doi/10.1103/PhysRevLett.80.2245},
	number = {10},
	journal = {Physical Review Letters},
	author = {Wootters, William K.},
	year = {1998},
	pages = {2245--2248},
}

@misc{jaderberg_quantum_2022,
      title={Quantum self-supervised learning}, 
      author={Ben Jaderberg and Lewis W. Anderson and Weidi Xie and Samuel Albanie and Martin Kiffner and Dieter Jaksch},
      year={2022},
      eprint={2103.14653},
      archivePrefix={arXiv},
      primaryClass={quant-ph},
      url={https://arxiv.org/abs/2103.14653}, 
}

@inproceedings{giurgica-tiron_digital_2020,
	title = {Digital zero noise extrapolation for quantum error mitigation},
	doi = {10.1109/QCE49297.2020.00045},
	booktitle = {2020 {IEEE} {International} {Conference} on {Quantum} {Computing} and {Engineering} ({QCE})},
	author = {Giurgica-Tiron, Tudor and Hindy, Yousef and LaRose, Ryan and Mari, Andrea and Zeng, William J.},
	year = {2020},
	pages = {306--316},
}

@article{mitiq,
  title     = {Mitiq: A software package for error mitigation on noisy quantum computers},
  author    = {Ryan LaRose and Andrea Mari and Sarah Kaiser and Peter J. Karalekas and Andre A. Alves and Piotr Czarnik and Mohamed El Mandouh and Max H. Gordon and Yousef Hindy and Aaron Robertson and Purva Thakre and Misty Wahl and Danny Samuel and Rahul Mistri and Maxime Tremblay and Nick Gardner and Nathaniel T. Stemen and Nathan Shammah and William J. Zeng},
  journal   = {Quantum},
  year      = {2022},
  -month     = {Aug},
  doi       = {10.22331/q-2022-08-11-774},
  url       = {https://doi.org/10.22331/q-2022-08-11-774},
  publisher = {Verein zur Forderung des Open Access Publizierens in den Quantenwissenschaften},
  volume    = {6},
  pages     = {774},
}

@article{endo_practical_2018,
	title = {Practical quantum error mitigation for near-future applications},
	volume = {8},
	url = {https://link.aps.org/doi/10.1103/PhysRevX.8.031027},
	number = {3},
	journal = {Physical Review X},
	author = {Endo, Suguru and Benjamin, Simon C. and Li, Ying},
	year = {2018},
	pages = {031027},
}

@article{jaynes_comparison_1963,
	title = {Comparison of quantum and semiclassical radiation theories with application to the beam maser},
	volume = {51},
	issn = {1558-2256},
	url = {https://ieeexplore.ieee.org/document/1443594},
	doi = {10.1109/PROC.1963.1664},
	abstract = {This paper has two purposes: 1) to clarify the relationship between the quantum theory of radiation, where the electromagnetic field-expansion coefficients satisfy commutation relations, and the semiclassical theory, where the electromagnetic field is considered as a definite function of time rather than as an operator; and 2) to apply some of the results in a study of amplitude and frequency stability in a molecular beam maser. In 1), it is shown that the semiclassical theory, when extended te take into account both the effect of the field on the molecules and the effect of the molecules on the field, reproduces almost quantitatively the same laws of energy exchange and coherence properties as the quantized field theory, even in the limit of one or a few quanta in the field mode. In particular, the semiclassical theory is shown to lead to a prediction of spontaneous emission, with the same decay rate as given by quantum electrodynamics, described by the Einstein A coefficients. In 2), the semiclassical theory is applied to the molecular beam maser. Equilibrium amplitude and frequency of oscillation are obtained for an arbitrary velocity distribution of focused molecules, generalizing the results obtained previously by Gordon, Zeiger, and Townes for a singel-velocity beam, and by Lamb and Helmer for a Maxwellian beam. A somewhat surprising result is obtained; which is that the measurable properties of the maser, such as starting current, effective molecular Q, etc., depend mostly on the slowest 5 to 10 per cent of the molecules. Next we calculate the effect of amplitude and frequency of oscillation, of small systematic perturbations. We obtain a prediction that stability can be improved by adjusting the system so that the molecules emit all their energy h Ω to the field, then reabsorb part of it, before leaving the cavity. In general, the most stable operation is obtained when the molecules are in the process of absorbing energy from the radiation as they leave the cavity, most unstable when they are still emitting energy at that time. Finally, we consider the response of an oscillating maser to randomly time-varying perturbations. Graphs are given showing predicted response to a small superimposed signal of a frequency near the oscillation frequency. The existence of "noise enhancing" and "noise quieting" modes of operation found here is a general property of any oscillating system in which amplitude is limited by nonlinearity.},
	number = {1},
	urldate = {2023-10-17},
	journal = {Proceedings of the IEEE},
	author = {Jaynes, E.T. and Cummings, F.W.},
	-month = jan,
	year = {1963},
	-note = {Conference Name: Proceedings of the IEEE},
	pages = {89--109},

}

@article{tavis_exact_1968,
	title = {Exact solution for an {$N$}-molecule---radiation-field {Hamiltonian}},
	volume = {170},
	url = {https://link.aps.org/doi/10.1103/PhysRev.170.379},
	doi = {10.1103/PhysRev.170.379},
	number = {2},
	journal = {Physical Review},
	author = {Tavis, Michael and Cummings, Frederick W.},
	year = {1968},
	pages = {379--384},
}

@misc{ibm_quantum_2023,
    author = {{IBM Quantum Platform}},
    year = {2023},
    howpublished = {\url{https://quantum-computing.ibm.com/}},
    urldate = {2023-11-02},
}

@misc{ibm_quantum_transpiler,
    author = {{IBM Quantum Documentation}},
    year = {2024},
    howpublished = {\url{https://docs.quantum.ibm.com/api/qiskit/transpiler}},
    urldate = {2024-09-09}
}

@article{byrnes_simulating_2006,
	title = {Simulating lattice gauge theories on a quantum computer},
	volume = {73},
	url = {https://link.aps.org/doi/10.1103/PhysRevA.73.022328},
	doi = {10.1103/PhysRevA.73.022328},
	abstract = {We examine the problem of simulating lattice gauge theories on a universal quantum computer. The basic strategy of our approach is to transcribe lattice gauge theories in the Hamiltonian formulation into a Hamiltonian involving only Pauli spin operators such that the simulation can be performed on a quantum computer using only one- and two-qubit manipulations. We examine three models, the U(1), SU(2), and SU(3) lattice gauge theories, which are transcribed into a spin Hamiltonian up to a cutoff in the Hilbert space of the gauge fields on the lattice. The number of qubits required for storing a particular state is found to have a linear dependence on the total number of lattice sites. The number of qubit operations required for performing the time evolution corresponding to the Hamiltonian is found to be between a linear to quadratic function of the number of lattice sites, depending on the arrangement of qubits in the quantum computer. We remark that our results may also be easily generalized to higher SU(N) gauge theories.},
	number = {2},
	urldate = {2023-11-13},
	journal = {Physical Review A},
	author = {Byrnes, Tim and Yamamoto, Yoshihisa},
	-month = feb,
	year = {2006},
	-note = {Publisher: American Physical Society},
	pages = {022328},

}

@misc{ibm_documentation_2023,
    author = {{IBM Quantum Documentation}},
    title = {Retired systems},
    year = {2023},
    howpublished = {\url{https://docs.quantum.ibm.com/run/retired-systems/}},
    urldate = {2023-12-03}
}

@article{van_den_berg_probabilistic_2023,
	title = {Probabilistic error cancellation with sparse {Pauli}–{Lindblad} models on noisy quantum processors},
	volume = {19},
	copyright = {2023 The Author(s), under exclusive licence to Springer Nature Limited},
	issn = {1745-2481},
	url = {https://www.nature.com/articles/s41567-023-02042-2},
	doi = {10.1038/s41567-023-02042-2},
	-language = {en},
	number = {8},
	urldate = {2023-12-06},
	journal = {Nature Physics},
	author = {van den Berg, Ewout and Minev, Zlatko K. and Kandala, Abhinav and Temme, Kristan},
	-month = aug,
	year = {2023},
	-note = {Number: 8
Publisher: Nature Publishing Group},
	keywords = {Quantum information, Quantum physics},
	pages = {1116--1121},
}

@incollection{altepeter_4_2004,
	series = {Lecture {Notes} in {Physics}},
	title = {Qubit quantum state tomography},
	isbn = {978-3-540-44481-7},
	url = {https://doi.org/10.1007/978-3-540-44481-7_4},
	booktitle = {Quantum {State} {Estimation}},
	publisher = {Springer},
	author = {Altepeter, Joseph B. and James, Daniel F.V. and Kwiat, Paul G.},
	editor = {Paris, Matteo and Řeháček, Jaroslav},
	year = {2004},
	doi = {10.1007/978-3-540-44481-7_4},
	pages = {113--145},
}

@misc{ahaukis,
  author = {Haukisalmi, A. and Paz, D.},
  title = {Spin-boson simulation},
  year = {2025},
  publisher = {GitHub},
  howpublished = {GitHub repository, \url{https://github.com/Dpazramos/spin-boson-sim}},
}

@misc{islcode,
   author={Jaderberg, B and Agarwal, A and Leonhardt, K and Kiffner, M and Jaksch, D and Haukisalmi, A and Paz, D},
    title = {{ISL}},
    year = {2025},
    publisher = {GitHub},
    howpublished = {GitHub repository, \url{https://github.com/Dpazramos/isl/}},
    urldate = {2025-05-08}
}

@article{fontana_non-trivial_2022,
	title = {Non-trivial symmetries in quantum landscapes and their resilience to quantum noise},
	volume = {6},
	url = {https://quantum-journal.org/papers/q-2022-09-15-804/},
	journal = {Quantum},
	author = {Fontana, Enrico and Cerezo, M. and Arrasmith, Andrew and Rungger, Ivan and Coles, Patrick J.},
	year = {2022},
	pages = {804},
}

@article{heisenberg_model_1928,
  author = {Heisenberg, W.},
  title = {Zur Theorie des Ferromagnetismus},
  journal = {Zeitschrift für Physik},
  year = {1928},
  volume = {49},
  number = {9},
  pages = {619--636},
  doi = {10.1007/BF01328601},
  bibcode = {1928ZPhy...49..619H},
  s2cid = {122524239},
  language = {German}
}

@misc{candu2013spin_chains,
  author = {Candu, Constantin and de Leeuw, Marius},
  title = {Spin Chains, Chapter in Introduction to Integrability – FS 2013},
  howpublished = {Lecture notes, ETH Zurich},
  year = {2013},
  note = {Available at: https://edu.itp.phys.ethz.ch/fs13/int/},
}

@misc{jiang2022,
  author       = {Yunfeng Jiang},
  title        = {Heisenberg Spin Chain and Bethe Ansatz},
  note         = {Lecture notes from *An Introduction to Quantum Integrable Systems*, Chinese Academy of Sciences, Fall 2022},
  howpublished = {\url{https://www.koushare.com/courseTopicInfo/i/AIQIS/0}},
  year         = {2022}
}

@misc{triton_cluster,
author = {{Aalto Scientific Computing}},
title = {Triton - Aalto University Computing Cluster},
year = {2024},
url = {https://scicomp.aalto.fi/triton/},
note = {Accessed: 2025-03-30},

}

@article{Cai_2023,
   title={Quantum error mitigation},
   volume={95},
   url={http://dx.doi.org/10.1103/RevModPhys.95.045005},
   number={4},
   journal={Reviews of Modern Physics},
   author={Cai, Zhenyu and Babbush, Ryan and Benjamin, Simon C. and Endo, Suguru and Huggins, William J. and Li, Ying and McClean, Jarrod R. and O’Brien, Thomas E.},
   year={2023},
}

@ARTICLE{Shafique_2024,
  author={Shafique, Muhammad Ali and Munir, Arslan and Latif, Imran},
  journal={IEEE Access}, 
  title={Quantum computing: circuits, algorithms, and applications}, 
  year={2024},
  volume={12},
  pages={22296-22314},
}

@article{Montanaro_2016,
   title={Quantum algorithms: an overview},
   volume={2},
   url={http://dx.doi.org/10.1038/npjqi.2015.23},
   number={1},
   journal={npj Quantum Information},
   publisher={Springer Science and Business Media LLC},
   author={Montanaro, Ashley},
   year={2016},}

@INPROCEEDINGS{Ramezani_2020,
  author={Ramezani, Somayeh Bakhtiari and Sommers, Alexander and Manchukonda, Harish Kumar and Rahimi, Shahram and Amirlatifi, Amin},
  booktitle={2020 International Joint Conference on Neural Networks (IJCNN)}, 
  title={Machine learning algorithms in quantum computing: a survey}, 
  year={2020},
  pages={1-8},
}

@article{abbas_2024,
  author = {Amira Abbas and Andris Ambainis and Brandon Augustino and Andreas Bärtschi and Harry Buhrman and Carleton Coffrin and Giorgio Cortiana and Vedran Dunjko and Daniel J. Egger and Bruce G. Elmegreen and Nicola Franco and Filippo Fratini and Bryce Fuller and Julien Gacon and Constantin Gonciulea and Sander Gribling and Swati Gupta and Stuart Hadfield and Raoul Heese and Gerhard Kircher and Thomas Kleinert and Thorsten Koch and Georgios Korpas and Steve Lenk and Jakub Marecek and Vanio Markov and Guglielmo Mazzola and Stefano Mensa and Naeimeh Mohseni and Giacomo Nannicini and Corey O’Meara and Elena Peña Tapia and Sebastian Pokutta and Manuel Proissl and Patrick Rebentrost and Emre Sahin and Benjamin C. B. Symons and Sabine Tornow and Víctor Valls and Stefan Woerner and Mira L. Wolf-Bauwens and Jon Yard and Sheir Yarkoni and Dirk Zechiel and Sergiy Zhuk and Christa Zoufal},
  title = {Challenges and opportunities in quantum optimization},
  journal = {Nature Reviews Physics},
  volume = {6},
  pages = {718--735},
  year = {2024},
  doi = {10.1038/s42254-024-00770-9},
  url = {https://doi.org/10.1038/s42254-024-00770-9},
}

@article{Smith_2019,
  author       = {Adam Smith and M. S. Kim and Frank Pollmann and Johannes Knolle},
  title        = {Simulating quantum many-body dynamics on a current digital quantum computer},
  journal      = {npj Quantum Information},
  year         = {2019},
  volume       = {5},
  pages        = {106},
  doi          = {10.1038/s41534-019-0217-0},
  url          = {https://doi.org/10.1038/s41534-019-0217-0},
  publisher    = {Nature Publishing Group},
  received     = {2019-06-14},
  accepted     = {2019-10-01},
  published    = {2019-11-28}
}

@BOOK{NAP_2019,
  author    = "National Academies of Sciences, Engineering, and Medicine",
  editor    = "Emily Grumbling and Mark Horowitz",
  title     = "Quantum Computing: Progress and Prospects",
  isbn      = "978-0-309-47969-1",
  doi       = "10.17226/25196",
  url       = "https://nap.nationalacademies.org/catalog/25196/quantum-computing-progress-and-prospects",
  year      = 2019,
  publisher = "The National Academies Press",
  address   = "Washington, DC"
}

@article{Alexeev_2021,
  title = {Quantum Computer Systems for Scientific Discovery},
  author = {Alexeev, Yuri and Bacon, Dave and Brown, Kenneth R. and Calderbank, Robert and Carr, Lincoln D. and Chong, Frederic T. and DeMarco, Brian and Englund, Dirk and Farhi, Edward and Fefferman, Bill and Gorshkov, Alexey V. and Houck, Andrew and Kim, Jungsang and Kimmel, Shelby and Lange, Michael and Lloyd, Seth and Lukin, Mikhail D. and Maslov, Dmitri and Maunz, Peter and Monroe, Christopher and Preskill, John and Roetteler, Martin and Savage, Martin J. and Thompson, Jeff},
  journal = {PRX Quantum},
  volume = {2},
  issue = {1},
  pages = {017001},
  numpages = {19},
  year = {2021},
  month = {Feb},
  publisher = {American Physical Society},
  doi = {10.1103/PRXQuantum.2.017001},
  url = {https://link.aps.org/doi/10.1103/PRXQuantum.2.017001}
}

@ARTICLE{Corcoles_2020,
  author={Córcoles, Antonio D. and Kandala, Abhinav and Javadi-Abhari, Ali and McClure, Douglas T. and Cross, Andrew W. and Temme, Kristan and Nation, Paul D. and Steffen, Matthias and Gambetta, Jay M.},
  journal={Proceedings of the IEEE}, 
  title={Challenges and opportunities of near-term quantum computing systems}, 
  year={2020},
  volume={108},
  number={8},
  pages={1338-1352},
}

@phdthesis{Kolotouros_2024,
  author       = {Ioannis Kolotouros},
  title        = {Designing and Improving Quantum Algorithms for the NISQ Era},
  school       = {University of Edinburgh},
  year         = {2024},
  type         = {PhD dissertation},
  doi          = {10.7488/era/5434},
  url          = {https://hdl.handle.net/1842/42877}
}

@INPROCEEDINGS{Khanal_2023,
  author={Khanal, Bikram and Rivas, Pablo},
  booktitle={2023 Congress in Computer Science, Computer Engineering, \& Applied Computing (CSCE)}, 
  title={Evaluating the impact of noise on variational quantum circuits in NISQ era devices}, 
  year={2023},
  pages={1658-1664},
  doi={10.1109/CSCE60160.2023.00272},
}

@article{Burdine_2024,
  author       = {Colin Burdine and Enrique P. Blair},
  title        = {Trotterless simulation of open quantum systems for NISQ quantum devices},
  journal      = {Advanced Quantum Technologies},
  year         = {2024},
  volume       = {8},
  number       = {1},
  pages        = {2400240},
  url          = {https://doi.org/10.1002/qute.202400240},
}

@article{Fontana_2021,
   title={Evaluating the noise resilience of variational quantum algorithms},
   volume={104},
   url={http://dx.doi.org/10.1103/PhysRevA.104.022403},
   number={2},
   journal={Physical Review A},
   author={Fontana, Enrico and Fitzpatrick, Nathan and Ramo, David Muñoz and Duncan, Ross and Rungger, Ivan},
   year={2021},
}

@misc{Kluber_2025,
      title={Trotterization in quantum theory}, 
      author={Physics Claire Kluber},
      year={2025},
      eprint={2310.13296},
      archivePrefix={arXiv},
      primaryClass={quant-ph},
      url={https://arxiv.org/abs/2310.13296}, 
}

@article{Yang_2022,
   title={Improved quantum computing with higher-order Trotter decomposition},
   volume={106},
   url={http://dx.doi.org/10.1103/PhysRevA.106.042401},
   number={4},
   journal={Physical Review A},
   author={Yang, Xiaodong and Nie, Xinfang and Ji, Yunlan and Xin, Tao and Lu, Dawei and Li, Jun},
   year={2022},
}

@misc{sarkar_2024,
      title={Scalable quantum circuits for exponential of {P}auli strings and {H}amiltonian simulations}, 
      author={Rohit Sarma Sarkar and Sabyasachi Chakraborty and Bibhas Adhikari},
      year={2024},
      eprint={2405.13605},
      archivePrefix={arXiv},
      primaryClass={quant-ph},
      url={https://arxiv.org/abs/2405.13605}, 
}

@misc{Dion_2024,
      title={Efficiently manipulating Pauli strings with PauliArray}, 
      author={Maxime Dion and Tania Belabbas and Nolan Bastien},
      year={2024},
      eprint={2405.19287},
      archivePrefix={arXiv},
      primaryClass={quant-ph},
      url={https://arxiv.org/abs/2405.19287}, 
}

@article{Zhang_2020,
   title={Error-mitigated quantum gates exceeding physical fidelities in a trapped-ion system},
   volume={11},
   url={http://dx.doi.org/10.1038/s41467-020-14376-z},
   number={1},
   journal={Nature Communications},
   author={Zhang, Shuaining and Lu, Yao and Zhang, Kuan and Chen, Wentao and Li, Ying and Zhang, Jing-Ning and Kim, Kihwan},
   year={2020},
}

@article{Maciejewski_2020,
   title={Mitigation of readout noise in near-term quantum devices by classical post-processing based on detector tomography},
   volume={4},
   ISSN={2521-327X},
   url={http://dx.doi.org/10.22331/q-2020-04-24-257},
   DOI={10.22331/q-2020-04-24-257},
   journal={Quantum},
   publisher={Verein zur Forderung des Open Access Publizierens in den Quantenwissenschaften},
   author={Maciejewski, Filip B. and Zimborás, Zoltán and Oszmaniec, Michał},
   year={2020},
   month=apr, pages={257} }

@article{Bravyi_2021,
   title={Mitigating measurement errors in multiqubit experiments},
   volume={103},
   url={http://dx.doi.org/10.1103/PhysRevA.103.042605},
   number={4},
   journal={Physical Review A},
   author={Bravyi, Sergey and Sheldon, Sarah and Kandala, Abhinav and Mckay, David C. and Gambetta, Jay M.},
   year={2021},
}

@INPROCEEDINGS{Perez_2024,
  author={Pérez-Guijarro, Jordi and Pagès-Zamora, Alba and Fonollosa, Javier R.},
  booktitle={ICASSP 2024 - 2024 IEEE International Conference on Acoustics, Speech and Signal Processing (ICASSP)}, 
  title={Extension of Clifford Data Regression Methods for Quantum Error Mitigation}, 
  year={2024},
  volume={},
  number={},
  pages={9691-9695},
  doi={10.1109/ICASSP48485.2024.10446476}}

@article{Lowe_2021,
   title={Unified approach to data-driven quantum error mitigation},
   volume={3},
   url={http://dx.doi.org/10.1103/PhysRevResearch.3.033098},
   number={3},
   journal={Physical Review Research},
   author={Lowe, Angus and Gordon, Max Hunter and Czarnik, Piotr and Arrasmith, Andrew and Coles, Patrick J. and Cincio, Lukasz},
   year={2021},
}

@article{Czarnik_2021,
   title={Error mitigation with Clifford quantum-circuit data},
   volume={5},
   ISSN={2521-327X},
   url={http://dx.doi.org/10.22331/q-2021-11-26-592},
   DOI={10.22331/q-2021-11-26-592},
   journal={Quantum},
   publisher={Verein zur Forderung des Open Access Publizierens in den Quantenwissenschaften},
   author={Czarnik, Piotr and Arrasmith, Andrew and Coles, Patrick J. and Cincio, Lukasz},
   year={2021},
   month=nov, pages={592} }

@article{Russo_2023,
   title={Testing platform-independent quantum error mitigation on noisy quantum computers},
   volume={4},
   url={http://dx.doi.org/10.1109/TQE.2023.3305232},
   journal={IEEE Transactions on Quantum Engineering},
   author={Russo, Vincent and Mari, Andrea and Shammah, Nathan and LaRose, Ryan and Zeng, William J.},
   year={2023},
}

@article{Song_2019,
   title={Quantum computation with universal error mitigation on a superconducting quantum processor},
   volume={5},
   url={http://dx.doi.org/10.1126/sciadv.aaw5686},
   number={9},
   journal={Science Advances},
   author={Song, Chao and Cui, Jing and Wang, H. and Hao, J. and Feng, H. and Li, Ying},
   year={2019},
}

@article{Moradi_2023,
  author       = {S. Moradi and Clemens Spielvogel and Denis Krajnc and C. Brandner and S. Hillmich and R. Wille and T. Traub-Weidinger and X. Li and M. Hacker and W. Drexler and L. Papp},
  title        = {Error mitigation enables PET radiomic cancer characterization on quantum computers},
  journal      = {European Journal of Nuclear Medicine and Molecular Imaging},
  year         = {2023},
  volume       = {50},
  pages        = {3826--3837},
  url          = {https://doi.org/10.1007/s00259-023-06362-6},
}

@misc{Larose_2022,
      title={Error mitigation increases the effective quantum volume of quantum computers}, 
      author={Ryan LaRose and Andrea Mari and Vincent Russo and Dan Strano and William J. Zeng},
      year={2022},
      eprint={2203.05489},
      archivePrefix={arXiv},
      primaryClass={quant-ph},
      url={https://arxiv.org/abs/2203.05489}, 
}

@article{Takagi_2022,
   title={Fundamental limits of quantum error mitigation},
   volume={8},
   url={http://dx.doi.org/10.1038/s41534-022-00618-z},
   number={1},
   journal={npj Quantum Information},
   author={Takagi, Ryuji and Endo, Suguru and Minagawa, Shintaro and Gu, Mile},
   year={2022},
}

@article{De_Palma_2023,
   title={Limitations of variational quantum algorithms: a quantum optimal transport approach},
   volume={4},
   url={http://dx.doi.org/10.1103/PRXQuantum.4.010309},
   number={1},
   journal={PRX Quantum},
   author={De Palma, Giacomo and Marvian, Milad and Rouzé, Cambyse and França, Daniel Stilck},
   year={2023},
}

@article{Cramer_2010,
   title={Efficient quantum state tomography},
   volume={1},
   url={http://dx.doi.org/10.1038/ncomms1147},
   number={1},
   journal={Nature Communications},
   author={Cramer, Marcus and Plenio, Martin B. and Flammia, Steven T. and Somma, Rolando and Gross, David and Bartlett, Stephen D. and Landon-Cardinal, Olivier and Poulin, David and Liu, Yi-Kai},
   year={2010},
}

@article{Ayral_2021,
   title={Quantum divide and compute: Exploring the effect of different noise sources},
   volume={2},
   url={http://dx.doi.org/10.1007/s42979-021-00508-9},
   number={3},
   journal={SN Computer Science},
   author={Ayral, Thomas and Régent, François-Marie Le and Saleem, Zain and Alexeev, Yuri and Suchara, Martin},
   year={2021},
}

@article{Virtanen_2020,
   title={SciPy 1.0: fundamental algorithms for scientific computing in Python},
   volume={17},
   ISSN={1548-7105},
   url={http://dx.doi.org/10.1038/s41592-019-0686-2},
   DOI={10.1038/s41592-019-0686-2},
   number={3},
   journal={Nature Methods},
   publisher={Springer Science and Business Media LLC},
   author={Virtanen, Pauli and Gommers, Ralf and Oliphant, Travis E. and Haberland, Matt and Reddy, Tyler and Cournapeau, David and Burovski, Evgeni and Peterson, Pearu and Weckesser, Warren and Bright, Jonathan and van der Walt, Stéfan J. and Brett, Matthew and Wilson, Joshua and Millman, K. Jarrod and Mayorov, Nikolay and Nelson, Andrew R. J. and Jones, Eric and Kern, Robert and Larson, Eric and Carey, C J and Polat, İlhan and Feng, Yu and Moore, Eric W. and VanderPlas, Jake and Laxalde, Denis and Perktold, Josef and Cimrman, Robert and Henriksen, Ian and Quintero, E. A. and Harris, Charles R. and Archibald, Anne M. and Ribeiro, Antônio H. and Pedregosa, Fabian and van Mulbregt, Paul and Vijaykumar, Aditya and Bardelli, Alessandro Pietro and Rothberg, Alex and Hilboll, Andreas and Kloeckner, Andreas and Scopatz, Anthony and Lee, Antony and Rokem, Ariel and Woods, C. Nathan and Fulton, Chad and Masson, Charles and Häggström, Christian and Fitzgerald, Clark and Nicholson, David A. and Hagen, David R. and Pasechnik, Dmitrii V. and Olivetti, Emanuele and Martin, Eric and Wieser, Eric and Silva, Fabrice and Lenders, Felix and Wilhelm, Florian and Young, G. and Price, Gavin A. and Ingold, Gert-Ludwig and Allen, Gregory E. and Lee, Gregory R. and Audren, Hervé and Probst, Irvin and Dietrich, Jörg P. and Silterra, Jacob and Webber, James T and Slavič, Janko and Nothman, Joel and Buchner, Johannes and Kulick, Johannes and Schönberger, Johannes L. and de Miranda Cardoso, José Vinícius and Reimer, Joscha and Harrington, Joseph and Rodríguez, Juan Luis Cano and Nunez-Iglesias, Juan and Kuczynski, Justin and Tritz, Kevin and Thoma, Martin and Newville, Matthew and Kümmerer, Matthias and Bolingbroke, Maximilian and Tartre, Michael and Pak, Mikhail and Smith, Nathaniel J. and Nowaczyk, Nikolai and Shebanov, Nikolay and Pavlyk, Oleksandr and Brodtkorb, Per A. and Lee, Perry and McGibbon, Robert T. and Feldbauer, Roman and Lewis, Sam and Tygier, Sam and Sievert, Scott and Vigna, Sebastiano and Peterson, Stefan and More, Surhud and Pudlik, Tadeusz and Oshima, Takuya and Pingel, Thomas J. and Robitaille, Thomas P. and Spura, Thomas and Jones, Thouis R. and Cera, Tim and Leslie, Tim and Zito, Tiziano and Krauss, Tom and Upadhyay, Utkarsh and Halchenko, Yaroslav O. and Vázquez-Baeza, Yoshiki},
   year={2020},
   month=feb, pages={261–272} }

@misc{watanabe_2023,
      title={Optimizing parameterized quantum circuits with free-axis selection}, 
      author={Hiroshi C. Watanabe and Rudy Raymond and Yu-ya Ohnishi and Eriko Kaminishi and Michihiko Sugawara},
      year={2023},
      eprint={2104.14875},
      archivePrefix={arXiv},
      primaryClass={quant-ph},
      url={https://arxiv.org/abs/2104.14875}, 
}

@misc{pankkonen_2025_freezing,
      title={Gate freezing method for gradient-free variational quantum algorithms in circuit optimization}, 
      author={Joona V. Pankkonen and Lauri Ylinen and Matti Raasakka and Andrea Marchesin and Ilkka Tittonen},
      year={2025},
      eprint={2507.07742},
      archivePrefix={arXiv},
      primaryClass={quant-ph},
      url={https://arxiv.org/abs/2507.07742}, 
}

@misc{pankkonen_2025_improving,
      title={Improving variational quantum circuit optimization via hybrid algorithms and random axis initialization}, 
      author={Joona V. Pankkonen and Lauri Ylinen and Matti Raasakka and Ilkka Tittonen},
      year={2025},
      eprint={2503.20728},
      archivePrefix={arXiv},
      primaryClass={quant-ph},
      url={https://arxiv.org/abs/2503.20728}, 
}

@article{Cai_2020,
   title={Mitigating coherent noise using Pauli conjugation},
   volume={6},
   url={http://dx.doi.org/10.1038/s41534-019-0233-0},
   number={1},
   journal={npj Quantum Information},
   publisher={Springer Science and Business Media LLC},
   author={Cai, Zhenyu and Xu, Xiaosi and Benjamin, Simon C.},
   year={2020},
}

@article{Garcia_2020,
  title = {Pairwise tomography networks for many-body quantum systems},
  author = {Garc\'{\i}a-P\'erez, Guillermo and Rossi, Matteo A. C. and Sokolov, Boris and Borrelli, Elsi-Mari and Maniscalco, Sabrina},
  journal = {Phys. Rev. Res.},
  volume = {2},
  issue = {2},
  pages = {023393},
  numpages = {9},
  year = {2020},
  month = {Jun},
  publisher = {American Physical Society},
  doi = {10.1103/PhysRevResearch.2.023393},
  url = {https://link.aps.org/doi/10.1103/PhysRevResearch.2.023393}
}

@misc{filippov2023,
      title={Scalable tensor-network error mitigation for near-term quantum computing}, 
      author={Sergei Filippov and Matea Leahy and Matteo A. C. Rossi and Guillermo García-Pérez},
      year={2023},
      eprint={2307.11740},
      archivePrefix={arXiv},
      primaryClass={quant-ph},
      url={https://arxiv.org/abs/2307.11740}, 
}

@article{Jorge_2022,
    author = {Campos-Gonzalez-Angulo, Jorge A. and Yuen-Zhou, Joel},
    title = {Generalization of the Tavis–Cummings model for multi-level anharmonic systems: Insights on the second excitation manifold},
    journal = {Journal of Chemical Physics},
    volume = {156},
    number = {19},
    pages = {194308},
    year = {2022},
    doi = {10.1063/5.0087234},
    url = {https://doi.org/10.1063/5.0087234},
}
